\pdfoutput=1
\documentclass[pdflatex]{elsarticle}
\usepackage[a4paper, includeheadfoot, margin = 2.5 cm]{geometry}
\usepackage{amsmath, amsfonts, amssymb, hyperref, color, graphicx, xspace}
\usepackage[usenames]{xcolor}
\definecolor{nicered}{rgb}{.7,.1,.1}
\definecolor{nicegreen}{rgb}{.1,.5,.1}
\definecolor{darkblue}{rgb}{0,0,.5}
\usepackage{nicefrac}

\usepackage{mmacells}

\hypersetup{colorlinks, citecolor=nicegreen, linkcolor=nicered, urlcolor=darkblue}

\newcommand{\fb}{\texttt{FindBounce}\@\xspace}
\newcommand{\mth}{\texttt{Mathematica}\@\xspace}
\newcommand{\bfn}{\texttt{BounceFunction}\@\xspace}

\newcommand{\fbrelease}{%
  \url{https://github.com/vguada/FindBounce/releases}\xspace}

\journal{Computer Physics Communications}

\begin{document}

\begin{frontmatter}

\title{\Large \bf \fb: package for multi-field bounce actions}

\author[IJS]{Victor Guada}
\ead{victor.guada@ijs.si}
\author[IJS]{Miha Nemev\v{s}ek}
\ead{miha.nemevsek@ijs.si}
\author[C3M]{Matev\v{z} Pintar}
\ead{matevz.pintar@c3m.si}

\address[IJS]{Jo\v{z}ef Stefan Institute, Jamova 39, 1000 Ljubljana, Slovenia}
\address[C3M]{C3M d.o.o., Tehnolo\v{s}ki park 21, 1000 Ljubljana, Slovenia}

\begin{abstract}
We are launching \fb, a \mth package for the evaluation of the Euclidean bounce action
that enters the decay rate of metastable states in quantum and thermal field theories.
It is based on the idea of polygonal bounces, which is a semi-analytical approach 
to solving the bounce equation by discretizing the potential into piecewise linear segments.
This allows for a fast and robust evaluation of arbitrary potentials with specified precision 
and any number of scalar fields.
Time cost grows linearly with the number of fields and/or the number of segments.
Computation with 20 fields takes $\sim 2$ seconds with 0.5\% accuracy of the action.
The \fb function is simple to use with the native \mth look and feel, it is easy
to install, and comes with detailed documentation and physical examples, such
as the calculation of the nucleation temperature.
We also provide timing benchmarks with comparisons to existing tools, where applicable.

\end{abstract}

\begin{keyword}
 Quantum tunneling, Instantons, False vacuum decay, Vacuum stability, Phase transitions, 
 Cosmology, Baryogenesis, Gravitational waves.
\end{keyword}
\end{frontmatter}

\clearpage
\section*{Program Summary}
\noindent
{\em Program title:} \fb \\[0.5em]
{\em Program obtainable from:} \url{https://github.com/vguada/FindBounce} \\[0.5em]
{\em Licensing provisions:} GNU General Public License 3 (GPLv3) \\[0.5em]
{\em Distribution format:} .paclet \\[0.5em]
{\em Programming language:} \mth \\[0.5em]
{\em Computer:} Personal computer \\[0.5em]
{\em Operating system:} Tested on Mac OS X and Windows, should work wherever \mth is installed. \\[0.5em]
{\em Typical running time:} Less than 1 (2) seconds for 10 (20) fields. Linear increase with number of fields. \\[0.5em]
{\em Nature of problem:} Evaluation of the Euclidean bounce action that controls the decay rate of metastable
local minima in thermal and quantum field theories. \\[0.5em]
{\em Solution method:} Semi-analytical solution of a system of coupled differential equations, based on the
polygonal bounce idea~\cite{Guada:2018jek}. \\[0.5em]
{\em Restrictions:} \mth version 10 or above, works in $D=3,4$. \\[0.5em]

\clearpage
\tableofcontents

\newpage

%
% Introduction
%
\section{Introduction} \label{secIntro}
False vacuum decay and phase transitions are pervasive topics in particle physics and 
cosmology. Most physical models of nature contain a portion of parameter space with a local
ground state that can transform into a different one with a lower energy, either by quantum 
or thermal fluctuations. The theory of computing the lifetime of such metastable states began 
developing some 30 years ago~\cite{Kobzarev:1974cp, Coleman:1977py, Callan:1977pt, Coleman:1985}
and still poses computational and conceptual challenges.

The task at hand is to find the so-called bounce solution, which is the dominant semi-classical 
contribution, that estimates the exponential factor $B$ in the false vacuum decay rate
\begin{equation} \label{eq:FVDecay}
  \Gamma \simeq A \, e^{-B}\left(1+\mathcal{O}(\hbar)\right).
\end{equation}
The pioneering work of Coleman~\cite{Coleman:1977py} proposed a general method that applies 
to any potential. It requires solving a partial differential equation (PDE) in Euclidean space with fixed 
boundary conditions. An analytical approximation was given in the thin wall regime and an 
under(over)-shooting argument was made that is typically applied in numerical studies.

One of the difficulties in automatic computation for arbitrary potentials resides in the numerical 
nature of the problem, particularly in the thin wall regime, where the two vacua are nearly degenerate.
There, the initial field value has to be chosen with sufficient precision and needs to be sustained 
throughout the calculation in order to track the stiff PDEs with any chosen integrator. Finally, the field 
evolution has to stop at the appropriate moment before it slips back into the true vacuum or diverges. 
Controlling this behavior requires special care and may lead to numerical inaccuracies.

Closed form analytical solutions are clearly not available for arbitrary potentials. The most common
approach is the thin-wall approximation~\cite{Coleman:1977py} that is applicable to a general class of 
potentials with nearly degenerate minima. Other cases of soluble problems include 
the Fubini-Lipatov instanton~\cite{Fubini:1976jm, Lipatov:1976ny} and its 
generalization~\cite{Loran:2006sf}, pure quartic~\cite{Lee:1985uv} and 
logarithmic~\cite{FerrazdeCamargo:1982sk,Aravind:2014pva} potentials.
For most other cases, numerical approaches are used. For example, single field 
renormalizable potentials can be reduced to a single parameter problem by rescaling and then solved 
numerically using the shooting procedure~\cite{Adams:1993zs, Sarid:1998sn}. In many cases, having 
bounds on the action may be useful for restricting some particular 
solutions~\cite{Dasgupta:1996qu, Aravind:2014aza, Sato:2017iga, Brown:2017cca}.
Finally, new approaches were proposed recently, based on the tunneling potential~\cite{Espinosa:2018hue}, 
machine learning techniques~\cite{Jinno:2018dek} and real time formalism~\cite{Andreassen:2016cff,
Ai:2019fri, Hertzberg:2019wgx, Braden:2018tky}.

The issues related to finding the bounce become more challenging when theories with multiple scalar
fields are considered. Here, the path in field space is not fixed a priori and in general traverses a non-trivial 
potential landscape. In such case, finding the initial field value via the shooting procedure becomes a highly 
non-linear problem in the multi-dimensional field space.

There are a number of approaches that tackle the problem of multi-field bounce calculation. These include 
an improved action method that converts the saddle point into a minimum~\cite{Kusenko:1995jv, Chigusa:2019wxb,
Sato:2019axv}, numerical functional minimization~\cite{John:1998ip}, path deformation and 
shooting~\cite{Cline:1999wi, Wainwright:2011kj}, frictionless dimensional 
continuation~\cite{Konstandin:2006nd, Park:2010rh}, semi-analytical
techniques~\cite{Akula:2016gpl}, multiple shooting~\cite{Masoumi:2016wot}, tunneling
potential~\cite{Espinosa:2018szu} and numerically solving coupled PDEs with variable
coefficients~\cite{Athron:2019nbd}, as well as machine learning techniques~\cite{Piscopo:2019txs}.
Existing publicly available tools are mostly based on numerical methods and rely on variations of path 
deformation~\cite{Wainwright:2011kj} or multiple shooting~\cite{Masoumi:2016wot, Athron:2019nbd}.

Apart from the approaches described above, perhaps the simplest class of potentials with constant and 
linear field dependence was studied in~\cite{Duncan:1992ai}. The box-like and triangular shaped potentials 
with two segments were solved in closed form for $D=4$. It turns out this approximation coincides
with the thin wall and is therefore quite accurate for nearly degenerate potentials. Its validity 
was assessed in~\cite{Dutta:2012qt} for a single field and in~\cite{Masoumi:2017trx} for multiple fields.
Moreover, the combination of linear and quadratic was solved in~\cite{Dutta:2011rc}, while
the analytical continuation to Minkowski space was considered in~\cite{Pastras:2011zr}.

Recently, the approach of~\cite{Duncan:1992ai} was extended~\cite{Guada:2018jek} from the triangular
bi-linear setting to an arbitrary polygonal shape, which allows for an arbitrarily precise evaluation. Moreover,  
both $D=3,4$ spacetime dimensions were worked out and, most non-trivially, an arbitrary number of fields
can be considered. We refer to this approach as the polygonal bounce method and is the methodological
fulcrum of the \fb package. It is a semi-analytical procedure with a simple analytical solution 
on each segment of the piece-wise linear potential. Any number of segments can be glued together into 
a single smooth field value function from which an arbitrarily precise bounce action follows. 

The purpose of this work is to implement the polygonal bounce method of~\cite{Guada:2018jek} in \mth 
and release a publicly available documented software package called \fb. It serves to compute the 
bounce action $B$ for theories with arbitrary number of fields in flat spacetime dimensions $D=3, 4$.
This is done by the main function \fb, which computes the bounce action and returns the associated field 
solution. The output of the \fb is given by a dedicated data structure called \texttt{BounceFunction}.
It stores the parameters of the field solution that can be used for further manipulation.

There are a number of phenomenological processes that require knowledge of the bounce action
and the associated field configuration. Apart from the requirement that a given beyond the Standard
Model (BSM) model should be long-lived enough, the bounce action at finite temperatures is
important as well. 
It is needed to extract the parameters related to the production of gravitational waves from a first-order phase 
transition, see e.g.~\cite{Witten:1984rs, Hogan:1986qda, Kosowsky:1992rz, Grojean:2006bp, 
Hindmarsh:2013xza, Cutting:2018tjt} and a recent review on LISA capabilities~\cite{Caprini:2019egz}, 
EW baryogenesis~\cite{Bochkarev:1990fx, Cohen:1990py, Turok:1990zg} with the review 
in~\cite{Morrissey:2012db}, as well as the production of primordial inter-galactic magnetic 
fields~\cite{Vachaspati:1991nm, Sigl:1996dm, DeSimone:2011ek, Tevzadze:2012kk, Ellis:2019tjf}.

In the upcoming section~\S\ref{sec:FVBasic}, we provide a telegraphic overview of the basics
regarding false vacuum decay, together with a short review of polygonal bounce features, 
needed for understanding the \fb implementation. The reader who is interested in a quick tryout of 
the \fb package can skip to~\S\ref{sec:Guide}, where the minimal short guide to installation 
is presented with the most basic examples.
More detailed inner workings of the package, with a description of the \texttt{FindBounce} function 
and available options is available in~\S\ref{sec:FBMethod}.
Examples with benchmarks, performance, timing and comparisons can be found in~\S\ref{sec:Examples}. 
We leave the concluding remarks and an outlook for future developments to~\S\ref{sec:Outlook}.

%
% Basics of false vacuum decay
%
\section{Basics of false vacuum decay} \label{sec:FVBasic}

\begin{figure}
  \centering
  \hfill \includegraphics[width=.48\textwidth]{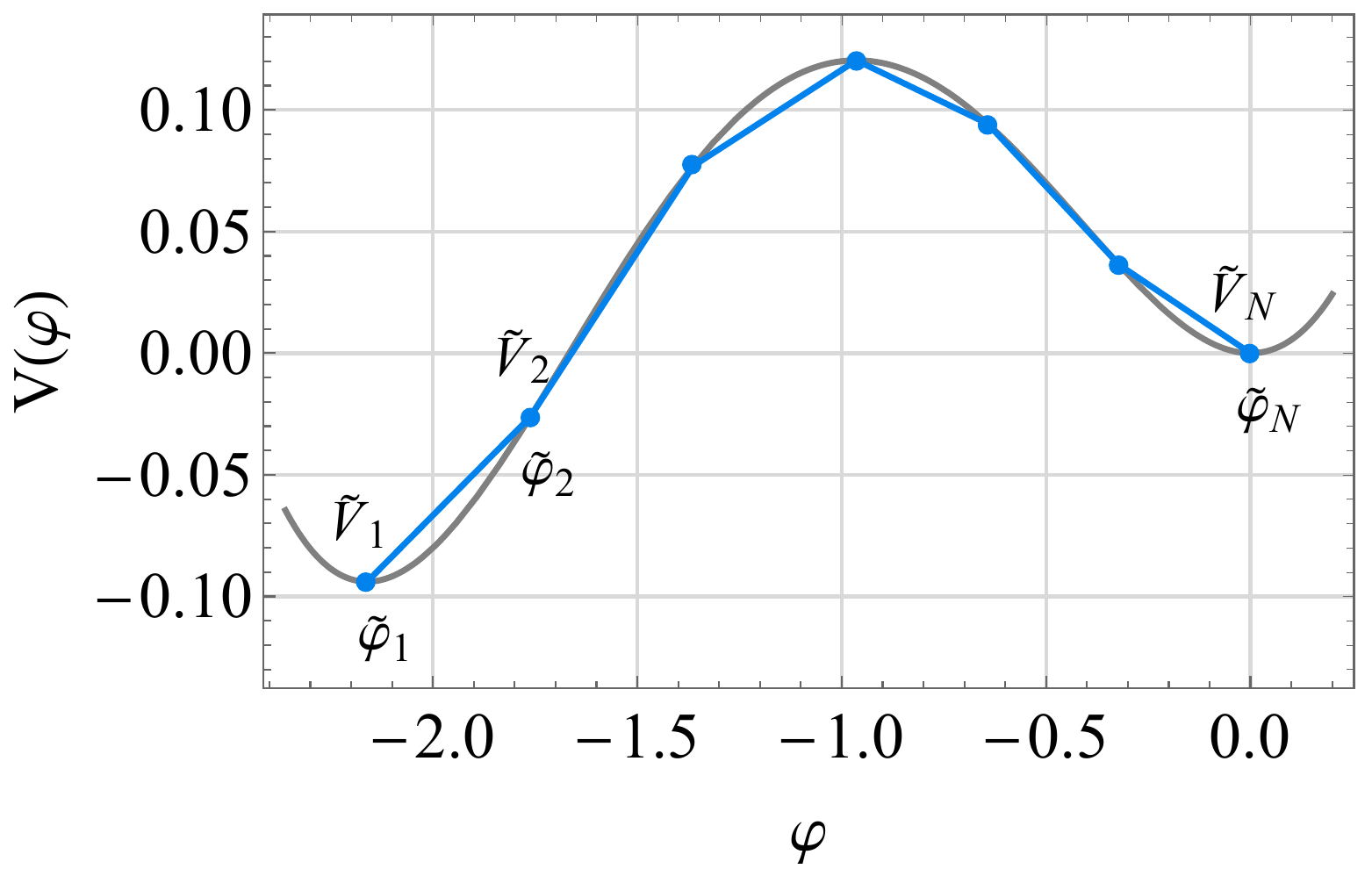}%
  \hspace{1mm}
  \hfill \includegraphics[width=.48\textwidth]{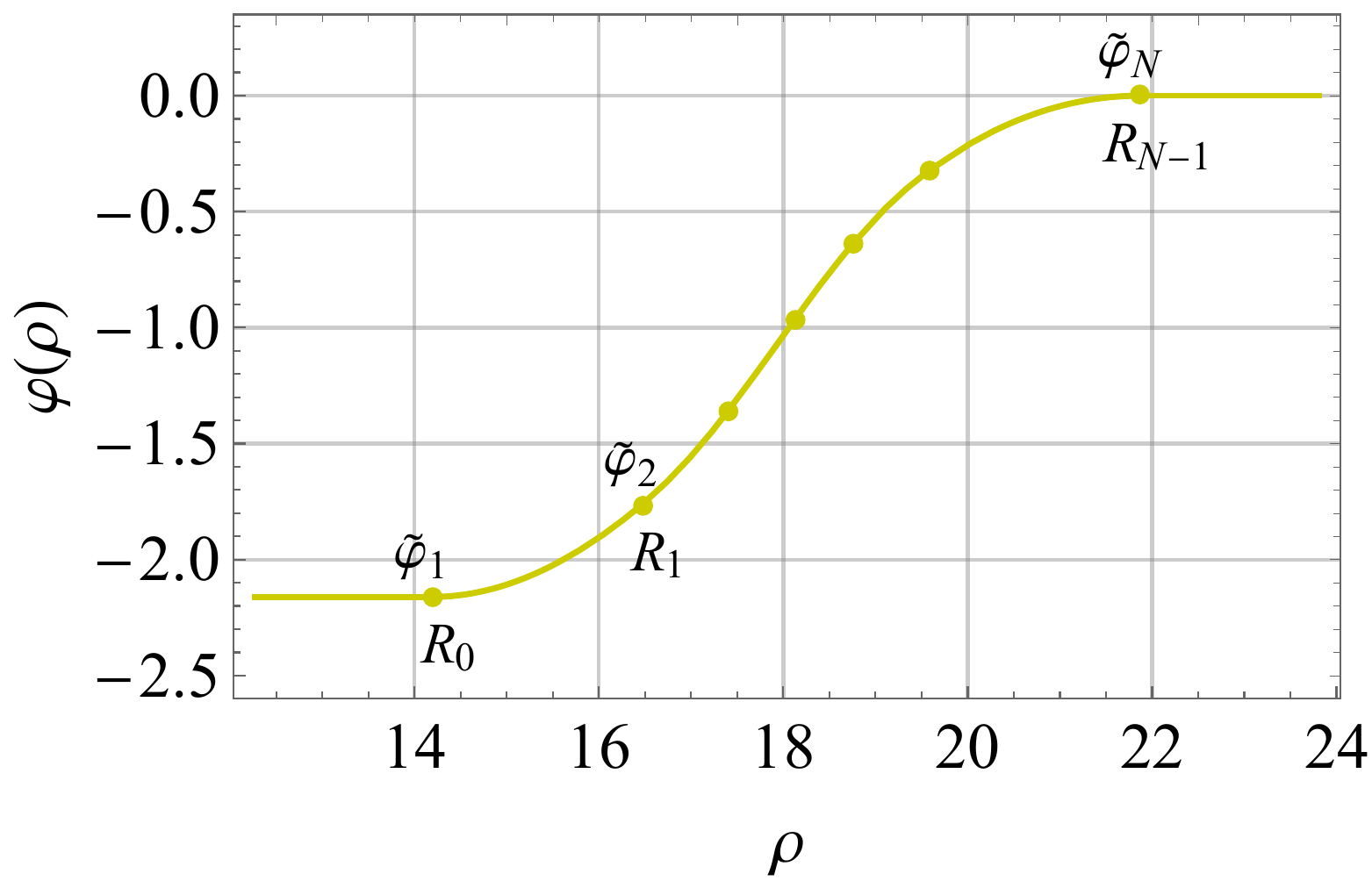} \hspace*{\fill}
  \caption{
    Left: Linearly off-set quartic potential in gray and the polygonal approximation with \(N=7\) in blue.
    Right: The bounce field configuration corresponding to the potential on the left, computed with the
    polygonal bounce approximation.}
  \label{fig:VPhi}
\end{figure}

The object of interest is the Euclidean action $S_D$ of a real scalar field $\varphi$, defined in $D$ 
dimensions as
\begin{equation} \label{eqSED}
  S_D = \frac{2 \pi^{D/2}}{\Gamma \left(D/2\right)} 
  \int_0^\infty \rho^{D-1} \text{ d} \rho \left( \frac{1}{2} \dot \varphi^2 + V(\varphi)  \right),
\end{equation}
and $V(\varphi)$ is the potential. We already assumed that the solution is $O(D)$ symmetric, as proven 
for single~\cite{Coleman:1977th} and multi-field theories~\cite{Blum:2016ipp}.
Therefore, $S_D$ and the field solution depend only on the Euclidean radius $\rho^2 = t^2 + \sum x_i^2$,
which determines the shape of the solution and the size of the appearing bubble. 
The bounce solution interpolates between the two (or more) minima of the potential $V(\varphi)$, as in 
Figure~\ref{fig:VPhi}, and obeys the following Euler-Lagrange equation with boundary conditions
\begin{align} \label{eqBounceEqD}
  \ddot \varphi + \frac{D-1}{\rho} \, \dot \varphi &= \frac{d V}{d \varphi}, 
  &
  \varphi (0) &= \varphi_0, 
  &
  \varphi(\infty) &= \tilde \varphi_N, 
  &
  \quad \dot \varphi(0, \infty) &= 0.
\end{align}
In the particle analogy~\cite{Coleman:1977py}, the field rolls down the inverted potential from $\varphi_0$ 
with zero velocity and comes to a stop in the false vacuum $\tilde \varphi_N$ at $\rho = \infty$. 
The usual shooting procedure involves numerically integrating the bounce Eq.~\eqref{eqBounceEqD} 
and varying $\varphi_0$ until the boundary conditions are met. 

Numerical approaches based on shooting require a very precise initial value at $\rho = 0$, which
has to be set carefully due to the $1/\rho$ friction term. Thereafter, high precision has to be kept throughout
the integration, especially in the vicinity of the thin wall regime, where the bubble radius becomes large. In addition,
the integration has to stop at a certain finite $\rho <  \infty$, before it diverges or oscillates back to the true vacuum.

The polygonal bounce approach, described below, sidesteps these issues. 
Because the potential is composed of linear segments, the Euclidean time support region of the field is automatically 
compactified from $[0, \infty]$ to a finite region $[R_0, R_{N-1}]$, as seen on the right
panel of Figure~\ref{fig:VPhi}. 
Outside of this range, the field value is constant. This compactification regularizes the 
solution and makes it numerically stable, especially in thin wall cases.

\subsection{Polygonal bounce method} \label{sec:PBMethod}
%
% - polygonal bounce: construction, output, lambda~error of the action, good for integrals and perturbation
%
The polygonal bounce idea was developed in~\cite{Guada:2018jek}, where further details regarding the construction
and its properties can be found. 
Here we review the salient features, necessary for understanding the \fb implementation. The 
basic idea is to extend the triangular case in~\cite{Duncan:1992ai} to an arbitrary number
of segments and space-time dimensions. 
Moreover, the work of~\cite{Guada:2018jek} shows how to go beyond the linear approximation and
compute the bounce with any number of fields. 

\paragraph{Polygonal construction}
To establish the notation, let us start with the single field case and define the potential $V_s$ on 
a linear segment $s$, with the corresponding field solution $\varphi_s(\rho)$
\begin{align} \label{eqVPhis}
  V_s(\varphi) &\simeq \underbrace{
  \left(\frac{\tilde V_{s+1} - \tilde V_s}{\tilde \varphi_{s+1} - \tilde \varphi_s}\right)}_{8 a_s}
  \left( \varphi_s - \tilde \varphi_s \right) + \tilde V_s  - \tilde V_N , 
  &
  \quad \varphi_s(\rho) &= v_s + \frac{4}{D} a_s \rho^2 + \frac{2}{D-2} \frac{b_s}{\rho^{D-2}},
\end{align}
with $s = 1, \ldots, N$, and $D > 2$. 
The list of field values $\{ \tilde \varphi \}$ is called the field segmentation, see the left panel of
Figure~\ref{fig:VPhi}. By default, the segmentation in \fb is taken to be homogeneous and consisting of 
$N=31$ points, or equivalently $N-1$ segments. 
The number of field points can be controlled by the option \texttt{"FieldPoints"}, as explained in~\S\ref{opt:FieldPoints}.
Once the segmentation is fixed, the potential is evaluated on these points. This determines
$\tilde V_s = V(\tilde \varphi_s)$ and thus the $a_s$ coefficients in~\eqref{eqVPhis} are known.  

The segmented solutions $\varphi_s(\rho)$ then need to be glued together into a continuously 
differentiable function. Thereby, the free parameters $v_s, b_s, R_s$ of each segment 
are fixed by the following matching conditions: $\varphi_{s}(R_{s}) = \tilde{\varphi}_{s+1} = 
\varphi_{s+1}(R_{s})$ and $\dot{\varphi}_{s}(R_{s}) = \dot{\varphi}_{s+1}(R_{s})$. 
This results in the following recursive relations 
\begin{align} \label{eqvbnDg2}
  v_s &= v_1 - \frac{4}{D-2} \sum_{\sigma=1}^{s-1} \left(a_{\sigma+1} - a_\sigma \right) R_\sigma^2,
  &
  b_s &= b_1 + \frac{4}{D} \sum_{\sigma = 1}^{s-1} \left(a_{\sigma+1} - a_\sigma \right) R_\sigma^D.
\end{align}
Fixing the solution onto the pre-determined segmentation at $R_s$, i.e. $\varphi_s(R_s) = \tilde \varphi_{s+1}$,
leads to
\begin{equation}
  v_s + \frac{4}{D} a_s R_s^2 + \frac{2}{D-2} \frac{b_s}{R_s^{D-2}} = \tilde \varphi_{s+1},
\end{equation}
which is a fewnomial equation for $R_s$ that admits the following simple solution
\begin{align} \label{eqRs}
  D = 3: R_s  &= \frac{1}{2 \sqrt{a_s}} \left( \frac{\delta_s}{\xi} + \xi\right), 
  &
  D = 4: R_s^2 &= \frac{1}{2 a_s} \left(\delta_s + \sqrt{\delta_s^2 - 4 a_s b_s}\right),
\end{align}
with $\delta_s = \tilde \varphi_{s+1} - v_s$ and $\xi^3 = \sqrt{36 a_s b_s^2 - \delta_s^3} - 6  \sqrt{a_s} b_s$. 
Therefore, the entire solution is constructed recursively from the starting values for $v_s$ and 
$b_s$, given by the initial Euclidean radius $R_{in}$.

The polygonal approach exhibits features that simplify the search for the bounce solution.
First, one replaces the field value $\varphi_0$, which is used in the numerical shooting 
algorithms, by the initial radius $R_{in}$. Thus the solution is constructed in terms of the Euclidean spacetime
variable and can be rescaled, e.g. $\varphi_s(\rho ; R_{in}) \to \varphi_s(\rho; \lambda R_{in})$. This can be seen 
by considering the initial condition in~\eqref{eqBounceEqD}, which gives either
\begin{align} \label{eqInitCondDCaseA}
  \text{case A}: v_1 &= \varphi_0 = \tilde \varphi_{in+1} - \frac{4}{D} a_{in} R_{in}^2, & b_1 &= 0;
  &
  \text{or case B}: v_1 &= \tilde \varphi_1 - \frac{4}{D-2} a_1 R_0^2, & b_1 &= \frac{4}{D} a_1 R_0^D.
\end{align}
The entire solution is determined by the initial radius, either $R_{in}$ in case A or $R_0$ in case B.

These two different cases are useful and come as a consequence of 
segmentation and the resulting lack of precision to perform the usual shooting. If the segmentation 
is sufficiently detailed, case A prevails and the field starts rolling from $\varphi_0$, otherwise it 
waits (as in the thin wall case) at the true minimum $\tilde \varphi_1$ until $R_0$ and then begins the 
evolution in $\rho$. The latter case is particularly suited for thin wall potentials. 

\begin{figure*}
  \centering
  \includegraphics[width=1\columnwidth]{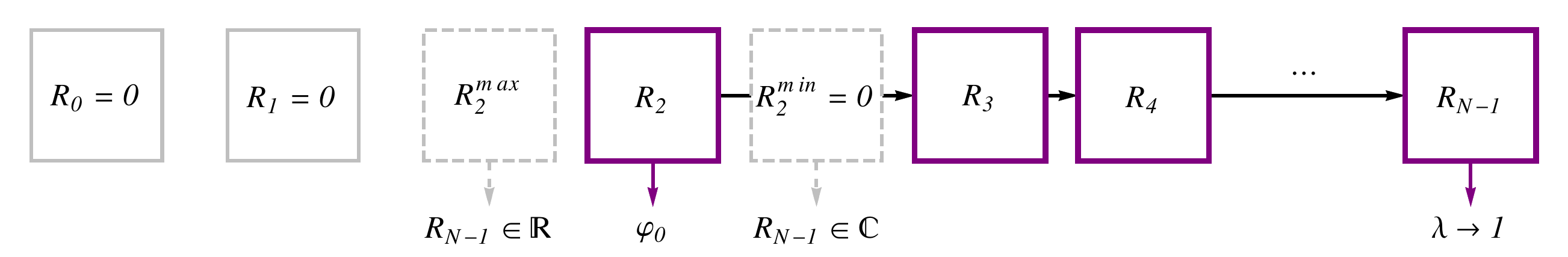}
  \caption{
  A schematic overview of finding the initial radius $R_{in}$. In this particular example, the solution 
  starts from segment $s = 2$, and is bounded by $R_2^{\min} = 0$ and $R_2^{\max}$, which
  can be computed from the segmentation, see~\cite{Guada:2018jek} for details. Note that starting from 
  these two boundary radii, the final radius $R_{N-1}$ changes from real to complex, which happens 
  only when starting from a segment, where a solution exists, in this example it is $s=2$. }
  \label{fig:Scheme}
\end{figure*}

Another useful feature of the method is that one can find the segment in field space, where the starting 
field value $\varphi_0$ has to be, prior to finding the exact solution. This can be done by scrolling through 
all the segments, starting from $\tilde \varphi_1$, bracketing the maximal and minimal values of the initial 
radius $R_{in}$ ($R_0$ on the minimum, $R_1$ on the first segment and so on), running the polygonal 
setup and checking the complexity of the final radius. 
This procedure is sketched on Figure~\ref{fig:Scheme}, see also~\cite{Guada:2018jek} for more details.

It turns out~\cite{Guada:2018jek} that the argument of the square root in~\eqref{eqRs} equals zero
when the final boundary conditions are met. Finding this transition point amounts to solving
\begin{align} \label{eqMatchbN}
  \sum_{\sigma = in}^{N-1} \left(a_{\sigma+1} - a_\sigma \right) R_\sigma^D &= 0
\end{align}
in terms of $R_{in}$. This can done by rescaling $R_{in}$ with Derrick's theorem or with a
favourite root finding algorithm, e.g. by using the native \texttt{FindRoot} method in \mth.
Notice that once the initial radius is found, the initial field value can be expressed directly
in terms of it
 \begin{equation} \label{eqRin}
  \varphi_0 = \tilde \varphi_{in+1} - \frac{4}{D} a_{in} R_{in}^2.
\end{equation}

% Derrick, lambda, checking the consistency
%
A convenient measure for checking the accuracy of the solution comes from Derrick's theorem
that relates the integrated kinetic and potential pieces of the Euclidean action
\begin{align} \label{eqTV}
  \mathcal T &= \frac{2 \pi^{D/2}}{\Gamma \left(D/2\right)} \int_0^\infty \rho^{D-1}
  \text{ d} \rho \frac{1}{2} \dot \varphi^2,
  &
  \mathcal V &= \frac{2 \pi^{D/2}}{\Gamma \left(D/2\right)} \int_0^\infty \rho^{D-1}
  \text{ d} \rho V(\varphi).
\end{align}
Upon rescaling the solution to $\varphi(\rho/\lambda)$ the action should remain at the minimum,
if the correct bounce solution was found, thereby relating $\mathcal T$ and $\mathcal V$
\begin{align} \label{eqSDerrickCont}
  S_D^{(\lambda)} &= \lambda^{D-2} \mathcal T + \lambda^D \mathcal V,
  &
  \frac{d S_{D}^{(\lambda)}}{d \lambda} \biggr|_{\lambda = 1} &= 0,
  &
  \lambda &= \sqrt{\frac{(2-D)\mathcal{T}}{D \mathcal{V}}} \to 1.
\end{align}
The second derivative $d^2 S_D/d \lambda$ is always negative, hence the action is unstable with respect
to rescaling, which may cause numerical issues. One way to solve the polygonal
bounce is to simply rescale $R_{in}$ by $\lambda$ and iterate until $\lambda = 1$ is reached.
In practice, finding the root of $\lambda - 1$ by varying $R_{in}$ also turns out to be convenient
and is implemented in the \fb package.

% Going beyond linear
%
\paragraph{Going beyond the linear approximation}
Once the basic polygonal bounce is obtained, it can be used for various purposes. For example, 
the $V(\varphi)$ can be expanded to higher orders and the bounce solution can be perturbatively
improved. Specifically, expanding to second order gives
\begin{equation}\label{eqVTaylor}
  V_s(\varphi) \simeq \tilde V_s -  \tilde V_N + \partial \tilde V_s \left(\varphi_s - \tilde \varphi_s \right) +
  \frac{\partial^2 \tilde V_s}{2} \left(\varphi_s - \tilde \varphi_s \right)^2.
\end{equation}
The $\partial \tilde V_s$ and $\partial^2 \tilde V_s$ parameters are determined by requiring the 
potential to be continuous and differentiable
\begin{align} \label{eqaalpha1}
  \partial \tilde V_s &= 8 \left(a_s + \alpha_s\right), 
  &
  8 \alpha_s &= 8 a_s - d \tilde V_{s+1},
  &
  \partial^2 \tilde V_s &\equiv \frac{d \tilde V_{s+1} - \partial \tilde V_s}{\tilde \varphi_{s+1} - \tilde \varphi_{s}} =
  \frac{d \tilde V_{s+1} - 8\left(a_s + \alpha_s\right)}{\tilde \varphi_{s+1} - \tilde \varphi_{s}}.
\end{align}
Here, $d \tilde V_{s} \equiv dV/d\varphi(\tilde \varphi_s)$ are the first derivatives evaluated on the 
segmentation points.
The functional form of the derivative of the potential $dV/d\varphi$ can be controlled by the \texttt{"Gradient"} 
option of the \fb function, as explained in~\S\ref{opt:Gradient}.

The complete bounce solution is then constructed on top of the polygonal one $\varphi = \varphi_{PB} + \xi$ 
with the addition of the non-homogeneous integral pieces
\begin{align}\label{eqxi}
 & \xi = \nu + \frac{4}{D} \alpha \rho^2 + \frac{2}{D-2} \frac{\beta}{\rho^{D-2}} + \mathcal I(\rho),
 &
 & \mathcal I= \int_{\rho_0}^\rho \text{d}y \, y^{1-D} \int_{\rho_1}^y \text{d} x  \, x^{D-1} \delta d V(x),
  \\ \label{eqIsD34}
  \mathcal I_s^{D=3} &= \partial^2 \tilde V_s \left(\frac{v_s - \tilde \varphi_s}{6} \rho^2 +
  b_s \rho + \frac{a_s}{15} \rho^4 \right),
  &
  \mathcal I_s^{D=4} &= \partial^2 \tilde V_s \left(\frac{v_s - \tilde \varphi_s}{8} \rho^2 +
  \frac{b_s}{2} \ln \rho + \frac{a_s}{24} \rho^4 \right).
\end{align}
Similarly to the purely poly-linear case, the $\alpha_s$ parameters are already known and so are the 
constants that enter in $\mathcal I_s$. What remains to be solved are the $\nu_s, \beta_s$ and the new
matching radii $R_s'$.

At first sight this construction might look somewhat cumbersome. 
In particular, it may seem that the matching is not easy to perform due to the non-linear terms in~\eqref{eqIsD34}. 
Keep in mind though that this is a perturbative expansion on top of the polygonal one, therefore 
we expect the new matching radii
to be close to the polygonal solution $R_s' = R_s(1 + r_s)$ with $r_s  < 1$. Thus the matching can be performed 
to the linear order in $r_s$ and one ends up with a single linear equation~\cite{Guada:2018jek}, which is easy 
to solve and evaluates quickly.

Such $2^\text{nd}$ order extension increases the precision of the polygonal approximation, as demonstrated 
in~\cite{Guada:2018jek}. The \fb function performs it by default and evaluates 
the derivatives of $V(\varphi)$ automatically. As discussed in~\S\ref{opt:Gradient}, this behavior can
be controlled by the \texttt{"Gradient"->Automatic} option (default value) and can be turned off with \texttt{"Gradient"->None}, if the simplest polygonal 
output is required by the user.

%We also consider the possibility to include $3^\text{nd}$ 
%
One may also consider including higher orders in the expansion of the potential. However, the matching 
conditions in this case cannot be solved in closed form and would require computationally expensive 
numerical root finding at each segment when constructing the bounce solution. It thus turns out that 
it is practically more convenient to obtain precise bounce configurations by increasing the number field 
points using the \texttt{"FieldPoints"} option.

% Multi-field solution
%
\paragraph{Multi-field solution}
The calculation of the bounce solution in the presence of multiple scalar fields is a non-trivial computational 
issue. It is technically involved because one needs to find the initial field value in a higher dimensional
field space and then integrate the coupled system of potentially stiff PDEs, which is usually done numerically.

There are a number of works~\cite{Kusenko:1995jv, John:1998ip, Cline:1999wi, Wainwright:2011kj,
Konstandin:2006nd, Park:2010rh, Akula:2016gpl, Masoumi:2016wot, Espinosa:2018szu, Athron:2019nbd}
that tackle this issue in different ways. In general, these numerical approaches work either by field
path deformation, combined with the single field numerical integration or with multiple shooting. 
The path deformation approach decouples
the single field shooting, freezes the solution and minimizes the potential on this background. This procedure
is then iterated. The drawback of this approach is that at each step either kinetic or potential piece of the 
Euclidean action is not completely extremized, which leads to oscillatory iterations and slows down the 
convergence. Conversely, direct multi-field shooting faces a highly non-linear increase with the number of 
fields. Both typically suffer from difficulties in the thin wall regime and provide a purely numerical output.

The polygonal approach to multi-fields addresses these common shortcomings by constructing a semi-analytical
solution with the following features.
\begin{enumerate}
  \item The functional $\rho$ dependence of the multifield field solution remains as it was in the single
  field case in Eq.~\eqref{eqVPhis}. 
  This enables a fast numerical evaluation as well as the possibility to iterate. The final result 
  has a closed analytical form, allowing for further manipulation and parameter extraction.

  \item Although the solution is built iteratively, a single iteration takes into account both the kinetic 
  and potential minimization simultaneously when deforming the path in field space via the 
  explicit $\rho$ dependence.
  This remedies the oscillatory behavior and the solution converges quickly, with the number 
  of iterations reported in the \texttt{"PathIterations"}  output of the \fb method.
  
   \item As in the single field case, the solution is found in terms of the Euclidean radii variables and exists
   on a compactified region. Therefore it enjoys all the regularities of the single field case and works well
   within the thin wall regime, as well as outside.

  \item Similarly to the extended case above, the multi-field approach is based on a semi-analytical perturbative
  expansion. By expanding the solution around a particular path, a linear system for path deformation is obtained.
  Such systems are easily solvable with fast numerical methods and scale linearly with the number of segments
  and most importantly with the number of fields.
\end{enumerate}

The starting point for the multifield bounce construction is an initial path in field space, denoted by
 $\bar \varphi_{i s}$ where $i$ is the field index $i=1,\ldots,n_f$ and $s = 1,\ldots,N$. This ansatz
is fairly arbitrary and can be controlled by the \texttt{"FieldPoints"} option to be ether a straight line 
connecting the two minima, a triangle through an intermediate (e.g. saddle) point, or any path 
given by the user, see~\S\ref{opt:FieldPoints}.
%
%
%(done by setting
%\texttt{"FieldPoints"->N}), a triangle through an intermediate (e.g. saddle) point with 
%\texttt{"MidFieldPoint"->}\(\tilde{\varphi}_{sp}\), or some other path given by the 
%user \texttt{"FieldPoints"->}\(\{\tilde \varphi \}\).
%

Once the initial path is given, the polygonal $\bar \varphi_{i s}(\rho)$ is computed along this 
longitudinal direction and provides the basis for perturbation $\varphi_{i s}(\rho) = \bar \varphi_{i s} +
\zeta_{i s}$. The bounce equation in Eq.~\eqref{eqBounceEqD} now becomes
\begin{equation}
\begin{split} \label{eqBounceEqMulti}
  &\underbrace{\ddot {\bar \varphi}_{i s} + \frac{D-1}{\rho} \dot{\bar \varphi}_{i s}}_{8 \bar a_{i s}} +
  \underbrace{\ddot \zeta_{i s} + \frac{D-1}{\rho} \dot \zeta_{i s}}_{8 a_{i s}} = \frac{d V}{d \varphi_i} 
  \left(\bar \varphi + \zeta \right).
\end{split}
\end{equation}
In order to stick to the polygonal form, we have to expand the right-hand size around a deformed 
path $\tilde \varphi_{i s} + \tilde \zeta_{i s}$ up to the leading constant order in $\zeta_{i s}$.
Such an expansion keeps the $\rho$ dependence of the field perturbation $\zeta$ in the polygonal form
\begin{align} \label{eqPBZeta}
   \zeta_{i s} &= v_{i s} + \frac{4}{D} a_{i s} \rho^2 + \frac{2}{D-2} \frac{b_{i s}}{\rho^{D-2}},
\end{align}
which evaluates quickly and, most importantly, can be iterated.

In contrast to the single field case, the determination of the $a_{i s}$ requires some care,
because the segmentation changes upon iteration. Subtracting the initial longitudinal 
path $\bar a_{i s}$, we get
\begin{align} \label{eqExpDer}
  8 a_{i s} &\simeq \frac{d V}{d \varphi_i}\left( \tilde \varphi_{i s} + \tilde \zeta_{i s} \right) - 8 \bar a_{i s},
  &
  \frac{d V}{d \varphi_i} &\simeq \frac{1}{2} \left(d_i \tilde V_s + d_i \tilde V_{s+1} +
  d^2_{ij}\tilde V_s \tilde \zeta_{j s} + d^2_{ij} \tilde V_{s+1} \tilde \zeta_{j s+1} \right).
\end{align}
The \fb package evaluates the gradient on segmentation points $d_i \tilde V_s \equiv dV/d\varphi_i(\tilde \varphi_s)$
using the vector of functions given by the \texttt{"Gradient"} option. By default, the gradient is extracted
analytically from $V(\varphi)$, but it can also be pre-computed and given by the \texttt{"Gradient"} option. 

Similarly, the mixed second derivatives are given by the \texttt{"Hessian"} matrix 
$d^2_{ij}\tilde V_s \equiv d^2V/d\varphi_i d\varphi_j(\tilde \varphi_s)$, see~\S\ref{opt:Gradient}.
The essential reason why the Hessian is needed is because the path in field space is typically curved, which 
requires the inclusion of second order derivatives that couple the path deformation displacements 
$\tilde \zeta_i$ in different field directions, which produces a coupled linear system for $\tilde \zeta_i$.

The rest of the construction proceeds as in the extended polygonal case by matching
the field values and derivatives
\begin{align} \label{eqRecRelBeta}
  b_{i s} = b_{i 1} & + \sum_{\sigma = 1}^{s-1} \frac{4}{D} \left(a_{i \sigma+1} - a_{i \sigma} \right) R_\sigma^D
 + \frac{1}{2} \left( \dot{\bar \varphi}_{i \sigma+1} - \dot{\bar \varphi}_{i \sigma} \right) R_\sigma^{D-1},
\\
 \label{eqRecRelNu}
  v_{i s} = v_{i 1} &- \sum_{\sigma = 1}^{s-1}  \frac{4}{D-2} \left(a_{i \sigma+1} - a_{i \sigma} \right) R_\sigma^2
- \frac{1}{D - 2} \left( \dot {\bar \varphi}_{i \sigma+1} - \dot {\bar \varphi}_{i \sigma} \right) R_\sigma.
\end{align}
These recursion relations closely parallel the ones in~\eqref{eqvbnDg2}, apart from the addition of field 
indices and the subtraction of the longitudinal ansatz $\bar \varphi$ that has discrete discontinuities
in derivatives between segments.

The rest of the setup proceeds similarly to the single field case; we refer the reader 
to~\cite{Guada:2018jek} for a detailed discussion. 
Again, we distinguish between the two cases. Case A has a moving initial point $\varphi_{i0}$,
while in case B it is fixed in the true vacuum. The matching conditions are derived by requiring the 
fields to be continuously differentiable along the deformed path, i.e. 
$\zeta_{i s}(R_s) = \tilde \zeta_{i s+1} = \zeta_{i s+1}(R_s)$
and $\dot \zeta_{i s}(R_s) = \dot \zeta_{i s+1}(R_s)$.

The linear system for $\tilde \zeta_{is}$ is then solved with the \mth's native \texttt{LinearSolve} function. 
This procedure is iterated until any of the following requirement is achieved: maximum number of 
iterations controlled by  \texttt{"MaxPathIterations"}, the amount of path deformation set 
by \texttt{"PathTolerance"} or the precision of the action set by \texttt{"ActionTolerance"}, 
as explained in~\S\ref{opt:PathTolerance}. 

%
% Quick guide
%
\section{Quick installation and running guide} \label{sec:Guide}

% Download and installation
\subsection{Download and installation}

The \fb package is released in the \texttt{.paclet} file format, which contains all the code, documentation and other 
necessary resources. 
The latest version of \texttt{.paclet} file can be downloaded from the repository ``releases" page (\fbrelease ) 
and can be installed by evaluating the following code in \mth.

% Explicit loading of "PacletManager`" context is not required because it is always automatically
% loaded at kernel startup. Also in MMA 12.1 PacletInstall will be moved to "System`" context.
\begin{mmaCell}[moredefined={PacletInstall}]{Input} 
PacletInstall["full/path/to/FindBounce-X.Y.Z.paclet"]
\end{mmaCell}

This will permanently install the \fb package to the \texttt{\$UserBasePacletsDirectory}. To update the documentation, 
it may be necessary to restart \mth. \mth will always use the latest installed version. All the previously installed versions 
of \fb can be enumerated by evaluating \texttt{PacletFind["FindBounce"]}.
More detailed information about the \fb package can be found with \texttt{PacletInformation["FindBounce"]}.
All the versions can be uninstalled with \texttt{PacletUninstall["FindBounce"]}.

% Running
%
\subsection{Running}

Once the package is installed, load it with \texttt{Needs}.
\begin{mmaCell}[index=1]{Input}
Needs["FindBounce`"]
\end{mmaCell}
To access the documentation, open the notebook interface help viewer and search for \fb.
Let us show how \fb can be used on a simple example
\begin{mmaCell}[morepattern={x_, x}]{Input}
V[x_]:= 0.5 x^2 + 0.5 x^3 + 0.12 x^4;
\end{mmaCell}
\begin{mmaCell}[moredefined={V}]{Input}
extrema = x/.Sort@Solve[D[V[\mmaFnc{x}],\mmaFnc{x}]==0];
\end{mmaCell}
The bounce is obtained with the \fb function
\begin{mmaCell}[moredefined={FindBounce,V,extrema},morefunctionlocal={x}]{Input}
bf = FindBounce[V[x],x,\{extrema[[1]],extrema[[3]]\}]
\end{mmaCell}
where the order of the minima is arbitrary.
\begin{mmaCell}[moregraphics={moreig={scale=.7}}]{Output}
\mmaGraphics{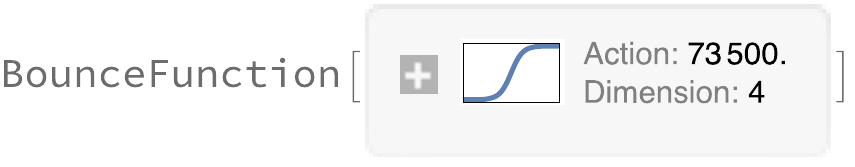}
\end{mmaCell}
Here, all the options have their default values and the results can be extracted as follows.
\begin{mmaCell}[moredefined={bf}]{Input}
bf["Action"]
\end{mmaCell}
\begin{mmaCell}{Output}
73496.
\end{mmaCell}
\begin{mmaCell}[moredefined={bf}]{Input}
bf["Dimension"]
\end{mmaCell}
\begin{mmaCell}{Output}
4
\end{mmaCell}
Notice that the summary box outputs rounded values, e.g. for the action, while directly accessing 
values from the \texttt{BounceFunction} object is done with default precision.

This concludes the simplest demonstration for single field use. More details regarding the other 
available options and their use are available in~\S\ref{sec:FBOptions}. 
Before moving on, let us briefly comment on \fb argument overloading.
It is possible to study arbitrary purely polylinear potentials by providing a set
of points and their potential values $\{ \tilde \varphi, V(\tilde \varphi)\}$:
\begin{mmaCell}[moredefined={FindBounce},label ={}]{Input}
FindBounce[\{\{x1,V1\},\{x2,V2\},...\}]
\end{mmaCell}
An example of such use is given in~\S\ref{subsec:DisappEx}. Finally, the \fb function is overloaded 
for use with multi-field potentials, in which case the evaluation is done by the following syntax:
\begin{mmaCell}[moredefined={FindBounce},morefunctionlocal={x,y},label = {}]{Input}
FindBounce[V[x,y,...],\{x,y,...\},\{m1,m2\}]
\end{mmaCell}
where $m_{1,2}$ are the two multi-field minima; see~\S\ref{subsec:TwoField} for definite examples.

% Contributing
\subsection{Contributing}

Please use the \texttt{issues} (\url{https://github.com/vguada/FindBounce/issues}) 
page on the \texttt{GitHub} repository to submit bugs or feature ideas. 
Pull requests are welcome, however in case of major changes, please open an issue first to discuss 
what you would like to change. 
For developers, the instructions on how to run the tests, build the package \texttt{.paclet} file from the source code,
and create the documentation, can be found in the \texttt{CONTRIBUTING.md} file.

%
% Bounce function
%
\section{Using the \fb and its output} \label{sec:FBMethod}

%Bounce function
\subsection{\fb options} \label{sec:FBOptions}

%Intro: list and describe options/ default values/
In this section we list and describe the available options of the \fb function. They
%These are given as rules at the end of the function, similar to other native \mth functions. 
%Further information find in:
are directly accessible within \mth using the \fb function documentation, which includes detailed descriptions and 
examples. The options and default values can be listed with the syntax
\begin{mmaCell}[moredefined={FindBounce}, label={}]{Input}
Options[FindBounce]
\end{mmaCell}
returning the options described below.

% - Full description of the option of the package.
%
\begin{itemize}

  % ActionTolerance
  %  
  \item \texttt{"ActionTolerance"} controls the relative variation of the bounce action $|\Delta S|/S$ between iterations 
  of the initial radius $R_{in}$ while solving the boundary conditions in~\eqref{eqvbnDg2}. 
  \fb also supervises the change of the action after each deformation of the path. 
  The default tolerance value is $10^{-6}$.  See for example the right panel of Figure~\ref{fig:tolerance}.

  % "BottomlessPotential"
  %  
  \item \texttt{"BottomlessPotential"} is a Boolean variable used to specify whether the combination
  of polygonal and quartic should be used. This option can be used to deal with single field potentials 
  unbounded from below, as discussed in~\S\ref{sec:Unbound}. The default value is \texttt{False}.

  % Dimension
  %    
  \item \texttt{"Dimension"} defines the number of space-time dimensions, given by the
  integer \(D\). The default value is $D=4$ for zero temperature tunneling via quantum fluctuations.
  In thermal field theory, the $D=3$ action is needed, which is obtained by the following
  evaluation.
\begin{mmaCell}[moredefined={FindBounce,V,extrema},morefunctionlocal={x}, index=7]{Input}
bf3 = FindBounce[V[x],x,\{extrema[[1]],extrema[[3]]\},"Dimension"-> 3];
\end{mmaCell}
  
  % FieldPoints
  %  	
  \item \label{opt:FieldPoints} 
  \texttt{"FieldPoints"} controls the segmentation in field space. 
  It is specified either by the integer number of field values or by an explicit list of field 
  values $\{ \tilde \varphi_1, \ldots , \tilde \varphi_N \}$ given by the user.
    \begin{itemize}
      \item \texttt{"FieldPoints"-> 31} by default. When specified by a single
      integer $N$, the segmentation is homogeneously split into $N$ equidistant field points.
      For multiple scalar fields, the initial trajectory is taken to be a straight line connecting
      the two minima, e.g. the black dot-dashed line in Figure~\ref{fig:OptionValue}.
      
      \item \texttt{"FieldPoints"-> \( \{ \tilde \varphi \} \)}. Arbitrary segmentations can be 
      given with an explicit set of field coordinates as an input. 
      The order of the minima is not important, \fb recognizes the higher one as the 
      false vacuum state. For single field potentials field points remain fixed, while for 
      multi-fields they move when the path is being updated.
    \end{itemize}

  % Gradient
  %  
  \item  \label{opt:Gradient}  \texttt{"Gradient"} controls the evaluation of the bounce beyond the 
  poly-linear approximation of $V(\varphi)$, as in~\eqref{eqVTaylor}. There are three available
  option values.
    \begin{itemize}
       \item \texttt{"Gradient"-> Automatic} by default. \fb computes the bounce by taking into account the 
       $2^{\text{nd}}$ order approximation to $V(\varphi)$ in~\eqref{eqVTaylor}. The gradient function(s) of
       the potential needed in~\eqref{eqExpDer} are obtained analytically by 
       running \texttt{Grad[V[x,y,...],\{x,y,...\}]}. 
              
       \item \texttt{"Gradient"-> "FiniteDifference"}. The set of gradient functions required in~\eqref{eqExpDer} 
       are approximated by finite differences, where the small variations of the fields $\Delta \varphi$, are 
       proportional to the total length of the path as $\Delta \varphi \equiv 10^{-4} L_{path}$.
       This option is suitable for non-analytic potentials, or when the default derivative, given by \mth, may
       be complicated and leads to delays in evaluation.
       
       \item \texttt{"Gradient"-> $\{dV/d \varphi_i\}$}. The set of gradient functions can 
       also be pre-computed, stored and given as an input with this option. This can be used in multiple 
       evaluations and scanning to save the computation time. See for example~\S\ref{subsec:TwoField}. 
              
       \item \texttt{"Gradient"-> None}. With this setting, the $2^{\text{nd}}$ order extension is turned off and 
       the polygonal method is implemented in the poly-linear approximation.  
       This may be necessary when the the derivative of the potential is discontinuous, 
       as in~\S\ref{subsec:QuarticEx}.
    \end{itemize}

  % Hessian
  %  	
  \item \texttt{"Hessian"} option for multi-field bounce calculations regulates the 
  evaluation of the second derivatives $d^2V/d \varphi_i d \varphi_j$ in~\eqref{eqExpDer}. 
    \begin{itemize}
      \item \texttt{"Hessian"-> Automatic} is the default behavior, where \mth computes the Hessian matrix
      analytically by running \texttt{Grad[Grad[V[x,y,\ldots],\{x,y,\ldots\}],\{x,y,\ldots\}]}.
      % \texttt{\{\{D[V[x,y,...],\{x,x\}],D[V[x,y,...],\{x,y\},...\},...\}}.
  
      \item \texttt{"Hessian"-> "FiniteDifference"} approximates the Hessian matrix required in~\eqref{eqExpDer} 
      with numerical finite differences. Similarly to the \texttt{"Gradient"} option, the variations of the fields are 
      computed from the path length and given by $\Delta \varphi \equiv 10^{-4} L_{path}$.
      \item \texttt{"Hessian"-> $\{d^2V/d \varphi_i d \varphi_j\}$}. Similarly to the \texttt{"Gradient"} option above, 
      the Hessian matrix of functions can be provided externally by the user to speed up the calculation.
    \end{itemize}

  % MaxPathPathIterations
  %     
  \item \texttt{"MaxPathIterations"} can be used to control the maximum number of times the path can be 
  iterated after starting from the initial ansatz; the default value is 3. See for example~\S\ref{subsec:TwoField}.

  % MaxRadiusIterations
  %    
  \item \texttt{"MaxRadiusIterations"}
  sets the maximum number of iterations to compute the initial radius $R_{in}$ that satisfies Eq.~\eqref{eqSDerrickCont}; 
  the default value is 100. Generically, \fb takes about $O(1)$ iterations to compute the action up to the 
  default tolerance value. However, this option may be overridden by the \texttt{ActionTolerance} requirement,
  which prioritizes the precision of the action and thus allows for a larger number of iterations.  
  
  % MidFieldPoint
  %  	
  \item \texttt{"MidFieldPoint"}
  allows the user to control the segmentation by setting a single arbitrary intermediate field point between
  the two minima, such as saddle points.
  \begin{itemize}
  	\item \texttt{"MidFieldPoint"-> None} by default: the segmentation is a homogeneously discretized straight line 
	in field space connecting the two minima.
	\item \texttt{"MidFieldPoint"-> Automatic}. The straight line connecting the two minima is divided by 
	$\tilde{\varphi}_{max}$ into two 	homogeneous segmentations. 
	The intermediate field point $\tilde{\varphi}_{max}$ is the local maximum of the $V$ on the straight line, 
	and is computed with \texttt{FindMaximum}.
	This option is suitable for nearly flat or very asymmetric potentials, where automatic segmentation may
	not detect the maximum unless a large value for \texttt{"FieldPoints"} is used.
  	\item \texttt{"MidFieldPoint"-> $\tilde{\varphi}_{int}$} The segmentation is divided by $\tilde{\varphi}_{int}$ 
	into two homogeneous segmentations. It consists of two straight lines that connect the two minima with the 
	intermediate field point $\tilde{\varphi}_{int}$.
  \end{itemize}
  See~\S\ref{subsec:QuarticEx} for single field and section~\S\ref{subsec:TwoField} for a multi-field example.

  % PathTolerance
  %  
  \item  \label{opt:PathTolerance} \texttt{"PathTolerance"}, controls the search with multiple scalar
  fields, where the path in field space changes with each iteration. Its value specifies the 
  maximal allowed deviation of the path from one iteration to another. 
  It is defined as the maximum length of deformation of any field point after each iteration,
  normalized to the total length of the path: \(\max_s \left |\tilde \zeta_s \right|/L_{\text{path}}\) with default value $10^{-2}$. 
\end{itemize}

The \fb stops if either \texttt{"MaxPathIterations"}, \texttt{"PathTolerance"} 
or \texttt{"ActionTolerance"} is satisfied. 

% miha: this is a repetition of the above?
%In addition to those options one can use the \texttt{"FieldPoints"}, \texttt{"Gradient"}, \texttt{"Hessian"} and \texttt{"MaxRadiusIterations"} options to optimize time performance. 
%The path in field space can be controlled with a single point by \texttt{"MidFieldPoint"} or more precisely with \texttt{"FieldPoints"}. The order of the minima of the potential can be specified either way since \fb
%recognizes the higher one as the false vacuum state.

%
%  Bounce function output and manipulation
%
\subsection{Bounce function output and manipulation} \label{sec:BfnOutput}

The results of the calculation are bundled in the \bfn container, which stores various parameters 
and other properties of the solution. The \bfn is rendered in \mth notebooks with a summary box that contains 
some minimal amount of information: the shape of the field solution, Euclidean action and the 
number of space-time dimensions.
The list of all the available properties can be accessed with the following syntax:
\begin{mmaCell}[moredefined={bf}]{Input}
bf["Properties"]
\end{mmaCell}
\begin{mmaCell}{Output}
\{Action, BottomlessPotential, Bounce, Coefficients, CoefficientsExtension, 
 Dimension, Path, PathIterations, Radii\}
\end{mmaCell}
%
%where:
%
\begin{itemize}
  % Action
  %
  \item \texttt{"Action"} gives the value of the Euclidean bounce action.
 
  % BottomlessPotential
  % 
  \item \texttt{"BottomlessPotential"} returns the constant factor of the quartic potential $V_{0}$ of 
  Eq.~\eqref{eq:quartic}. See~\S\ref{sec:Unbound} for an example.
   
  % Bounce
  % 
  \item \texttt{"Bounce"} returns the piece-wise smooth function that characterizes the bounce 
  solution $\varphi(\rho)$. It can be evaluated as a continuous function, see 
  section~\S\ref{subsec:TwoField} for an example.
  
  % Coefficients
  % 	  
  \item \texttt{"Coefficients"} provides the constant factors $\{v_{is}, a_{is}, b_{is}\}$ in Eqs.~\eqref{eqVPhis}
  and \eqref{eqPBZeta} that define the multi-field (polygonal) bounce solution in each segment. 
  See~\S\ref{sec:Unbound} for an example.
  
   % Coefficients
  % 	  
  \item \texttt{"CoefficientsExtension"} provides the constant factors $\{\nu_{is}, \alpha_{is}, \beta_{is},\partial^2\tilde V_{s}\}$ in Eqs.~\eqref{eqaalpha1}
  and \eqref{eqxi} that define the extension of the multi-field (polygonal) bounce solution $\xi$ and $\mathcal{I}$ in each segment. 
  
  % Dimension
  % 	  
  \item \texttt{"Dimension"} returns the number of space-time dimensions in which the bounce was computed, 
  where finite (zero) temperature corresponds to $D = 3$ ($D = 4$).

  % PathIterations
  % 
  \item \texttt{"PathIterations"} reports the number of times the path in field space was deformed from the 
  initial ansatz and an upper limit set by the \texttt{"MaxPathIterations"} option.

  % Path
  %   
  \item \texttt{"Path"} gives a list of points $\tilde \varphi_s = \varphi(R_{s})$ 
  that defines the trajectory of the bounce in field space. This output can be used
  as an initial path ansatz to save time when finding the bounce solution for similar 
  potentials, see section~\S\ref{subsec:TwoField} for an example.
  \item \texttt{"Radii"} returns the list of radii $R_{s}$ where the segments are joined from $R_{in}$ to $R_{N-1}$.
  
\end{itemize}

{\em \texttt{BouncePlot}.}
In addition to the \bfn described above, a plotting function wrapper \texttt{BouncePlot} is available, such that the 
field configuration(s) can be plotted with ease. 
The \texttt{BouncePlot} behaves similarly to the native \texttt{Plot}, where the default options can be changed as shown 
in the example below.
Multi-field bounce solutions, given as a list of functions, can also be plotted simultaneously.
\begin{mmaCell}[moredefined={BouncePlot,PlotLegends,Placed,bf3,bf}]{Input}
BouncePlot[\{bf3,bf\}, PlotLegends-> Placed[\{"D=3","D=4"\}, \{Right,Center\}]]
\end{mmaCell}
\begin{mmaCell}[moregraphics={moreig={scale=.45}}]{Output}
\mmaGraphics{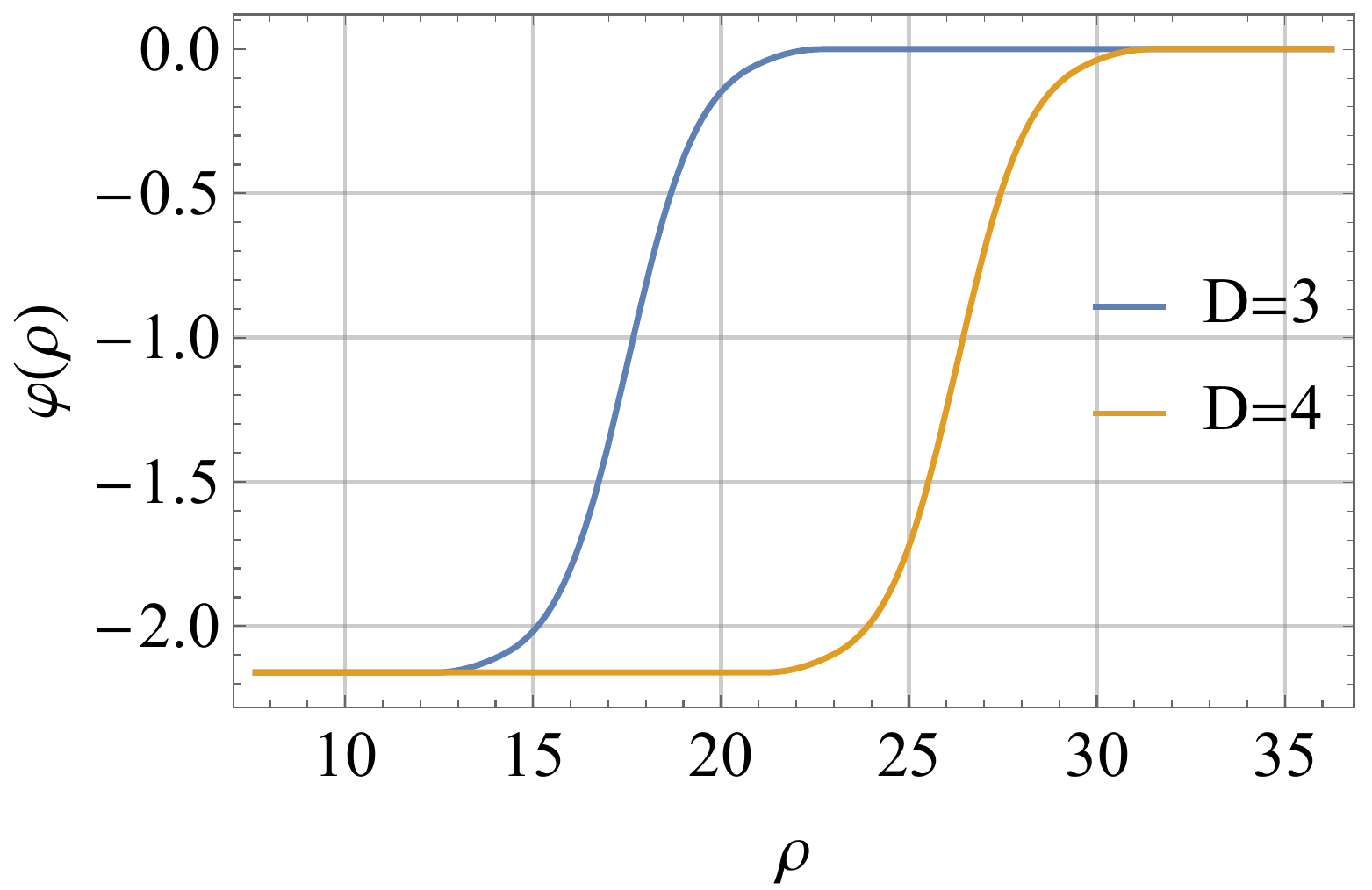}  
\end{mmaCell}
%As an alternative, one can also plot the above bounces field as
%\begin{mmaCell}[moredefined={bf3,bf},morefunctionlocal={r}]{Input}
%Plot[\{bf3["Bounce"][r],bf["Bounce"][r]\},\{r,bf3["Radii"][[1]],bf["Radii"][[-1]]\}]
%\end{mmaCell}
% ,
%  PlotLegends-> Placed[\{"D=3","D=4"\}, \{Right,Center\}]]
%\newpage

%
% Examples
%
\section{Examples} \label{sec:Examples}

This section contains a number of examples, test cases and demonstrations of the \fb method. Each subsection 
contains a simple self contained code that can be easily reproduced in \mth. All of the examples given here
(and more) are available also in the \mth documentation of \fb after installation.

We start with the single field benchmark in~\S\ref{subsec:SingleEx} that showcases the use of some available 
options and return values of \fb listed above in~\S\ref{sec:FBMethod}.
One of the main aspects is the performance in terms of the precision of the action and timing in thick 
and thin wall regimes. In particular, timing scales linearly with the number of field points.

We consider the exactly solvable quartic-quartic case in~\S\ref{subsec:QuarticEx} that shows how
\fb deals with such non-trivial cuspy potentials. In~\S\ref{subsec:DisappEx} we use the purely polygonal 
example with $N=5$ points to reproduce the curious case of the disappearing instanton in the presence of 
additional local minima.
We show how \fb can estimate the escape point $\varphi_0$ of unstable potentials in~\S\ref{sec:Unbound}
by combining the exact linear solution to an exact (unstable) quartic one.

For multi-field applications, we first perform the simplest study with two fields in~\S\ref{subsec:TwoField},
where we demonstrate how to control the precision and speed of the evaluation as well as the use of
an arbitrary path ansatz.
%pre-computed bounce solutions and scanning of parameters.
We also demonstrate the use of \fb on effective and thermal potentials and give a demonstration
on computing the nucleation temperature.
In the final sub-section~\S\ref{subsec:BeyondTwo}, we address the bounce calculation with an 
arbitrary number of fields and test the \fb method with up to twenty fields. 
We compare the results with other existing methods and demonstrate that the time demand of 
the \fb function scales {\em linearly} with the number of fields.

% Single field
%
\subsection{Single field benchmark} \label{subsec:SingleEx}

\begin{figure}[ht]
  \centering
  \hfill \raisebox{1mm}{\includegraphics[width=.47\textwidth]{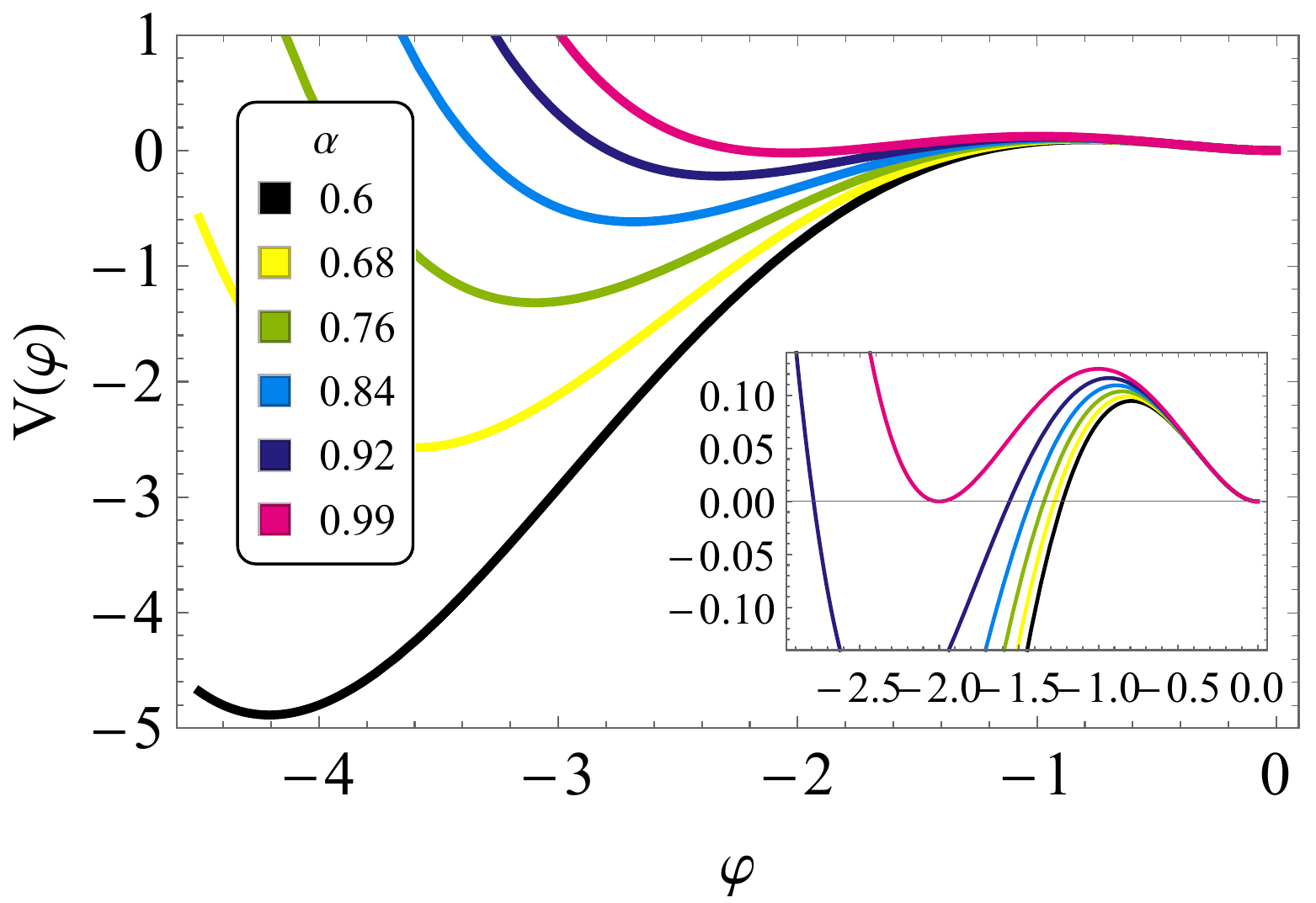}}%
    \hspace{2mm}
  \hfill \includegraphics[width=.51\textwidth]{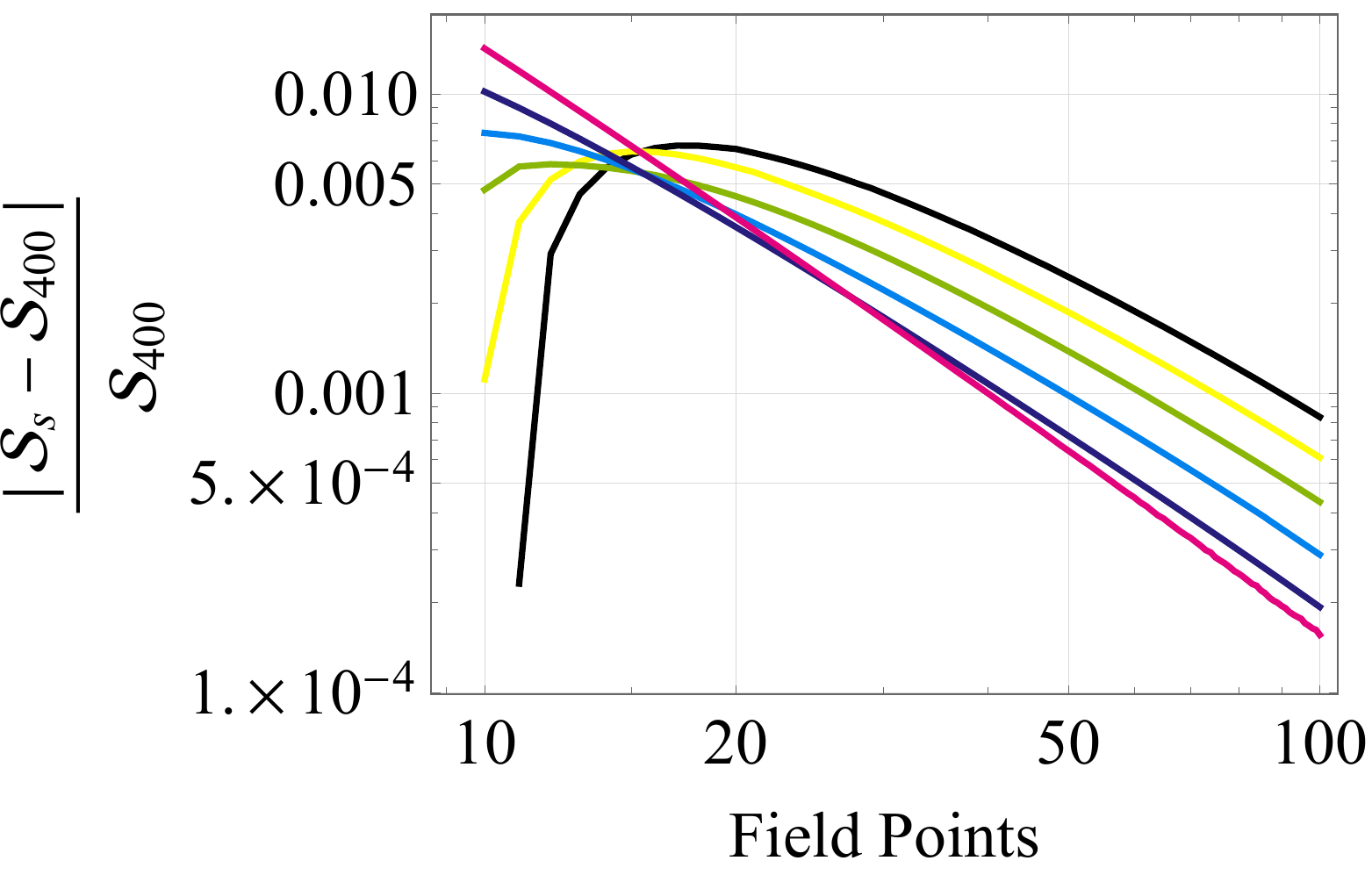}\label{fig:Action}
  \hspace*{\fill}
  \caption{
    Left: The benchmark potential from Eq.~\eqref{eq:BM} for different values of $\alpha$ going from thick 
    $\alpha = 0.6$ to thin wall $\alpha = 0.99$.    
    Right: The bounce action $\mathcal S_s$ for each potential configuration and a given number of 
    field points $s$, normalized to $s = 400$ and computed in $D=4$.}
  \label{fig:Potential1D}
\end{figure}

Let us consider a generic renormalizable scalar potential with terms up to the quartic power. Such
potentials in general feature an unstable ground state, as seen on the left panel of Figure~\ref{fig:Potential1D}.
Using the re-scaling properties of both the field $\varphi$ and the $\rho$ from~\eqref{eqBounceEqD}, one
can rewrite $V$ as a function of a single parameter $\alpha$, as~\cite{Dunne:2005rt}
\begin{align} \label{eq:BM}
  V(\varphi)\equiv  \frac{1}{2}\varphi^2 + \frac{1}{2} \varphi^3 + \frac{\alpha}{8} \varphi^4.
\end{align}
Here $0 \leq \alpha \leq 1$ covers all the possible scenarios going from thick to thin wall respectively.
The bounce configuration for a particular $\alpha$ and the number of field points $N$ in 3 space-time dimension can be obtained
by \fb with:
% 
%(*the potential*)
\begin{mmaCell}[pattern={x_,x,a,a_},index=1]{Input}
V[x_,a_]:= 0.5 x^2 + 0.5 x^3 + 0.125 a x^4;
\end{mmaCell}
\begin{mmaCell}[pattern={,a,a_},moredefined={FindBounce,V}]{Input}
extrema[a_]:= x /. Sort@NSolve[(D[V[\mmaFnc{x},a], \mmaFnc{x}]) == 0, \mmaFnc{x}];
\end{mmaCell}
\begin{mmaCell}[pattern={a,a_,n,n_Integer},moredefined={FindBounce,V,extrema},morefunctionlocal={x}]{Input}
bf[a_,n_Integer]:= FindBounce[V[x,a], {x}, extrema[a][[\{1,3\}]],
   "FieldPoints"-> n, "Dimension"-> 3];
\end{mmaCell}
%
%\vspace{1cm}
\begin{mmaCell}[moredefined={bf}]{Input}
bf[0.6,100]
\end{mmaCell}
\begin{mmaCell}[moregraphics={moreig={scale=.7}}]{Output}
\mmaGraphics{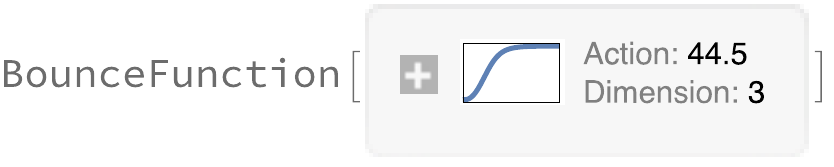}
\end{mmaCell}
\begin{mmaCell}[moredefined={bf}]{Input}
bf[0.6,100][\{"Action","Dimension"\}]
\end{mmaCell}
\begin{mmaCell}[moregraphics={moreig={scale=.7}}]{Output}
\{44.5098, 3\}
\end{mmaCell}

The resulting action for different values of $\alpha$ is plotted on the right panel of 
Figure~\ref{fig:Potential1D} and is normalized to $N=400$ field points, which is already very precise. 
The accuracy of the action improves with the number of field points and goes below the percent 
level with $N=30$ segmentation points in both, thin and thick wall.  As expected, the convergence is faster 
for thin walls since the minimal $N=3$ case already coincides with the thin wall solution.
Note also that in this case, the $2^\text{nd}$ order correction of the potential is taken into account by default
and the convergence is faster than the pure polygonal calculation.

\begin{figure}[t]
  \centering
  \hfill \includegraphics[width=.49\textwidth]{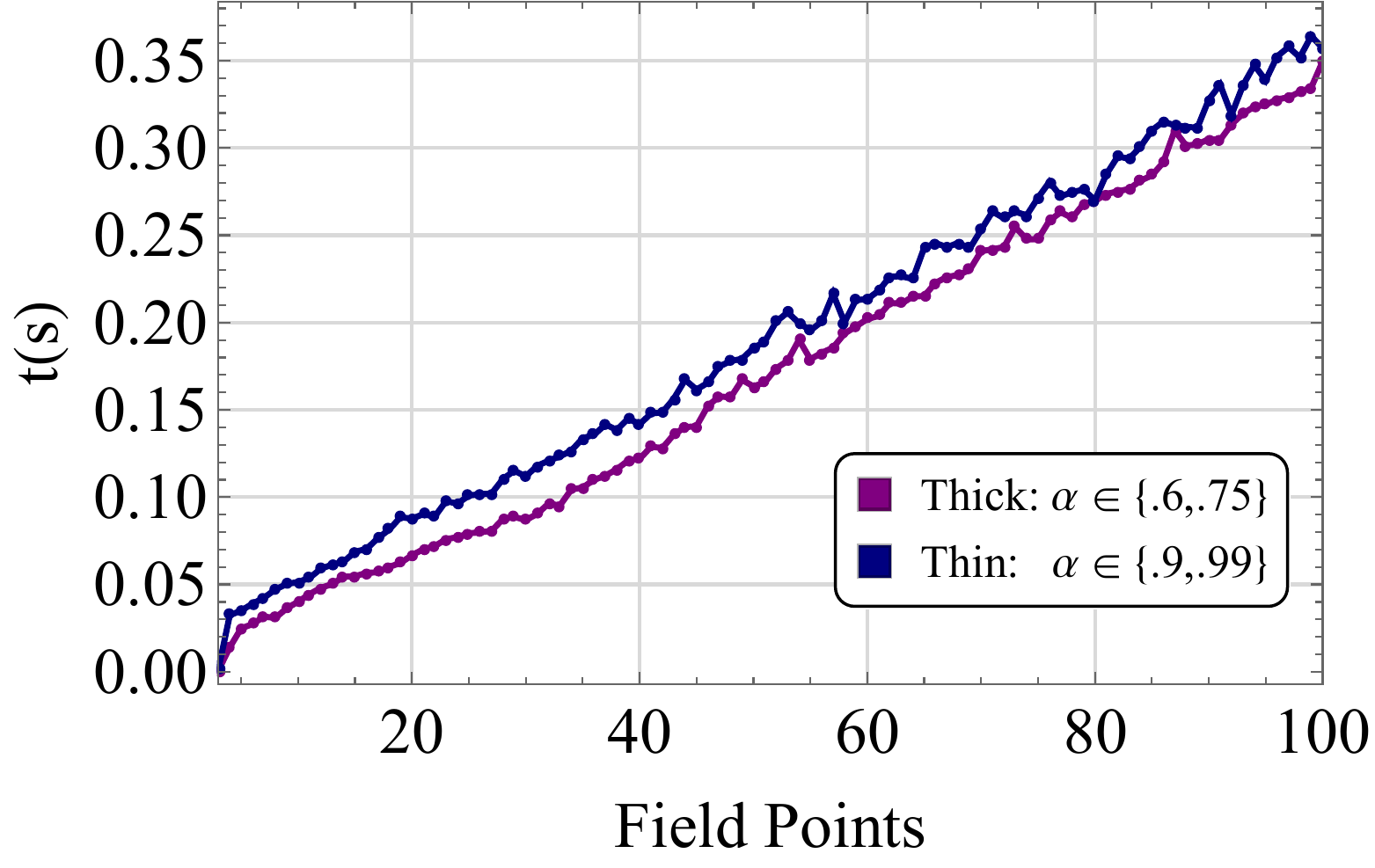}
    \hfill \includegraphics[width=.49\textwidth]{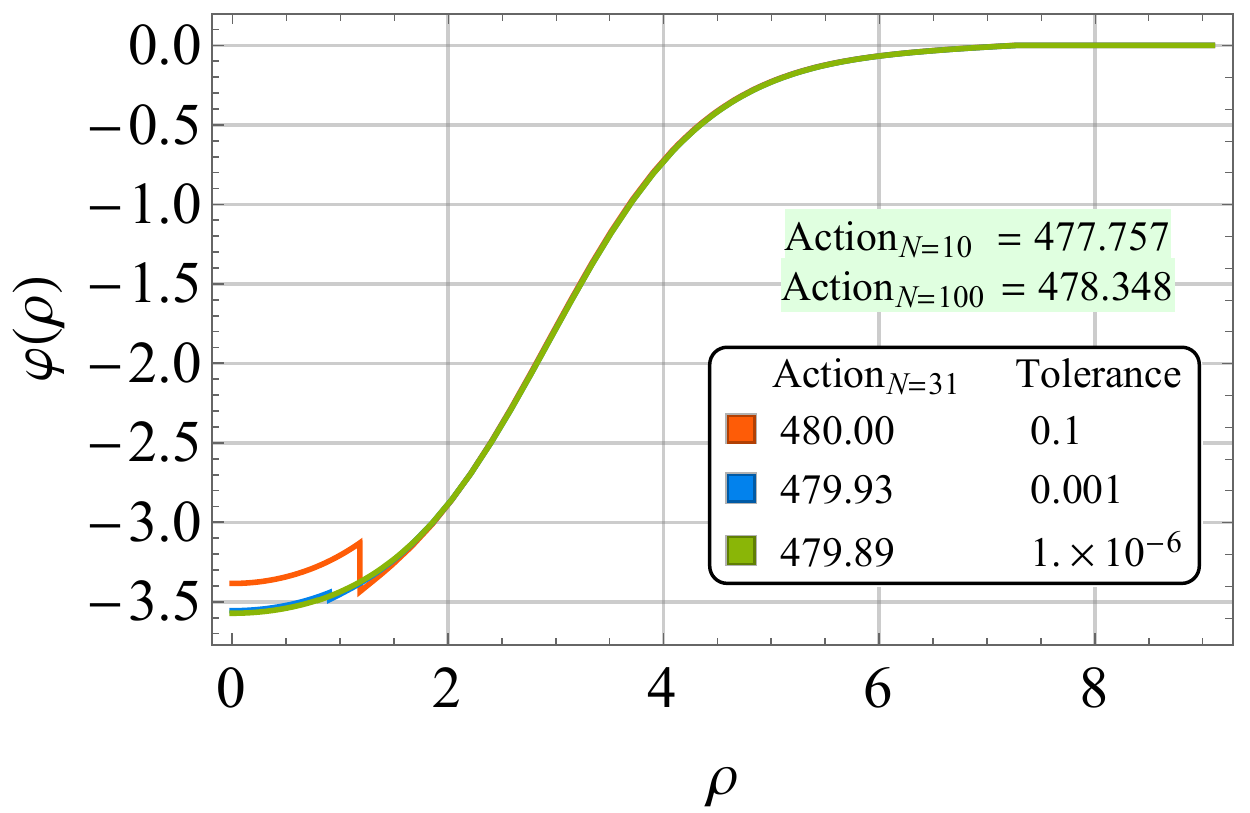}
  \hspace*{\fill}
  \caption{
    Left: Evaluation time with respect to the number of field points, averaged over two intervals of $\alpha$ corresponding
    to thin and thick wall regimes.
    Right: The bounce field configuration and action with $N=31$ (default) field points for different tolerance value of the action controlled by
     \texttt{"ActionTolerance"}. Reference values of the action for \texttt{"FieldPoints"->} 10 and 100 field points 
    with \texttt{"ActionTolerance"->} $10^{-6}$ (default) are shown on the green background.}
  \label{fig:tolerance}
\end{figure}

%
%Time performance
%
The construction of the bounce solution, in particular the $v_s, b_s$ parameters, already indicates that
adding more segments does not require significant additional computing cost: evaluation time of the 
sums in~\eqref{eqvbnDg2} grows linearly. 
Thus it is to be expected that the total time should grow linearly with the 
number of field points, which is indeed the case, as displayed in Figure~\ref{fig:tolerance}.
Note that the \fb performance is similar for both thick and thin walls, with a $\sim 10\%$
faster evaluation in the thin wall regime.

This behavior should be contrasted with numerical approaches based on under-over-shooting.
There, the very thin wall limit requires an exponential amount of precision in finding 
$\varphi_{0}$\footnote{As in~\cite{Coleman:1977py}, when $\varphi_0$ is very close to the true vacuum 
$\tilde{\varphi}_{1}$, the solution is given by $\varphi(\rho) - \tilde{\varphi}_{1} = 
2\left( \varphi_{0}-\tilde{\varphi}_{1} \right) I_{D/2-2} (m\rho)/(m\rho)$, 
where $m^2 \equiv V^{\prime\prime}(\tilde\varphi_{1})$. 
Thus in the thin wall limit $ R_{in} \gg m^{-1}$ the initial condition 
$\varphi( R_{in} )\equiv \varphi_{0}$ is exponentially tuned.}.
This may cause numerical approaches to fail or significantly reduce the speed of computation. 
\fb instead relies on the $R_{in}$, which is found by extremizing the action,
similar to the original thin wall approach~\cite{Coleman:1977py}. Such change of variables
thus provides a more stable universal behavior near the thin wall limit.

Speed and accuracy can be controlled as explained in~\S\ref{sec:FBOptions}. 
The bounce field configuration and the action can be computed with different requirements for 
\texttt{"ActionTolerance"}, as shown on the right panel of Figure~\ref{fig:tolerance}.
Even though the boundary conditions are not exactly satisfied at the first segment around $\rho \simeq 3$,
the rest are joined analytically and the solution is smooth. 
Despite the discontinuity, the bounce action is fairly precise and within the limits of required action tolerance.
For comparison, we also show the Euclidean action with different number of \texttt{"FieldPoints"-> 10, 100},
computed with the default \texttt{"ActionTolerance"-> $10^{-6}$}, which shows how the action converges 
with the number of field points.

As a final comment, one can use the \texttt{"FieldPoints"} option to specify a fixed custom segmentation
from which the bounce is obtained. 
This feature may be useful when dealing with non-homogeneous potentials that contain flat pieces, 
followed by local features. 
In such cases, rather than increasing the number of field points, constructing a custom segmentation
may be more beneficial.
In the following subsection we give one such example, where a bi-homogeneous segmentation, set 
by the \texttt{"MidFieldPoint"} option gives a more stable output.

%\newpage
%
% Quartic-quartic potential
%
\subsection{Quartic-quartic potential}\label{subsec:QuarticEx}

\begin{figure}[ht]
  \centering
  \hfill \includegraphics[width=.46\textwidth]{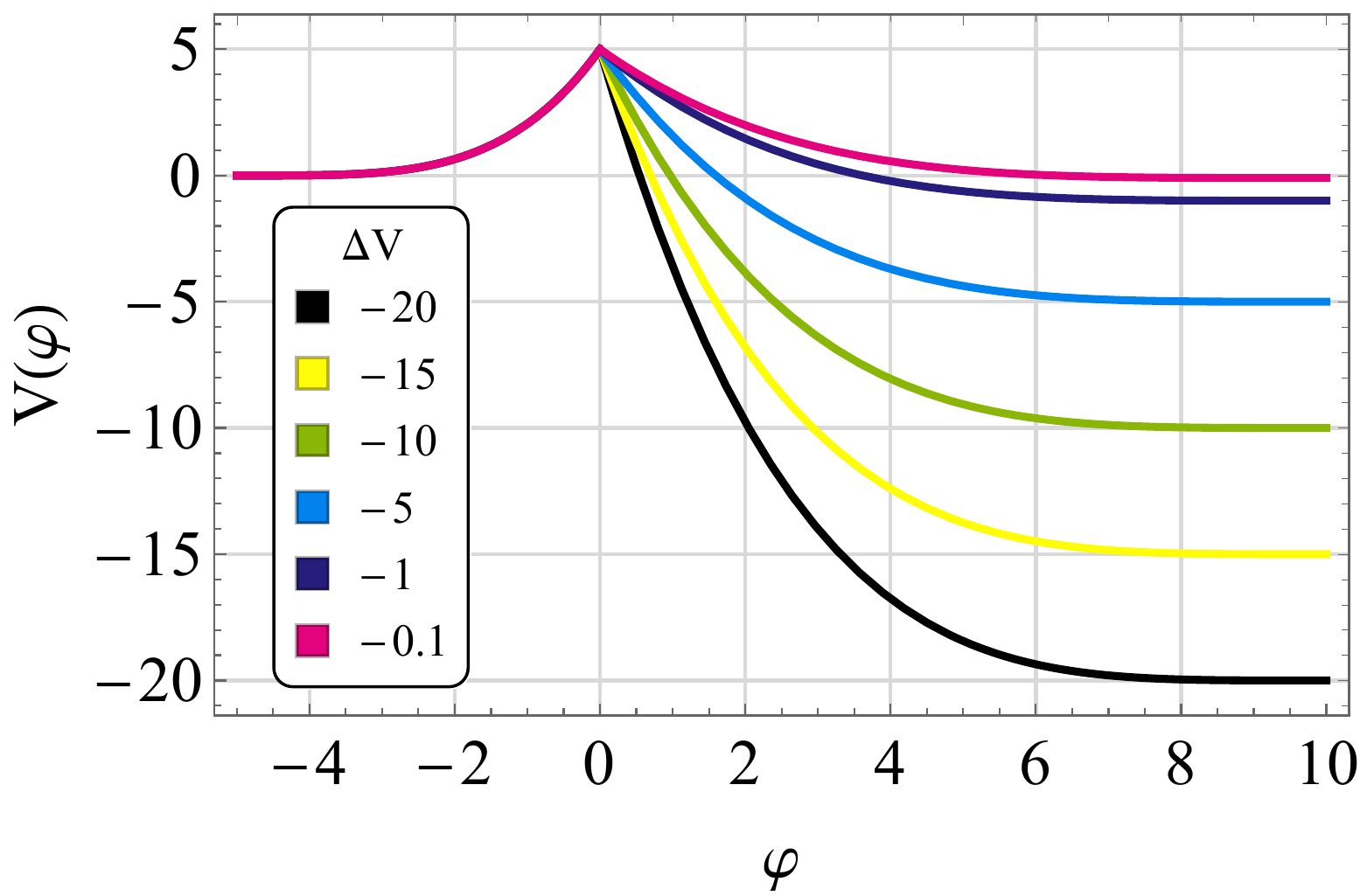}%
  \hspace{2mm}
  \hfill \includegraphics[width=.51\textwidth]{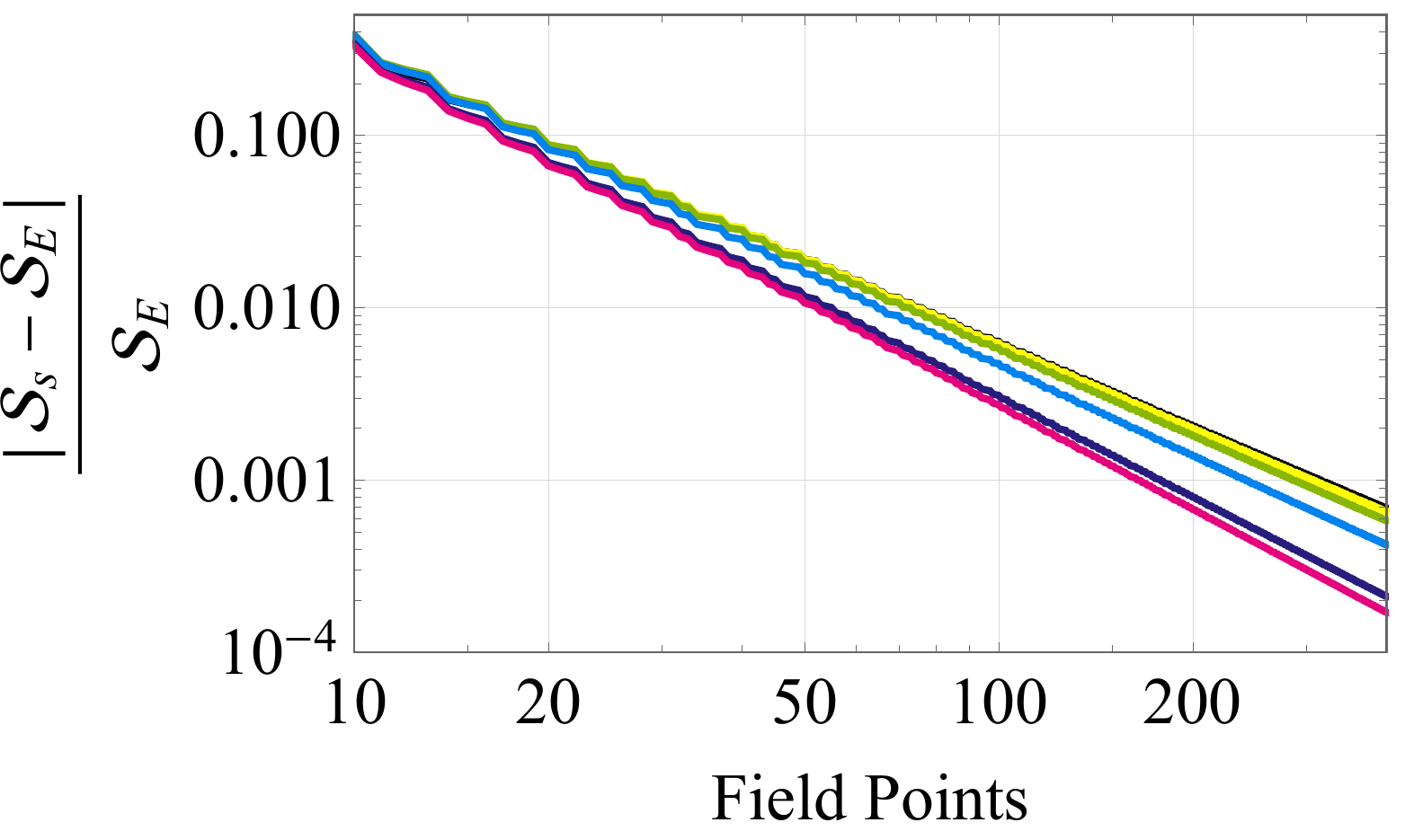} \label{fig:ActionE}
  \hspace*{\fill}
  \caption{
    Left: The piecewise quartic potential for different values of the potential difference between the vacua,
    going from the thin wall $\Delta V = -0.1$ to thick wall $\Delta V = -20$ regime. 
    Right: The bounce action $S_s$ for different 
    number of field points, normalized to the exact result $S_E$ of the quartic-quartic potential.}
  \label{fig:ExactPotential}
\end{figure}

Perhaps the cleanest way of testing the \fb package is to compare it to one of the few non-trivial analytical 
closed-form solutions that are available~\cite{Lee:1985uv,Duncan:1992ai,Dutta:2011rc,Coleman:1980aw}.
Here, we consider the two joined quartic potentials studied in~\cite{Dutta:2011rc}.

For the sake of illustration we choose a potential with fixed local extrema and leave a single parameter 
$\Delta V$, which controls the potential difference between the false and true vacua. As shown on the left of Figure~\ref{fig:ExactPotential}, we vary $\Delta V$ from -20 (thick wall) to $\Delta V =-0.1$ (thin wall). 
The corresponding action values are computed with different number of field points and
are shown on the right panel. They are evaluated at $D=4$ and normalized to the known exact value.

The \fb syntax used for this calculation is again simple.
\begin{mmaCell}[morepattern={x_, x},index=1]{Input}
DV = -20;
V[x_]:= Piecewise[ \{\{0.008 (x + 5)^4, x < 0\},
  \{DV + (5 - DV)(x/10 - 1)^4, x >= 0\}\} ];
\end{mmaCell}
  
\begin{mmaCell}[moredefined={FindBounce,V},morefunctionlocal={x}]{Input}
FindBounce[V[x], x, \{-5, 10\}, "Gradient"-> None, "MidFieldPoint"-> 0]
\end{mmaCell}
Notice that the first derivative of the potential, required in Eq.~\eqref{eqaalpha1}, is not well defined 
at the origin $\varphi = 0$, therefore the default evaluation of the gradient, as well as the automatic extension of 
the polygonal approach, was turned off with \texttt{"Gradient"-> None}.
If this were not the case, \fb would issue a warning message and automatically return the solution
computed without the gradient extension.
Due to the absence of the second order correction, the convergence is a bit slower compared to the 
previous example in~\S\ref{subsec:SingleEx}. Nevertheless, \fb finds the solution within 1\% accuracy 
for 50 (100) field points in thin (thick) wall regime. 

The other option used above is the \texttt{"MidFieldPoint"}, which was used to set the intermediate field 
point of the segmentation to $\varphi = 0$, see the left panel of Figure~\ref{fig:ExactPotential}.
This feature is optional but stabilizes and improves the accuracy of the bounce action, especially with a 
small number of field points. 

%\newpage

% Intermediate minima and disappearing instantons
%
\subsection{Intermediate minima and disappearing instantons}\label{subsec:DisappEx}

\begin{figure}[ht]
  \centering
  \hfill \includegraphics[width=.48\textwidth]{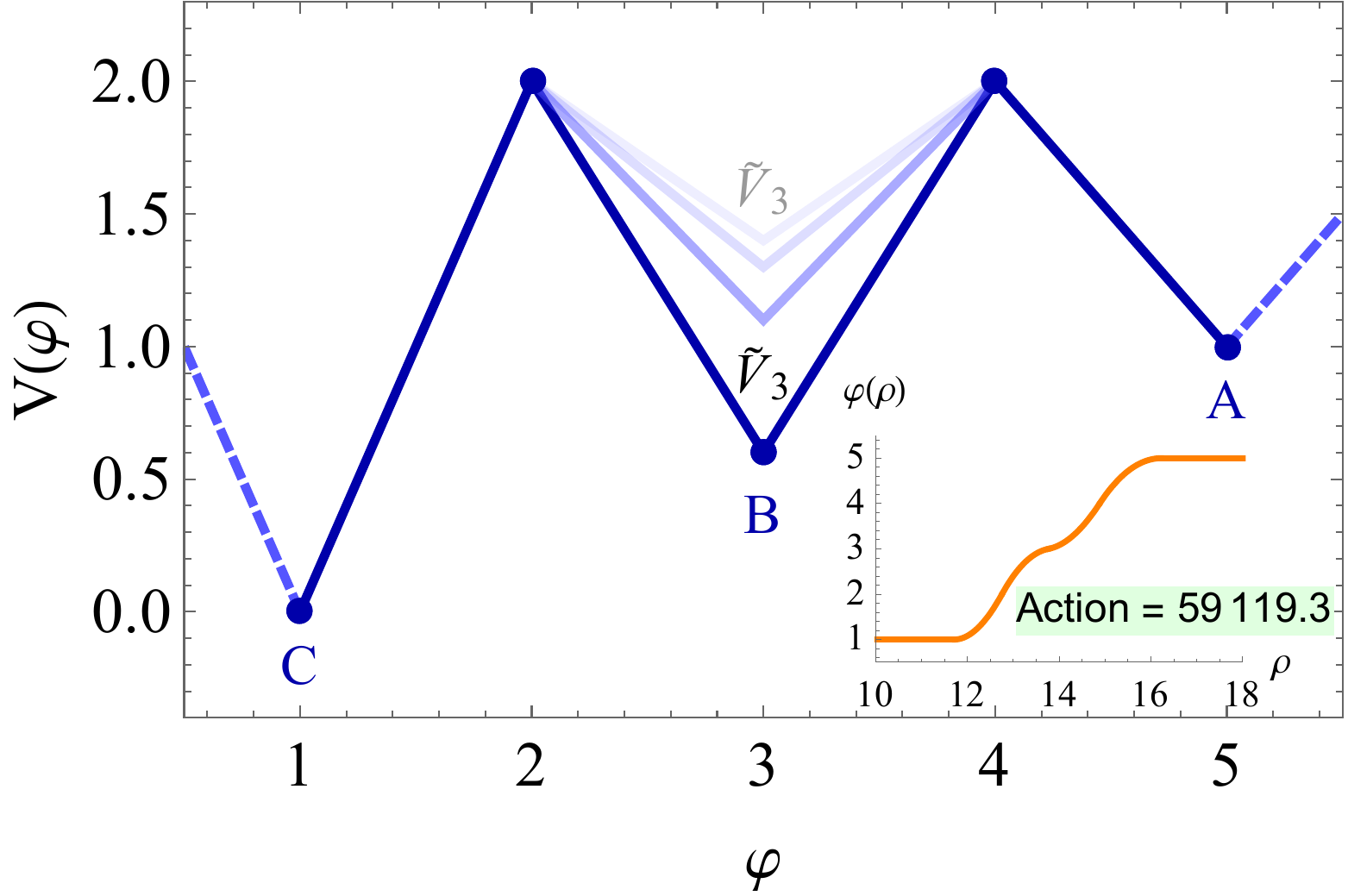}%
  \hspace{1mm}
  \hfill \includegraphics[width=.46\textwidth]{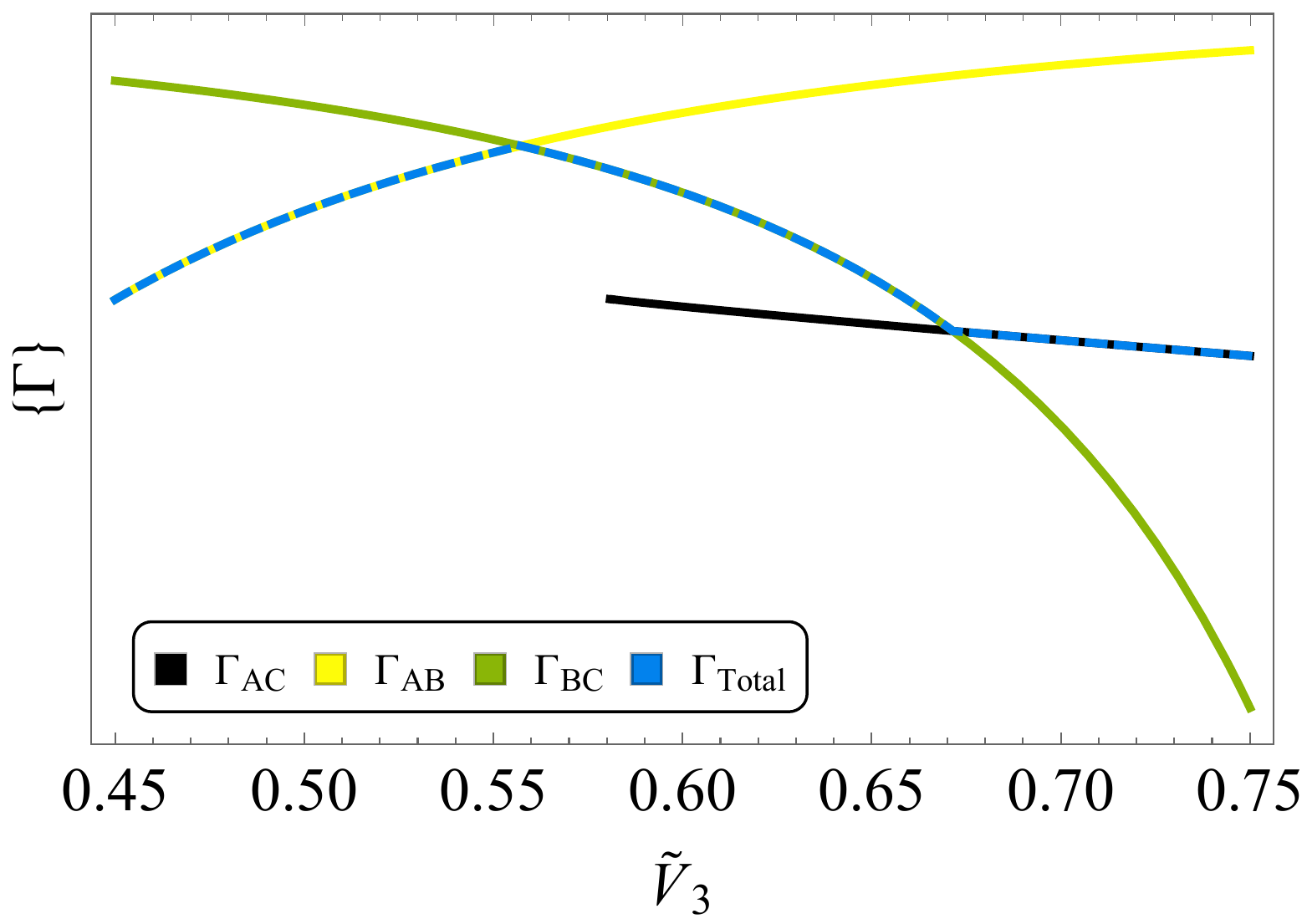} 
  \hspace*{\fill}
  \caption{
    Left: The minimal example of a potential with an intermediate minimum. The inset shows
    the bounce solution with the typical double bubble wall shape.
    Right: Decay rates for the direct (ABC) and the two subsequent (AB, BC) transitions, together
    with the total decay rate~\eqref{eq:GammaT}.}
  \label{fig:IntMinDisInst}
\end{figure}
 
A number of physically motivated models may feature a non-trivial potential with many local minima. 
Such situations appear in multiple axion and relaxion-type 
potentials~\cite{Graham:2015cka,Espinosa:2015eda,Hardy:2015laa,Patil:2015oxa,
Jaeckel:2015txa,Gupta:2015uea,Batell:2015fma,Jaeckel:2016qjp,Fonseca:2019lmc}.
Here we demonstrate the use of the polygonal approach to analyze the minimal
polygonal potential with an intermediate minimum, i.e. the two-triangle construction with $N=5$ 
field points, shown on the left of Figure~\ref{fig:IntMinDisInst}. 

The value of the potential at the mid-point $\tilde V_3$ controls the depth of the intermediate minima 
and the resulting bounce solution. As this is lowered below the highest extremum $\tilde V_3 < \tilde V_5$, two 
types of transitions are possible. The direct tunneling from $\tilde V_5$ to $\tilde V_1$ (the ABC) instanton, 
or a two-step transition first from $\tilde V_5$ to  $\tilde V_3$ (AB) and then from $\tilde V_3$ to 
$\tilde V_1$, the BC instanton. The right panel of Figure~\ref{fig:IntMinDisInst} shows the
associated rates and the total decay rate, defined by
\begin{align}\label{eq:GammaT}
{\Gamma_{total}} \approx \frac{\Gamma_{AB}\Gamma_{BC}}{\Gamma_{AB}+\Gamma_{BC}} + \Gamma_{AC}.
\end{align}

The point of emphasis is that the direct ABC transition exists only up to a certain
value of $\tilde V_3$ and then suddenly disappears. This behavior of a disappearing instanton
was pointed out in~\cite{Brown:2011um} and explained in the thin wall approximation. 
In order to construct the bounce of the direct solution, the field should traverse from 
$\tilde \varphi_1$ to $\tilde \varphi_3$ with a radius associated to the BC transition and finally
from $\tilde \varphi_3$ to $\tilde \varphi_5$ at the AB radius.
Clearly, to have a meaningful solution $R_{BC} < R_{AB}$. 
However, when we lower the intermediate minimum, $\tilde V_3$ comes increasingly close to 
$\tilde V_1$, thus $R_{AB}$ grows larger,
and thereby $\Gamma_{AB}$ decreases until direct tunneling via the ABC instanton becomes impossible. 
This is seen on the right panel of Figure~\ref{fig:IntMinDisInst}, where the $\Gamma_{ABC}$ in black
suddenly disappears. Of course, the two-step decay still exists, i.e. $\Gamma_{AB}$ and $\Gamma_{BC}$ 
are in fact non-zero.\footnote{Strictly speaking, one should perform the two-step analysis more carefully,
and allow the field to develop in real time after tunneling from $\tilde V_5$ in the vicinity of $\tilde V_3$.
This would lead to oscillations with subsequent decay, see e.g.~\cite{Darme:2017wvu,Sarangi:2007jb,
Saffin:2008vi,Tye:2009rb,Darme:2019ubo} and potential enhancement of the rate. We leave this 
interesting question for future studies~\cite{Guada:2020}.}

To study such particularly simple settings, \fb allows the user to manually set individual values of the potential 
as a list of points in field space $\left \{ \left\{ \tilde{\varphi}_1, \tilde{V}_{1} \right\}, \ldots ,\left\{ {\varphi}_N, 
\tilde{V}_{N} \right\} \right\} $. For example, the direct ABC instanton of Figure~\ref{fig:IntMinDisInst} is obtained with the 
following syntax.
\begin{mmaCell}[moredefined={FindBounce},index=1]{Input}
FindBounce[\{\{1,0\},\{2,2\},\{3,0.6\},\{4,2\},\{5,1\}\}]
\end{mmaCell}
\begin{mmaCell}[moregraphics={moreig={scale=.3}}]{Output}
\mmaGraphics{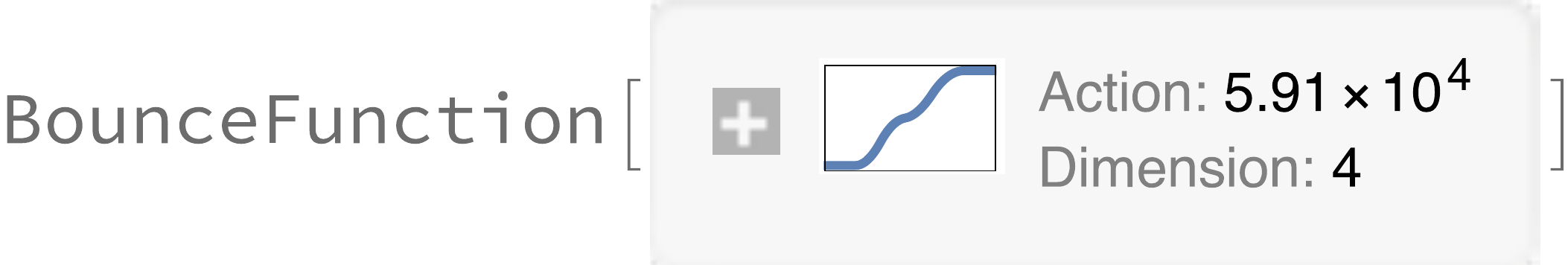}
\end{mmaCell}
Note also that the order of points in the list is arbitrary and can be given from left to right or vice versa.

%\newpage
%
% Potentials unbounded from below
%
\subsection{Potentials unbounded from below} \label{sec:Unbound}

In this section, we demonstrate how \fb can be used to deal with unstable potentials.
Many BSM theories contain portions of parameter space with unstable field directions,
such that local ground states may tunnel into the unstable region.
This seems to be the case for the Standard Model as well, with the estimated lifetime
significantly longer than the age of the universe, see e.g.~\cite{Degrassi:2012ry,Buttazzo:2013uya,Chigusa:2017dux,Andreassen:2017rzq,Chigusa:2018uuj}.

Note that the quartic-quadratic potentials do not admit the bounce solution at tree level 
due to scale invariance~\cite{Espinosa:2019hbm}, however the bounce may exists after 
the inclusion of quantum corrections. In any case, one might expect the unstable direction to 
be dominated by the negative quartic term at large field values. 

The escape point can be found either by the usual numerical shooting method or with the polygonal
approach. However, due to the steepness of the unstable direction, polygonal segments might
need to be extend to large values, which may require a large number of segments.
In order to provide a good estimate for large $\varphi_0$, we demonstrate how the poly-linear 
potential with many segments can be joined with an exact unstable quartic solution.
This functionality is implemented in \fb single field potentials in $D=4$. It can be turned on with the
\texttt{"BottomlessPotential"-> True}, option, together with the field values $\tilde \varphi_N$ and  
$\tilde \varphi_2$. The latter is the point where the PB is connected to the unstable quartic.

To understand how the estimate works, consider the quartic potential $V_q$ and the associated
solution
\begin{align} \label{eq:quartic}
  V_q(\varphi) &= V_{0} - a_{q} \left(\varphi - v_{q} \right)^4,
  &
  \varphi_{q}(\rho) &= v_q + \frac{b_{q}}{1+\frac{1}{2} a_{q} b^2_{q} \rho^2},
\end{align}
where $v_q$ and $b_q$ are constants of integration and $a_q > 0$ is a dimensionless parameter 
of the potential. Assuming the unstable $\varphi^4$ term dominates for large field values, $a_{q}$ is fixed 
by equating it to the $\varphi^4$ coefficient of the input potential $V(\varphi)$.

Similar to the pure polygonal construction explained in section~\S\ref{sec:FVBasic} above, the quartic 
piece of the potential can be matched to the polygonal ones. In particular, the coefficients $b_q, v_q$ are 
then determined by the boundary conditions $\varphi_{q}(R_{2}) =\varphi_{PB}(R_{2}) = \tilde \varphi_2$ 
and $\dot\varphi_{q}(R_{2}) =\dot\varphi_{PB}(R_{2})$. What remains to be determined
is the matching radius $R_2$, which can be found with \texttt{FindRoot}, similar to the polygonal case 
above. Finally, we fix $V_0$ by requiring the potential to be continuous.

\begin{figure}
  \centering
  \hfill \includegraphics[width=.48\textwidth]{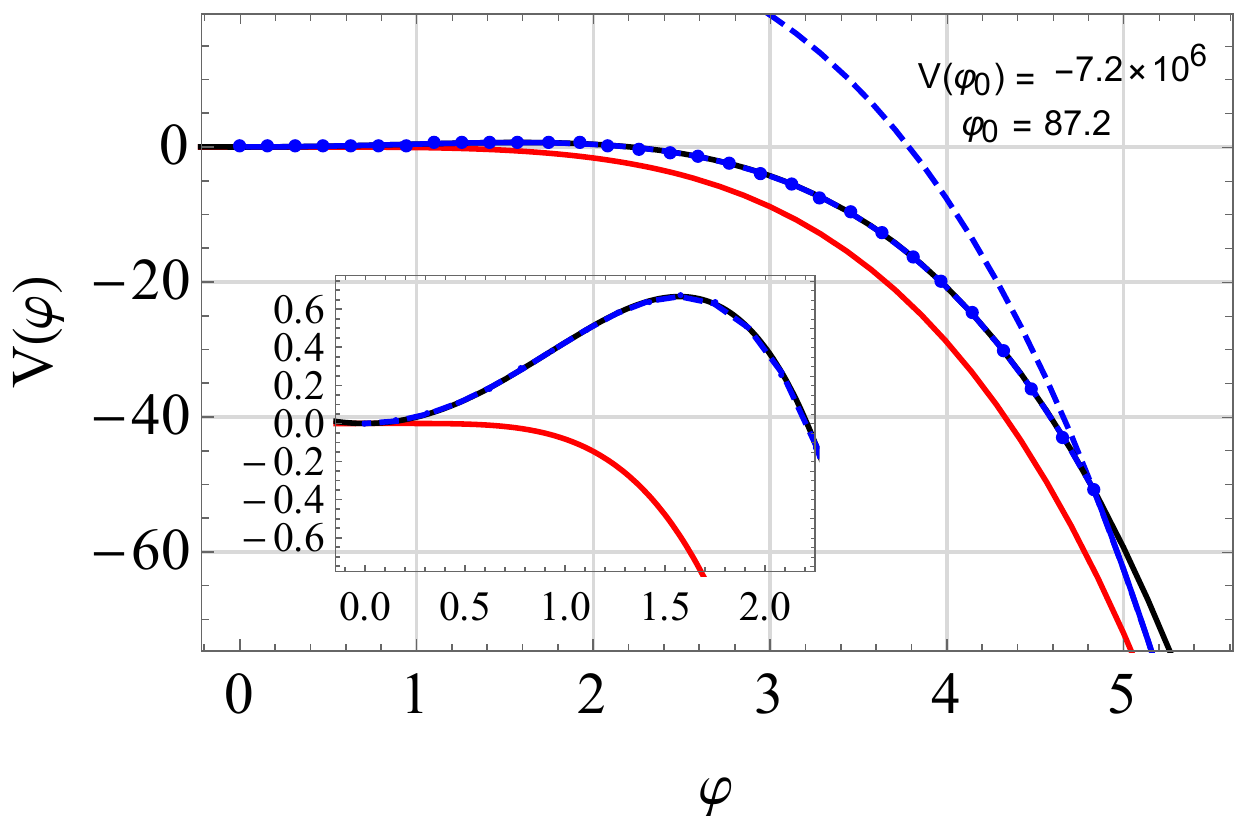}%
  \hspace{1mm}
  \hfill \includegraphics[width=.48\textwidth]{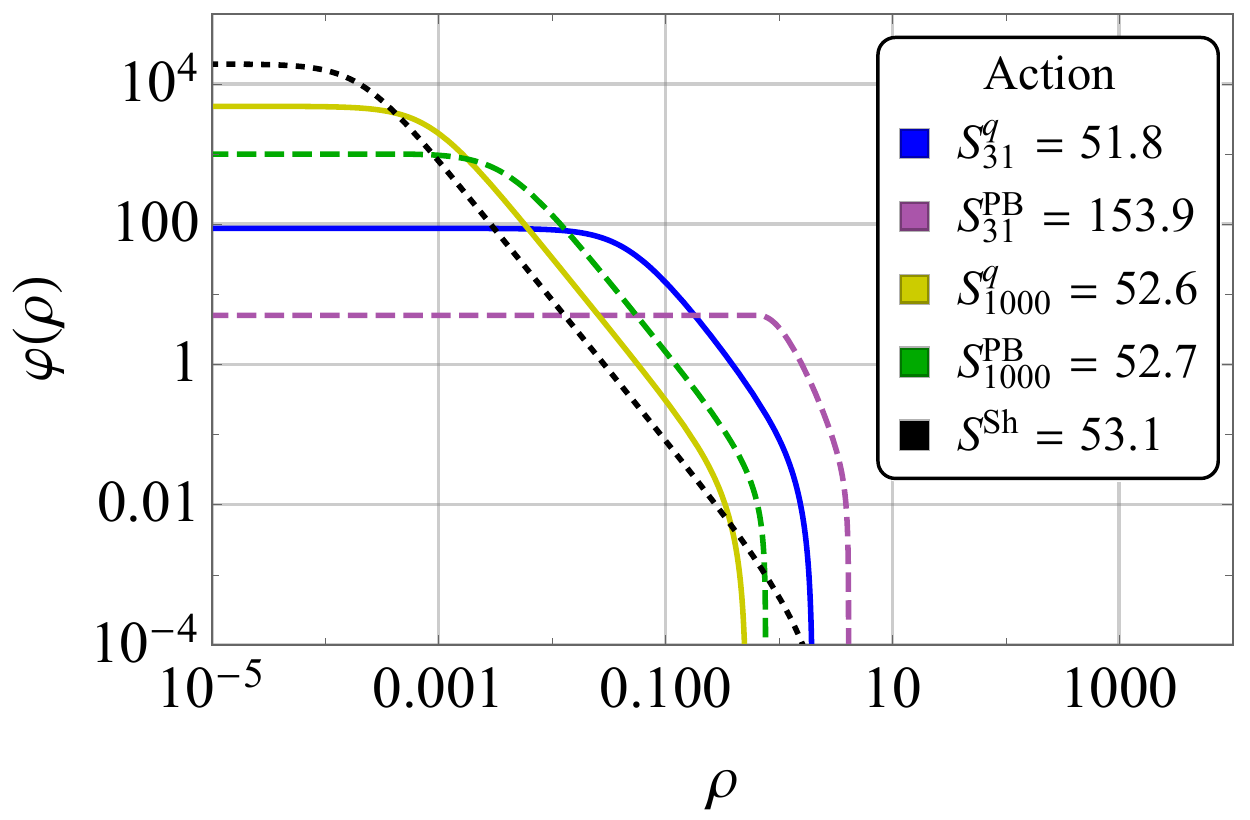} 
  \hspace*{\fill}
  \caption{
    Left: Potential unbounded from below. The pure quartic, polygonal-quartic and input potentials are 
    shown in red, solid blue and black, respectively. 
    The dashed blue line is the pure quartic potential that is joined to the piece-wise polygonal potential.
    Right: The bounce field configuration with its bounce action. 
    The solid lines show the field configuration with PB+quartic estimate, the dashed are the pure
    polygonal solutions and the dotted line is the numerical result obtained with over-under-shooting.
  }
  \label{fig:bottomless}  
\end{figure}

We demonstrate the use of \fb in such unbounded transitions with the following example,
where we specify the minimum at $\varphi = 0$ and the connecting point to the quartic at $\varphi = 5$.
\begin{mmaCell}[pattern={x_,x},index=1]{Input}
V[x_]:= 0.5 x^2 + 0.05 x^3 - 0.125 x^4;
\end{mmaCell}
\begin{mmaCell}[morefunctionlocal={x},moredefined={FindBounce,V}]{Input}
bf = FindBounce[V[x], x, \{0, 5\}, "BottomlessPotential"-> True];
\end{mmaCell}
\begin{mmaCell}[moredefined={bf}]{Input}
\{vq, aq, bq\} = bf["Coefficients"];
V0 = bf["BottomlessPotential"];
\end{mmaCell}
\begin{mmaCell}[moredefined={bf,v,a,b,V0},morefunctionlocal={x}]{Input}
Plot[V0 - aq[[1]](x - vq[[1]])^4, \{x, bf["Path"][[1,1]],bf["Path"][[2,1]]\}]
\end{mmaCell}
%
%(*Bounce coefficient and potential constant value of equation 23*)
%(*Plots the pure quartic potential (blue dashed line of figure 6)*)
%\vspace{2cm}
%
The potentials of interest are shown on the right side of Figure~\ref{fig:bottomless} where the red, solid 
blue and black lines represent the pure quartic $V_q$, polygonal-quartic and the input potential, respectively. 
The dashed blue line is the pure quartic potential~\eqref{eq:quartic} that was joined to the polygonal potential. 
Its parameters are given by \texttt{"Coefficients"} and \texttt{BottomlessPotential} as shown in the syntax below. 
The bounce solution given by \fb is then shown on the right of Figure~\ref{fig:bottomless}. Notice that
the materialization of the bounce happens at $\varphi_0 \sim 90$, much above the connecting point at $\varphi = 5$. 

%~\footnote{Units are ignored here due to the fact that one can always re-scale and make a change of variables.} 

%\newpage
%
% Multi field examples
%
\subsection{Two field benchmark} \label{subsec:TwoField}

Many extensions of the SM feature additional scalar fields, see~\cite{Branco:2011iw} for a review. 
Extra scalars can couple to the SM Higgs and may alter the vacuum structure, potentially 
triggering a first order phase transition.
The number of additional fields in generic SM extensions may be large. However, in many cases
it may be sufficient to consider the dynamics of two fields only.

As a simple multi-field example in \fb, let us consider such two field potential with parameters chosen
such that the field dynamics produces a large curvature in field space.
We start with the simplest tree-level potential at zero temperature, compute the bounce action and show some additional 
features of \fb. In the upcoming~\S\ref{sec:pheno} we show how to deal with finite temperature and 
thermal corrections.

Let us then consider the following example, where we call \fb on the two minima
and ask it to connect them with an intermediate point, set by \texttt{"MidFieldPoint"}.
We also show how the pre-computed gradient function can be specified.
\begin{mmaCell}[pattern={h_,h,s,s_},index=1]{Input}
V[h_,s_]:= -100 h^2 + 0.1 h^4 - 60 s^2 + 0.3 s^4 + 3 h^2 s^2;
minima = \{\{0.,10.\},\{22.4,0.\}\};
\end{mmaCell}
\begin{mmaCell}[moredefined={FindBounce,V,dV,minima,saddlePoint}]{Input}
bf = FindBounce[V[\mmaFnc{h},\mmaFnc{s}], \{\mmaFnc{h},\mmaFnc{s}\}, minima, "MidFieldPoint"-> \{6,6\},
  "Gradient"-> \{-200 h + 0.4 h^3 + 6 h s^2, -120 s + 1.2 s^3 + 6 s h^2\}];
\end{mmaCell}
\begin{mmaCell}[morefunctionlocal={h,s,r},moredefined={bf,V}]{Input}
\{Ri,Rf\} = bf["Radii"][[\{1,-1\}]];
Show[
  ContourPlot[V[h,s], \{h,-1,25\}, \{s,-1,11\},Contours->50],
  ParametricPlot[ Through@bf["Bounce"][r], \{r, Ri, Rf\}]]
\end{mmaCell}

\begin{figure}
  \centering
  \hfill \includegraphics[width=.48\textwidth]{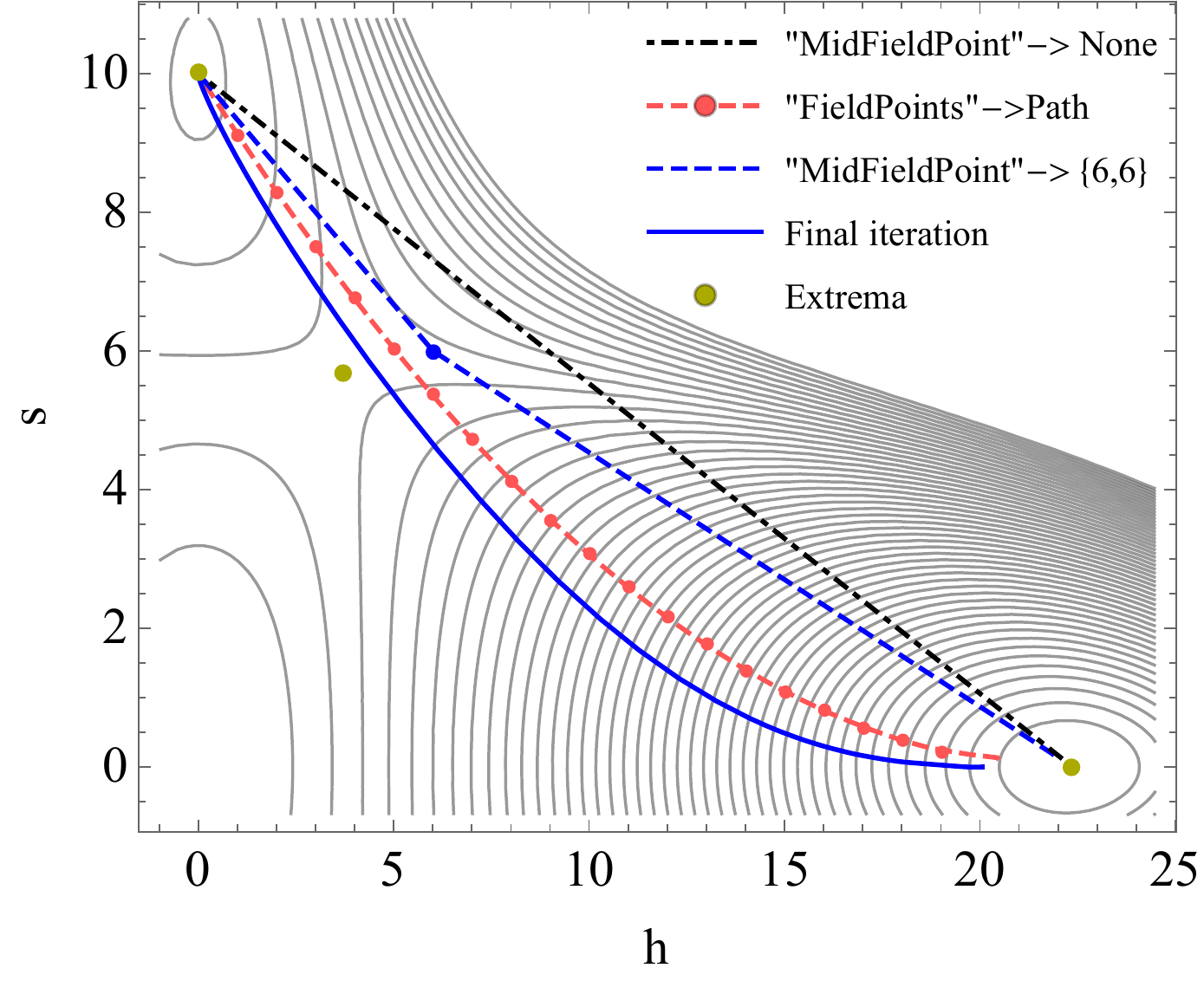}%
  \hspace{1mm}
  \hfill \includegraphics[width=.48\textwidth]{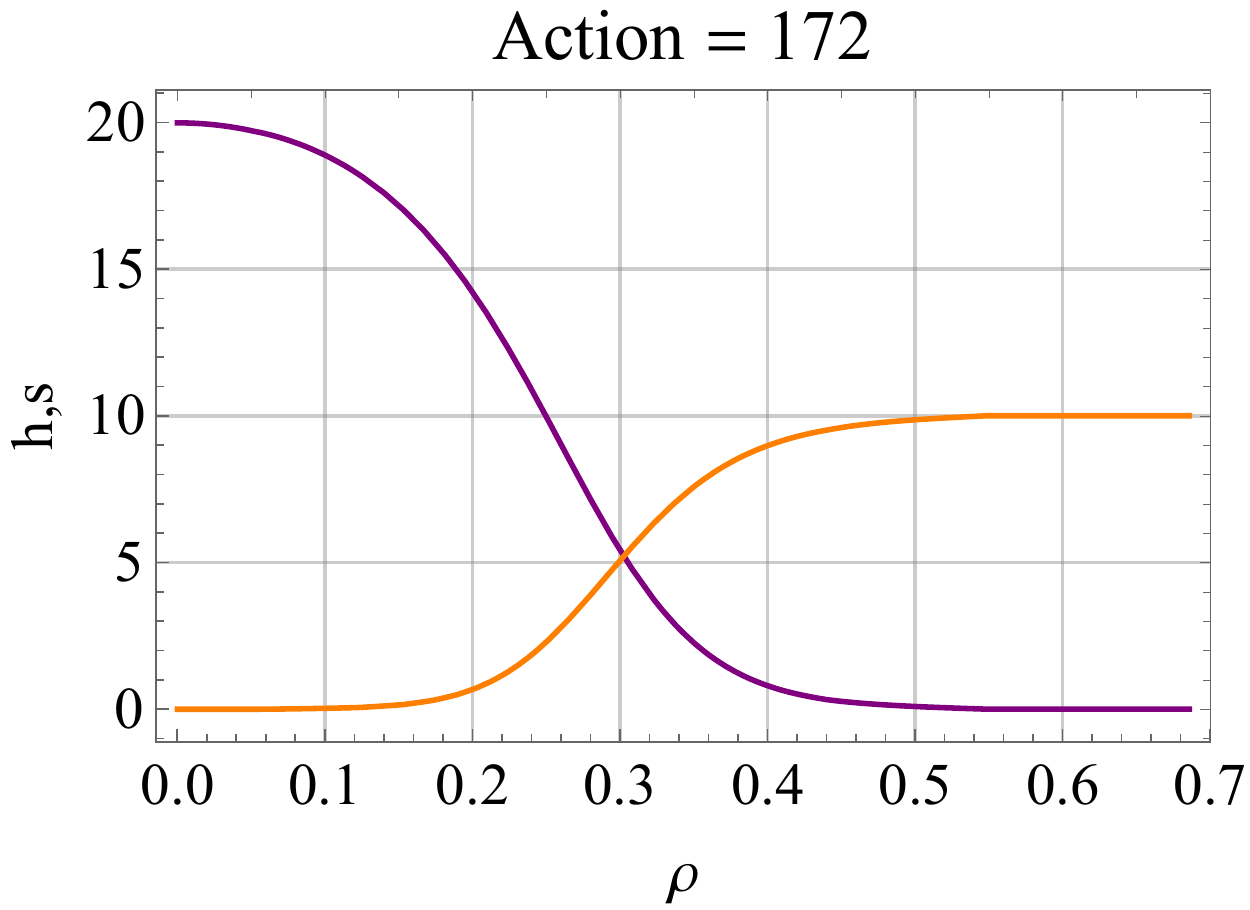} 
  \hspace*{\fill}
  \caption{
    Left: Potential contours in solid black lines, the final trajectory of the bounce field in solid blue line and 
    three different initial paths in dashed. Black dot-dashed line is the default straight line ansatz, 
    the straight dashed blue line includes the intermediate point at \{6,6\}, while the red dashed line is a
    parabola, set by hand.    
    Right: The field configuration $(h(\rho), s(\rho))$ of the final iteration and the associated bounce action
    in the caption.}
  \label{fig:OptionValue}
\end{figure}

% (*Returns the right panel of Figure 7*)
\begin{mmaCell}[moredefined={BouncePlot,bf}]{Input}
BouncePlot[bf, 
  PlotLabel->Row[\{"Action = ",Round@bf["Action"]\}],
  PlotStyle->\{Purple, Orange\}]
\end{mmaCell}
The code above returns the bounce field configuration corresponding to the solid blue line in
field space, shown on the left of Figure~\ref{fig:OptionValue}. The Euclidean time profiles 
($h(\rho),s(\rho)$) can also be plotted easily and are shown on the right of Figure~\ref{fig:OptionValue}.

The dashed lines in Figure~\ref{fig:OptionValue} represent the different choices of the initial path.
By default, \fb chooses a straight line from one minimum to the other, seen by the black dashed line.
In case there is a specific point that the segmentation should follow, such as the saddle point, or
an arbitrary point in the above example, it is specified with the \texttt{"MidFieldPoint"} option.
Finally, one can start with a completely arbitrary initial path, set by the \texttt{"FieldPoints"} option. 
In the example above, the dashed red line on the left of Figure~\ref{fig:OptionValue} was obtained
with a parabola connecting the two minima. The latter option is particularly useful when we already
have some idea about the path in field space, e.g. when performing potential parameter scans.

Whatever the choice of the initial path is, \fb iterates the path deformation procedure until it reaches at 
least one of the following three conditions. 
\begin{enumerate}
  \item The maximum number of iterations, controlled by \texttt{"MaxPathIterations"}. Here, zero 
  means no perturbation of the initial path; i.e. all the dashed lines in Figure~\ref{fig:OptionValue} 
  were obtained by setting \texttt{"MaxPathIterations"->} $0$.
  \item Path deformation measure, controlled set up by \texttt{"PathTolerance"}, as explained 
  in~\S\ref{opt:PathTolerance}. 
  \item \texttt{"ActionTolerance"} that directly measures the change of the Euclidean action between
  iterations.
\end{enumerate}
The resulting field configurations are shown in solid blue line on the left of Figure~\ref{fig:OptionValue}.

%
% Thermal corrections and nucleation temperature
%
\subsection{Thermal corrections and nucleation temperature} \label{sec:pheno}

As a phenomenological application of \fb, let us compute the nucleation temperature of a SM
extension with a real scalar singlet. 
This model may support a first order phase transition and might successfully explain phenomena 
such as baryogenesis~\cite{Huber:2000mg, Huber:2006wf, Patel:2011th, Cline:2012hg, 
Chala:2016ykx}, 
dark matter~\cite{Burgess:2000yq, McDonald:2001vt, Gonderinger:2009jp, Cline:2013gha} and 
act as a possible source of gravitational 
waves~\cite{Caprini:2009yp, Espinosa:2010hh, Binetruy:2012ze, Thrane:2013oya, Ellis:2019oqb}.

We consider the SM Higgs $h$ together with the singlet scalar singlet field $s$. For simplicity,
we assume an additional $\mathbb{Z}_2$ symmetry and define the tree-level renormalizable 
potential
\begin{align} \label{eq:PotentialTree}
V_{\mathrm{tree}}(h, s) = - \frac{1}{2} \mu_h^2 h^2 + \frac{1}{2} \mu_s^2 s^2 
      + \frac{1}{4} \lambda_h h^4 + \frac{1}{4} \lambda_s s^4 
      + \frac{1}{4} \lambda \,s^2 h^2.
\end{align}
One-loop thermal corrections to the potential above can be computed using the equations 
presented in the Appendix~\S\ref{Ap:EffectivePotential}. 
For excellent references, see~\cite{Dolan:1973qd,Weinberg:1974hy}, a review~\cite{Quiros:1999jp} 
or textbook~\cite{Kapusta:2006pm}.
The exact thermal one-loop functions $J_{\nicefrac{B}{F}}$ in~\eqref{ap:Jfunction} were implemented
efficiently in~\texttt{C++}~\cite{Fowlie:2018eiu} and can be employed in \mth with the provided interface.
However, we remain in the high-$T$ limit and use the closed form given in Eqs.~\eqref{ap:JBapprox}
and~\eqref{ap:JFapprox}. 
These are valid up to $\mathcal{O}(T^4)$, while neglecting the contributions of the quartic coupling 
of the potential for simplicity.

As shown in~\cite{Espinosa:2011ax}, the thermal one-loop corrections to the potential 
in~\eqref{eq:PotentialTree} at high-$T$ are given by
\begin{align}\label{eq:PotentialTemp}
  V(h, s, T) = \frac{1}{2} \left( c_h h^2 + c_s s^2 \right) T^2,
\end{align}
\begin{align}
  c_h &= \frac{2 M_W^2 + M_Z^2 + m_h^2 + 2 m_t^2}{4 v^2} + \frac{\lambda}{4!},
  &
  c_s  &= \frac{2 \lambda + 3 \lambda_s}{12}.
\end{align}
Here we considered the $W, Z$ gauge bosons and the top quark contributions to the effective
potential and neglected the other light fermions. 

Following~\cite{Athron:2019nbd, Espinosa:2011ax}, the tree level potential~\eqref{eq:PotentialTree} 
can be redefined as a function of the critical temperature and couplings of the singlet fields 
$\{T_{C},\lambda_{s},\lambda\}$ respectively.
Taking into account the constrains that lead to a first order phase transition and requiring that the 
deeper minimum becomes our Higgs vacuum as $T$ is lowered, the effective quadratic couplings 
turn into:
\begin{align}
  \mu_h^2(T) &= \lambda_h v_h^2(T),
  &
  \mu_s^2(T) &= -\lambda_s v_s^2(T),
  &
  \lambda_h  &= \frac{m^2_h}{2 v^2}.
\end{align}
The corresponding vacuum expectation values (vevs) are then
\begin{align}
  v_h^2(T) &= v^2\left(1-\frac{2 c_{h}}{m_{h}^2} T^2 \right),
  & 
  v_{s}^2(T) &= \frac{1}{\lambda_{s}} \left( \left(\frac{m_{h}}{2 v} v_{h}^2(T) \sqrt{2\lambda_{s}} 
  + c_{s} T_{C}^2 \right) - c_{s} T^2 \right),
\end{align}
where $v$ is the SM vev at zero temperature and the minima of the potential are
$\left\{v_h(T), 0\right\}$ and $\left\{0, v_s(T)\right\}$. 

With the thermal and quantum corrections in place, we can show how the nucleation temperature 
$T_N$ can be computed with \fb. The $T_N$ is defined as the temperature when the probability 
for a single bubble to be nucleated within one horizon volume is $\sim 1$~\cite{Anderson:1991zb}.
Assuming radiation domination and the SM degrees of freedom in the thermal plasma, the above
requirement roughly translates to $B = S_3/T_N \approx 140$.

Let us consider the benchmark in~\cite{Athron:2019nbd, Piscopo:2019txs} and use \fb to
determine the variation of $B$ with temperature.
\begin{mmaCell}[pattern={h_,h,s,s_,T_,T},index=1]{Input}
V[h_, s_, T_] := -uh2[T]/2*h^2+us2[T]/2*s^2+lh/4 h^4+ls/4*s^4+l/4*s^2 h^2;
ch = (2 mW^2 + mZ^2 + mh^2 + 2 mt^2)/(4v^2) + l/24;
cs = (2 l + 3 ls)/12;
uh2[T_] := mh^2/(2v^2) vh2[T];
us2[T_] := -ls*vs2[T];
vh2[T_] := v^2(1 - 2 ch/mh^2 T^2);
vs2[T_] := ((mh/(2 v) vh2[TC]*Sqrt[2 ls] + cs*TC^2) - cs*T^2)/ls;
\end{mmaCell}

\begin{mmaCell}[]{Input}
\{mW, mZ, mh, v, mt\} = \{80.4, 91.2, 125.1, 246.2, 173.2\};(*GeV*)
\{TC, l, ls, lh\} = \{110(*GeV*), 1.5, 0.65, mh^2/(2 v^2)\};
\end{mmaCell}

\begin{mmaCell}[moredefined={mh,V,FindBounce,vh2,vs2},pattern={T_,T,fp,fp_},morefunctionlocal={h,s}]{Input}
(*S3/T at finite temperature*)
BT[T_?NumericQ,fp_] := 1/T*FindBounce[
  V[h, s, T], \{h, s\},\{\{Sqrt@vh2[T], 0\},\{0,Sqrt@vs2[T]\}\},
  "Dimension" -> 3, "FieldPoints"-> fp ]["Action"];
\end{mmaCell}

\begin{mmaCell}[moredefined={BT,LogPlot,PerformanceGoal},morefunctionlocal={T}]{Input}
LogPlot[\{BT[T,10],BT[T,31], 140\}, \{T, 70, 100\},
   AxesLabel -> \{"T(GeV)", "S3/T"\}, PerformanceGoal -> "Speed"]
\end{mmaCell}
\begin{mmaCell}[moregraphics={moreig={scale=.45}}]{Output}
\mmaGraphics{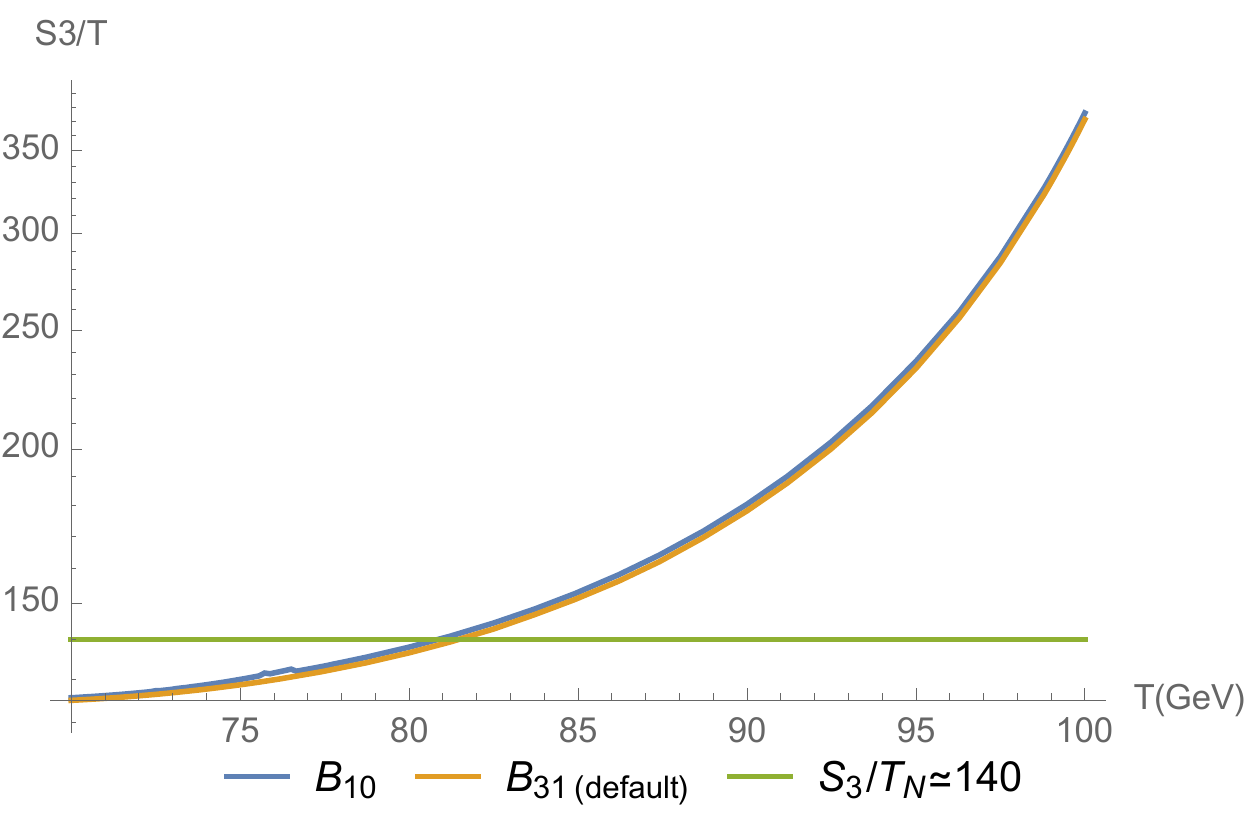}
\end{mmaCell}

\begin{mmaCell}[moredefined={BT}]{Input}
Round[T /.FindRoot[BT[\mmaFnc{T},10] == 140,\{\mmaFnc{T},100\}]]
\end{mmaCell}
\begin{mmaCell}[]{Output}
81
\end{mmaCell}
The nucleation temperature we get is approximately $81\, \mathrm{GeV}$, where $B_{10}(B_{31})$ 
is the normalized action, obtained with with 10(31) field points.

Clearly, one can improve the precision by increasing the number of field points \texttt{"FieldPoints"} and action 
tolerance with \texttt{"ActionTolerance"}. 
However, it should be kept in mind that one should also consistently consider quantum corrections, 
daisy resummation terms, the pre-factor $A$ in~\eqref{eq:FVDecay} and contributions from 
lighter fermions, among others. 
Taking the above leading contributions only, \fb with only a few field points (10) already gives a 
reasonably accurate estimate, while being computationally inexpensive.

%
% Beyond two fields
% 
\subsection{Beyond two fields} \label{subsec:BeyondTwo}
We devote this section to the estimates of the computational time of \fb with an arbitrary  
number of fields. 
We start with a simple example and show the \mth code to obtain the minima and compute the
bounce configurations.
Our main result is that the time requirement increases linearly with respect to the number of 
fields and that the calculation with 10 fields takes $\sim 1$ second. 
We compare the results to other available packages in the literature and test \fb with up to 20 fields.

From~\cite{Athron:2019nbd}, we consider the multi-field potential as a function of the number of fields, 
\begin{align} \label{eq:PotentialMultifield}
V(\varphi) = \left(\sum_{i=1}^{n_{\varphi}} c_i \left( \varphi_i - 1 \right)^2 - c_{n_{\varphi} + 1} \right) 
\sum_{i=1}^{n_\varphi}  \varphi_i^2,
\end{align}
where $c_i$ take values between 0 to 1. 
The position of the minima is a point in field space with components close to 0 and 1. 
Given the number of fields $n_\varphi = 6$ and a constant set of random parameters $c_i$, 
the code to compute the bounce is:
\begin{mmaCell}[morefunctionlocal={r,i},index = 1,moredefined={RandomReal}]{Input}
nf = 6;
SeedRandom[1];
c = RandomReal[1, nf + 1]; 
phi = Table[Symbol["phi" <> ToString[i]], \{i, nf\}];
\end{mmaCell}
\begin{mmaCell}[moredefined={nf,c},morefunctionlocal={i},pattern={phi_,phi}]{Input}
V[phi_] := (Sum[c[[i]](phi[[i]]-1)^2,\{i,nf\}]-c[[-1]])Sum[phi[[i]]^2,\{i,nf\}];
\end{mmaCell}
\begin{mmaCell}[moredefined={nf,V,ConstantArray,phi}]{Input}
extrema = Table[FindRoot[D[V[\mmaFnc{phi}] == 0, \{\mmaFnc{phi}\}],
  Transpose@\{phi, ConstantArray[\mmaFnc{phi0}, nf]\}], \{\mmaFnc{phi0},0,1\}];
\end{mmaCell}
\begin{mmaCell}[moredefined={nf,extrema,V,ConstantArray,phi,DeleteDuplicates},morefunctionlocal={i}]{Input}
\{minima, d2V\} = \{phi /. extrema, D[V[phi],\{phi\},\{phi\}] /. extrema\};
ei = Table[DeleteDuplicates@Sign[Eigenvalues@d2V[[i]]], \{i,2\}];
typeV = Table[If[Length@ei[[i]] > 1, 
   "Saddle", If[ei[[i, 1]] > 0, "Minimum", "Maximum"]], \{i,2\}]
\end{mmaCell}
\begin{mmaCell}{Output}
\{Minimum,Minimum\}
\end{mmaCell}
\begin{mmaCell}[morefunctionlocal={phi},moredefined={minima,V,FindBounce}]{Input}
bf = FindBounce[V[phi], phi, minima]
\end{mmaCell}
\begin{mmaCell}[moregraphics={moreig={scale=.7}}]{Output}
\mmaGraphics{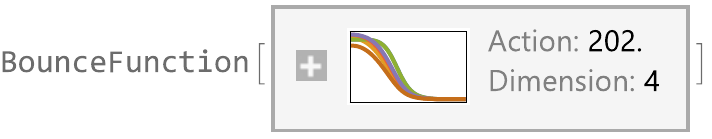}
\end{mmaCell}
\begin{mmaCell}[morefunctionlocal={r},moredefined ={bf,BouncePlot,PlotLegends,LineLegend,LegendLabel}]{Input}
BouncePlot[ bf, PlotLegends -> LineLegend[Automatic, LegendLabel-> "Fields"]]
\end{mmaCell}
\begin{mmaCell}[moregraphics={moreig={scale=.45}}]{Output}
\mmaGraphics{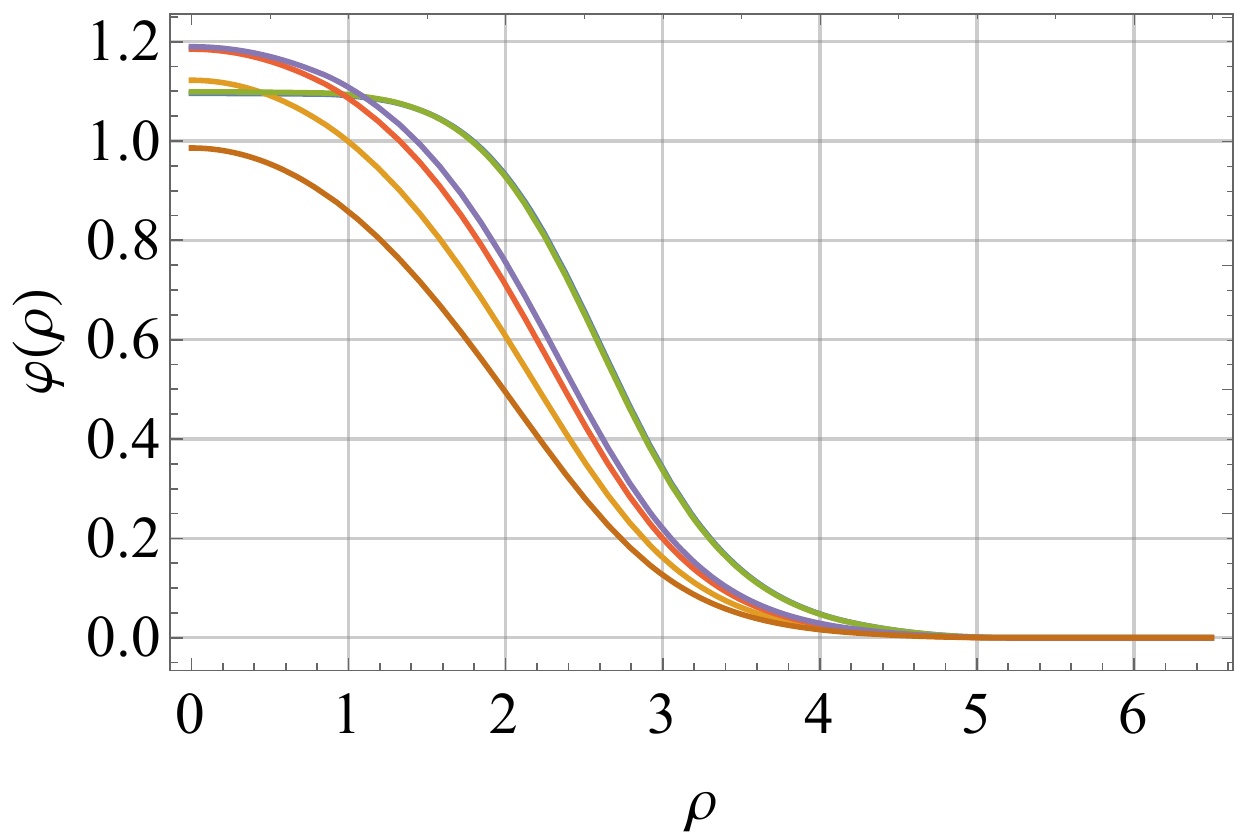}
\end{mmaCell}
Since $c_i$ were chosen at random, \texttt{FindRoot} might not find the minimum of the potential 
but a saddle point instead. In such case \fb returns \texttt{\$Failed}. 
For more general potentials, the extrema can be computed with 
\texttt{Vevacious}~\cite{Camargo-Molina:2013qva}, which is a tool that finds all the tree-level extrema 
of a generic one-loop effective potential.
 
\begin{figure}
  \centering
  \includegraphics[width=.53\textwidth]{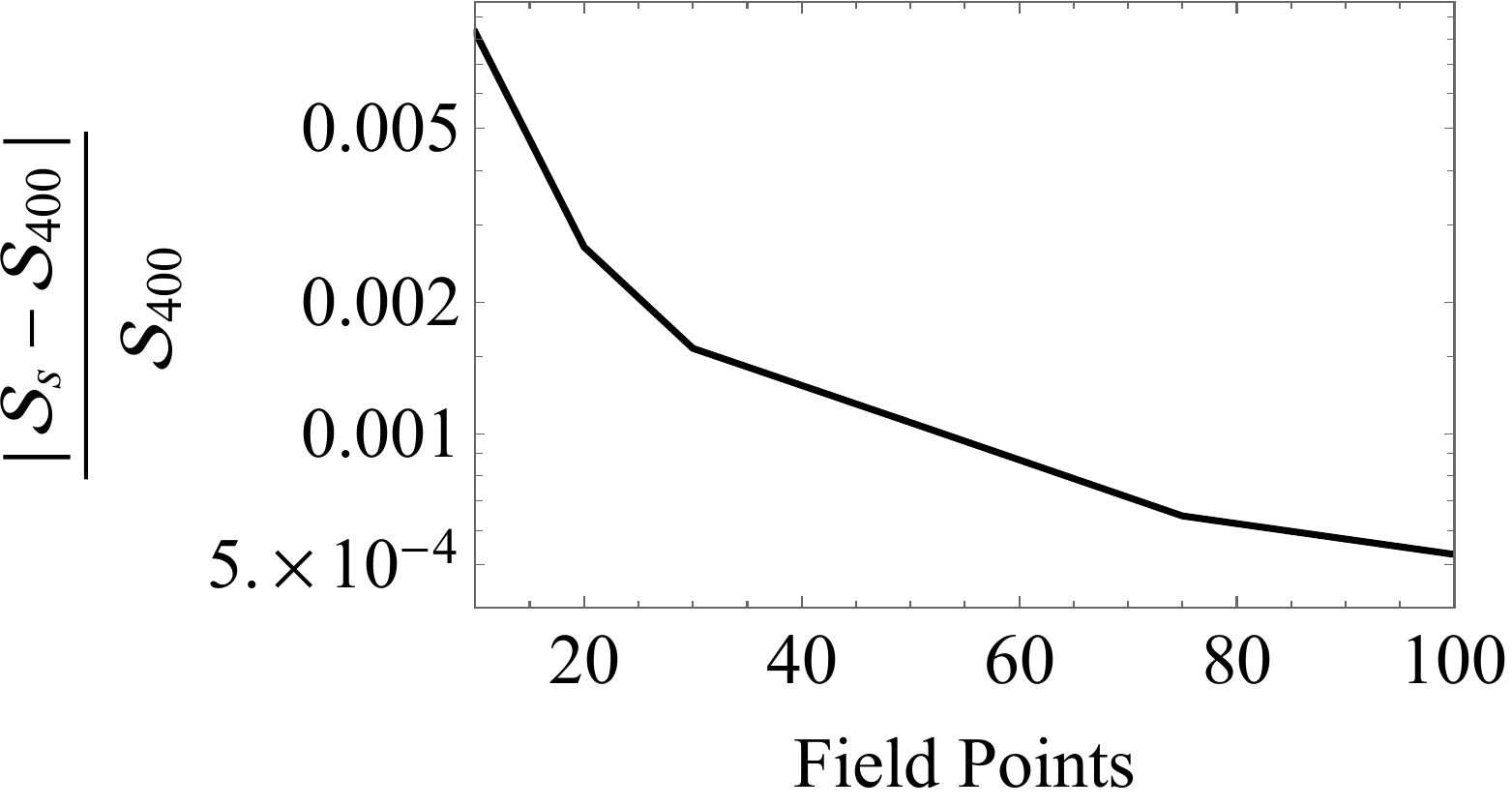}
  \includegraphics[width=.46\textwidth]{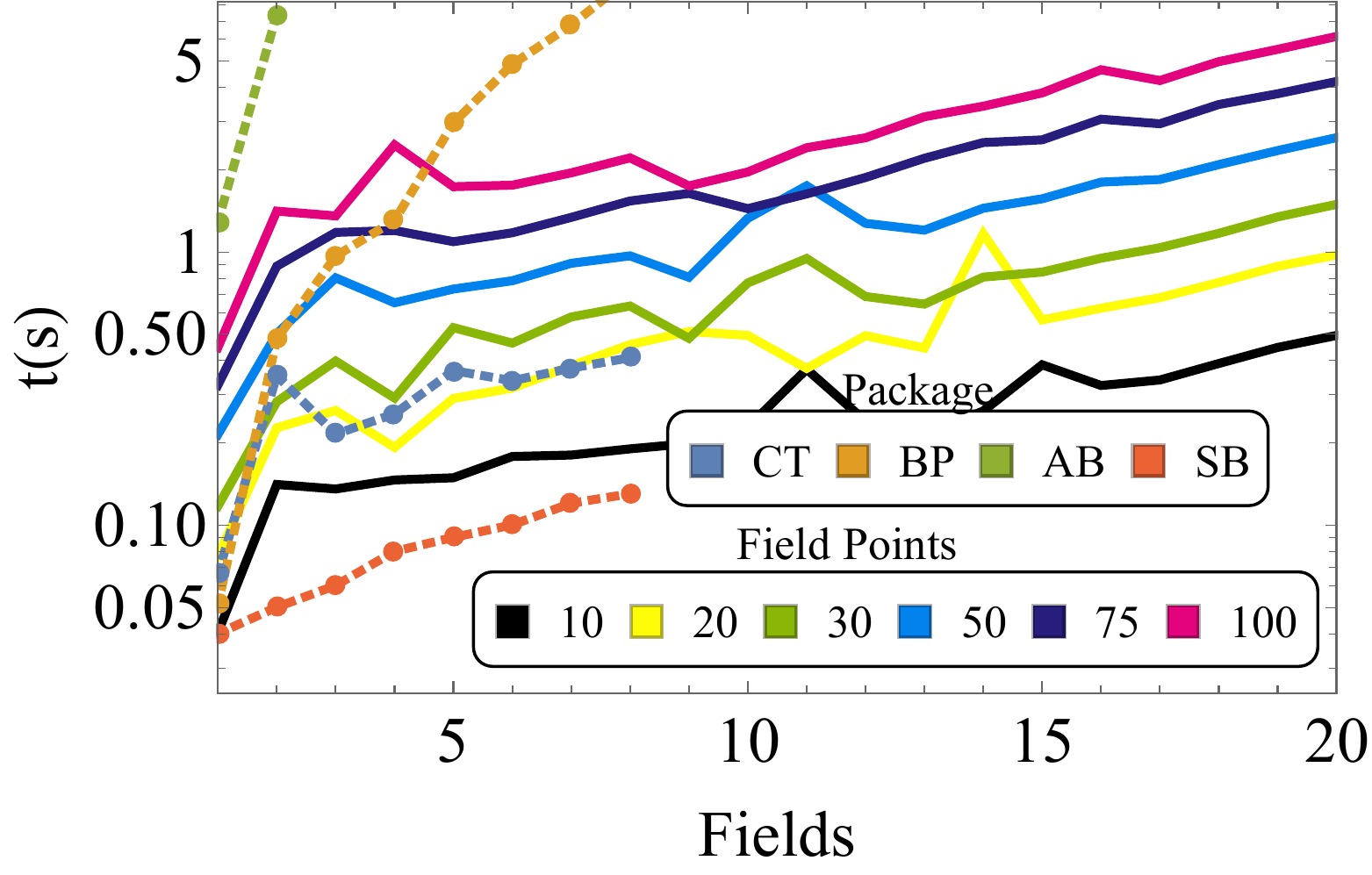}
  \caption{
    Left: The multi-field bounce action $S_s$ average, normalized to the one with \(N=400\) field points.
    Right: Multi-field time measure with respect to number of fields for several field points.
    The solid lines were obtained by \fb with \texttt{"ActionTolerance"~$\to 10^{-4}$}.
    The dashed lines with dots are the time measurements from other existing tools, see text
    for details.
    }
 \label{fig:TimeField}
\end{figure}

With the multi-field potential defined in~\eqref{eq:PotentialMultifield}, we are ready to compute the 
action, estimate the computation time and compare to the existing tools. We vary the number of fields
from the single field case going up to twenty fields.
The results are collected in Table~\ref{table:Action} and on the left of Figure~\ref{fig:TimeField}, together
with results obtained by other methods. The parameters for $n_\varphi \leq 8$ were taken 
from~\cite{Athron:2019nbd}, while for $n_\varphi > 8$ they were chosen randomly, 
as listed in the Appendix~\S\ref{Ap:MultiField}.  
The comparison includes the \fb with 10, 30 and 100 field points, \texttt{CosmoTransition}(CT)~\cite{Wainwright:2011kj}, 
\texttt{AnyBubble}(AB)~\cite{Masoumi:2016wot}, \texttt{BubbleProfile}(BP)~\cite{Athron:2019nbd} 
and \texttt{SimpleBounce}(SB)~\cite{Sato:2019wpo}.

\begin{table}[h]
  \centering
\begin{tabular}{| c | c c c c c c c |}
  \hline
  \multicolumn{8}{|c|}{Action}                                                                                         
  \\ \hline
  $n_\varphi$ 	& FB$_{10}$ 	& FB$_{30}$ 	& FB$_{100}$ 	& CT  	& AB 	& BP 	& SB   
  \\ \hline
  1      		& 52.1      		& 52.6      		& 52.4       	& 52.6 	& 52.4 	& 54.1 	& 52.4 \\
  2      		& 20.8      		& 20.8      		& 20.8       	& 21.1 	& 20.8 	& 20.8 	& 20.8 \\
  3      		& 20.8      		& 20.7      		& 20.7       	& 22.0 	& 22.0 	& 22.0 	& 22.0 \\
  4      		& 57.9      		& 56.2      		& 55.8       	& 55.9 	& 56.4 	& 55.9 	& 55.8 \\
  5      		& 16.3      		& 16.3      		& 16.3       	& 16.3 	& 16.3 	& 16.3 	& 16.4 \\
  6      		& 24.6      		& 24.5      		& 24.5       	& 24.5 	& 24.5 	& 24.4 	& 24.5 \\
  7      		& 36.9      		& 36.7      		& 36.7       	& 36.7 	& 36.6 	& 36.7 	& 36.7 \\
  8      		& 46.4      		& 46.1      		& 46.0       	& 46.1 	& 46.0 	& 46.0 	& 46.0 
  \\ \hline
\end{tabular}
  \caption{Comparison of the bounce action obtained by different methods in the literature for various 
  number of fields $n_\varphi$. 
  It includes \fb with 10, 30 and 100 field points and  \texttt{CosmoTransition}, \texttt{AnyBubble},
  \texttt{BubbleProfile} and \texttt{SimpleBounce}. 
  The action values for the other methods were adopted from~\cite{Athron:2019nbd, Sato:2019wpo}. 
  }
 \label{table:Action}
\end{table}

We find that the action computed with $N \lesssim 20$ field points is accurate up to 
roughly 1\%, as shown on the left panel of Figure~\ref{fig:TimeField}.
Similarly, bounces with \texttt{"FieldPoints"->} $10, 30$ in Table~\ref{table:Action} are 
accurate to 1\% level or below. 
Table~\ref{table:Action} also shows how the accuracy of the action improves with the
number of field points. In particular, the $N=100$ case reproduces the values of the 
action obtained by other methods.
Therefore, it is clear that one can use \fb with any type of potential to get the arbitrarily 
precise bounce action by adjusting the \texttt{"FieldPoints"} and \texttt{"ActionTolerance"}
settings.

Note that the parameters $c_i$ that regulate the multi-field potential for $n_\varphi \leq 8$ were
chosen in a way that the bounce solution belongs to the thick wall regime.
This is well suited for numerical shooting methods and typically gives the action of $\mathcal O(10)$,
as seen in Table~\ref{table:Action}.
For larger number of fields $n_\varphi > 8$, the parameters were chosen at random, 
see Table~\ref{tab:MultiFieldParameters} and the resulting action in Table~\ref{tab:MultiFieldAction} 
contains both small and large values, in particular $n_\varphi = 15, 19$ that belong to the thin wall
regime. Clearly, \fb can deal with both cases quite efficiently.

Let us turn to the timing performance of \fb. As explained in the PB method section~\S\ref{sec:PBMethod}, 
we expect this increase to be linear. 
Indeed, as shown on the right panel of Figure~\ref{fig:TimeField} in solid lines, the time consumption
of \fb scales linearly with the number of fields $n_\varphi$. 
This behavior is independent of the number of field points, i.e. for different colors of the solid
lines in Figure~\ref{fig:TimeField} and covers both thin and thick wall cases.
Moreover, the time demand of \fb with respect to the number of field points also scales linearly
in the multi-field scenarios.

A reliable timing comparison to other tools could be somewhat involved.
The running time clearly depends on the CPU capabilities and the optimization efficiency 
of the implementation in a given computer language. 
In our setup, the computation time of \fb for 10(20) fields turns out to be less than 1(2) seconds 
for the default value with 31 field points. This was computed on a desktop system 
using the \mth native functions \texttt{AboluteTiming} and \texttt{RepeatedTiming}
running on a desktop iMac 10.12.6, equipped with an Intel Core i7, processor speed 3.4 GHz and 
16 GB of DDR3 RAM clocked at 1.6 GHz.

For the other implementations, the timing reports were adopted from~\cite{Athron:2019nbd}
and~\cite{Sato:2019wpo}.
We find \fb to be comparable to these tools in terms of speed performance, as shown on the right 
side of Figure~\ref{fig:TimeField}.
In particular, the time demand with $N=10$ field points is comparable to the values quoted
by \texttt{SimpleBounce} and \texttt{CosmoTransition}.
It should be noted though that the above tools are implemented in \texttt{C++} and \texttt{Python},
while \fb was coded straightforwardly in \mth, with no significant numerical optimization.
%and benefits from its native symbolic capabilities.

%\newpage
%
% Conclusions and outlook
%
\section{Conclusions and outlook} \label{sec:Outlook}

The \fb package presented in this work performs the task of calculating the semi-classical contribution
to the false vacuum decay rate, the so-called bounce field configuration and the associated Euclidean
action.
We demonstrated the basic use of the \fb function implemented in \mth, as well as the
extended options and manipulation of the output. 
The current version of the package can deal efficiently with single and multi-field calculations,
ranging up to 20 fields in a matter of seconds.
% Improvement of the multifield, the rho part

% Continuation to Minkowski, oscillations and resonant tunneling
There are a number of physically relevant directions one can pursue, that may be implemented
in the future versions of the \fb framework.
An obvious question relates to the growth of the bubble after nucleation, which is governed by 
the bounce equations in Minkowski space-time~\cite{Coleman:1977py}. An initial step in this direction
was done in~\cite{Pastras:2011zr} and may be generalized straightforwardly to more segments and fields.
Moreover, matching such solutions to subsequent tunneling may be needed when potentials with
multiple minima are considered. We leave this fascinating subject for future studies~\cite{Guada:2020}.

Another ubiquitous phenomenon is the tunneling without barriers into an unstable region of the 
potential. In this version of the \fb package, we demonstrate how this is done for single field
theories by matching the polygonal potential onto an unstable quartic potential. Extension to 
multi-field case requires additional care and we leave it for an updated version.

% Pre-factor calculations - 2nd order and linear approx
The total decay rate in~\eqref{eq:FVDecay} depends on both the bounce and the pre-factor
$A$~\cite{Callan:1977pt}, which is in general not
easily calculable.
Numerical approaches were done in $D=3$~\cite{Strumia:1998nf, Munster:1999hr} and 
for $D=4$~\cite{Baacke:2003uw, Dunne:2005rt}, for recent developments in treating gauge fields,
see also~\cite{Andreassen:2016cvx,Andreassen:2017rzq, Chigusa:2018uuj}.
The use of the polygonal method in computing the pre-factor was demonstrated in~\cite{Guada:2018jek},
where the semi-analytical bounce background was used for numerical computation of the fluctuation 
operator eigenfunctions, while the WKB integrals could be performed analytically. 
It may be feasible to revisit the calculation of the pre-factor ab-initio by considering the polygonal
(or its second order extension) potential in a more direct way by calculating the eigenfunctions
of the second derivative of $V(\varphi)$. This would offer an extension beyond the thin wall 
case~\cite{Konoplich:1986zp}, an arbitrary precision with more segments and, perhaps most interestingly,
to multiple fields.

% Thermal potentials, gravitational wave alpha and beta parameters
The semi-analytical approach to computing the bounce field configuration and the action at finite
temperature is relevant for characterizing the strength of the potential phase transitions. In particular,
we saw how one can determine the nucleation temperature $T_N$, that happens below the critical 
temperature, using the output from the \fb result. Similarly, one can get an analytical insight in the 
gravitational wave spectra by computing the $\alpha$ and $\beta$ parameters~\cite{Caprini:2019egz}, 
which are related to the strength and the position of the maximum in the frequency range. 
We intend to return to these issues in future instalments of the \fb approach. 

% Inclusion of gravity, non-trivial curved backgrounds

\section*{Acknowledgements}
We would like to thank Vedran Brdar, Jose Espinosa, Yann Gouttenoire, Stefan Huber, 
Luca di Luzio, Eric Madge and Lorenzo Ubaldi for discussions, comments and feedback
on the beta version of the package. 
The work of VG was supported by the Slovenian Research Agency's young researcher 
program under the grant No. PR-07582. 
MN was supported by the Slovenian Research Agency under the research core funding 
No. P1-0035 and in part by the research grant J1-8137.
MN acknowledges the support of the COST actions 
CA15108 - ``Connecting insights in fundamental physics'' and 
CA16201 - ``Unraveling new physics at the LHC through the precision frontier''.
MN is grateful to the Mainz Institute for Theoretical Physics (MITP) of the DFG Cluster of
Excellence PRISMA+ (Project ID 33083149), for its hospitality and its partial support during
the completion of this work.

%
% Appendices
%
\appendix
\renewcommand*{\thesection}{\Alph{section}}

%
% One-Loop effective potential
%
\section{One-Loop effective potential} \label{Ap:EffectivePotential}
The calculation of the effective potential was first computed by Coleman and Weinberg~\cite{Coleman:1973jx} 
at one-loop and at higher loops by Jackiw and Iliopoulus, et al.~\cite{Jackiw:1974cv,Iliopoulos:1974ur} at zero 
temperature.
For finite temperatures excellent references are~\cite{Dolan:1973qd,Weinberg:1974hy}, a review~\cite{Quiros:1999jp} 
and the textbook~\cite{Kapusta:2006pm} 
For recent works on the consistent use of potentials, see~\cite{Andreassen:2014eha}.
The quantum and thermal results at one-loop order in the $\overline{\mathrm{MS}}$ scheme are:
\begin{align} \label{ap:quantum}
  \text{Quantum: } 
  \Delta V_1(\varphi) &= \sum_i \pm \frac{n_i m_i^4(\varphi)}{64 \pi^2} 
  \left( \log  \frac{m_i^2 \left(\varphi\right)}{\mu_R^2}  - C_i \right),
  \\  \label{ap:thermal}
  \text{Thermal: }  
  \Delta V_1(\varphi, T) &=  \sum_i \frac{n_i T^4}{2\pi^2}  J_{\nicefrac{B}{F}}
  \left(\frac{m_i^2 \left(\varphi \right)}{T^2}\right),
\end{align}
where $\mu_{R}$ is the renormalization scale and $C_i = 3/2 \left(5/6\right)$ for scalars and fermions (gauge bosons). 
The sum runs over all the species that couple to $\varphi$, where $n_i$ and $m_i^2(\varphi)$ are the number of 
degrees of freedom and the field-dependent squared masses of the species $i$. 
The upper and lower sign corresponds to bosons (B) and fermions (F), respectively. 
The relevant thermal functions are given by
\begin{align} \label{ap:Jfunction}
  J_{\nicefrac{B}{F}}\left( y^2 \right) = \mp \int_0^\infty \text{d}x \, x^2 \log\left(1 \pm e^{-\sqrt{x^2+y^2}}  \right),
\end{align}
and can be expanded at high temperatures, $\left(y^2 \ll 1\right)$ such that
\begin{align} \label{ap:JBapprox}
  J_B(y^2) &= -\frac{7 \pi}{360} +\frac{\pi^2}{24} y^2 + \frac{y^4}{32} \log \frac{y^2}{a} + \mathcal{O}\left(y^6\right),
  \\ \label{ap:JFapprox}
  J_F(y^2) &= -\frac{\pi^6}{45} + \frac{\pi^2}{12}y^2 - \frac{\pi}{6}\left( y^2\right)^{\frac{3}{2}} - 
  \frac{y^4}{32} \log \frac{y^2}{16 a} + \mathcal{O}\left(y^6\right),
\end{align}
with $a = \pi^2 e^{\frac{3}{2}-2\gamma_{E}}$.

%
%  Multi field Potential
%
\section{Multi field potential parameters} 
\label{Ap:MultiField}
The multi-field potential for comparison with other packages, shown in Table~\ref{table:Action}, was 
taken from~\cite{John:1998ip} and defined for each number of fields $n_\varphi$ as
\begin{equation} \label{eq:Potential1F}
  V(\varphi) = \frac{1}{10}\left(\varphi^4- 8\varphi^3+10\varphi^2+8  \right),
\end{equation}
for single field potentials and by Eq.~\eqref{eq:PotentialMultifield} for multifields. The relevant parameters $c_i$ for 
higher number of fields are given by the list in Table~\ref{tab:MultiFieldParameters} with the corresponding actions
listed in Table~\ref{tab:MultiFieldAction}.
Note that the last component of the parameters $c_{n_{\varphi}+1}$ controls the degeneracy of the vacua.
Tiny values of this parameter corresponds to thin wall scenarios, which is the case for the potentials with 15 and 19 
fields. There, the action is notably larger. Nevertheless, as shown in Figure~\ref{fig:TimeField}, the computational 
time of the bounce action is practically independent of whether the bounce is computed in thin or thick wall regime.

\begin{table}
\centering
\begin{tabular}{|c | p{15cm} |}
  \hline
  $n_\varphi$ & $c_i$
  \\  \hline
  3 & 0.68, 0.18, 0.30, 0.28
  \\
  4 & 0.53, 0.77, 0.84, 0.01, 0.26 
  \\
  5 & 0.47, 0.23, 0.57, 0.14, 0.52, 0.66
  \\
  6 & 0.34, 0.47, 0.23, 0.57, 0.14, 0.52, 0.66
  \\
  7 & 0.52, 0.34, 0.47, 0.23, 0.57, 0.14, 0.52, 0.66
  \\
  8 & 0.24, 0.52, 0.34, 0.47, 0.23, 0.57, 0.14, 0.52, 0.66
  \\
  9 & 0.21, 0.24, 0.52, 0.34, 0.47, 0.23, 0.57, 0.14, 0.52, 0.66
  \\
  10 & 0.12, 0.21, 0.24, 0.52, 0.34, 0.47, 0.23, 0.57, 0.14, 0.52, 0.66
  \\
  11 & 0.23, 0.21, 0.21, 0.24, 0.52, 0.34, 0.47, 0.23, 0.57, 0.14, 0.52, 0.66
  \\
  12 & 0.12, 0.11, 0.12, 0.21, 0.24, 0.52, 0.34, 0.47, 0.23, 0.57, 0.14, 0.52, 0.66
  \\
  13 & 0.54, 0.47, 0.53, 0.28, 0.35, 0.27, 0.42, 0.59, 0.33, 0.16, 0.38, 0.35, 0.17, 0.41
  \\
  14 & 0.39, 0.23, 0.26, 0.40, 0.11, 0.42, 0.41, 0.27, 0.42, 0.54, 0.18, 0.59, 0.13, 0.29, 0.58
  \\
  15 & 0.21, 0.22, 0.22, 0.23, 0.39, 0.55, 0.43, 0.12, 0.16, 0.58, 0.25, 0.50, 0.45, 0.35,
  0.45, 0.12
  \\
  16 & 0.42, 0.34, 0.43, 0.22, 0.59, 0.41, 0.58, 0.41, 0.26, 0.45, 0.16, 0.31, 0.39, 0.57, 
  0.43, 0.10, 0.46
  \\
  17 & 0.24, 0.35, 0.39, 0.56, 0.37, 0.41, 0.52, 0.31, 0.52, 0.22, 0.58, 0.39, 0.39, 0.17,
  0.46, 0.30, 0.37, 0.43
  \\
  18 & 0.18, 0.17, 0.30, 0.22, 0.38, 0.48, 0.11, 0.49, 0.43, 0.47, 0.21, 0.29, 0.32, 0.36,
  0.30, 0.56, 0.46, 0.42, 0.44
  \\
  19 & 0.40, 0.14, 0.10, 0.43, 0.39, 0.27, 0.33, 0.59, 0.48, 0.36, 0.24, 0.28, 0.51, 0.59,
  0.40, 0.39, 0.24, 0.35, 0.20, 0.14
  \\
  20 & 0.42, 0.11, 0.47, 0.13, 0.16, 0.24, 0.58, 0.53, 0.38, 0.44, 0.18, 0.46, 0.47, 0.27,
  0.53, 0.24, 0.33, 0.40, 0.32, 0.29, 0.44
\\
\hline
\end{tabular}
\caption{Multi-field potential parameters $c_i$ that define the potential in~\eqref{eq:PotentialMultifield}.}
\label{tab:MultiFieldParameters}
\end{table}

\begin{table}
\centering
\begin{tabular}{| c | c c c | c | c c c |}
\hline
\multicolumn{8}{|c|}{Action}                                                                     
\\ \hline
$n_\varphi$ & PB$_{10}$ & PB$_{30}$ & PB$_{100}$ & $n_\varphi$ & PB$_{10}$ & PB$_{30}$ & PB$_{100}$
\\ \hline
1      & 52.1         & 52.6         & 52.4      & 11     & 78.3         & 78.3         &  78.2   
\\
2      & 20.8         & 20.8         & 20.8      & 12     & 80.2         & 80.0         & 79.0             
\\
3      & 20.8         & 20.7         & 20.7      & 13     & 274          & 271          & 271   
\\
4      & 57.9         & 56.2         & 55.8     & 14     & 154          & 155          & 155      
\\
5      & 16.3         & 16.3         & 16.3     & 15     & $2.90 \times 10^3$ & $2.87 \times 10^3$ & $2.87 \times 10^3$     
\\
6      & 24.6         & 24.5         & 24.5     & 16     & 358          & 355          & 355     
\\
7      & 36.9         & 36.7         & 36.7   & 17     & 472          & 468          & 468      
\\
8      & 46.4         & 46.1         & 46.0    & 18     & 439          & 435          & 435     
\\
9      & 56.1         & 55.7         & 55.6    & 19     & $3.96 \times 10^3$ & $3.93 \times 10^3$ & $3.93 \times 10^3$      
\\
10     & 63.8         & 63.4         & 63.3  & 20     & 565          & 560          & 560       
\\
\hline
\end{tabular}
  \caption{The bounce action of the single and multi-field potential in Eq.~\eqref{eq:Potential1F} 
  and~\eqref{eq:PotentialMultifield}, respectively computed using \fb with 10, 30 and 100 number of 
  field points.}
\label{tab:MultiFieldAction}
\end{table}

%
% Bibliography
%
%\bibliographystyle{JHEP}

\addcontentsline{toc}{section}{References}

%\bibliography{FindBounce}

\begin{thebibliography}{99}

%\cite{Kobzarev:1974cp}
\bibitem{Kobzarev:1974cp}
  I.~Y.~Kobzarev, L.~B.~Okun and M.~B.~Voloshin,
  %``Bubbles in Metastable Vacuum,''
  Sov.\ J.\ Nucl.\ Phys.\  {\bf 20} (1975) 644
   [Yad.\ Fiz.\  {\bf 20} (1974) 1229].
  %%CITATION = SJNCA,20,644;%%

%\cite{Coleman:1977py}
\bibitem{Coleman:1977py}
  S.~R.~Coleman,
  %``The Fate of the False Vacuum. 1. Semiclassical Theory,''
  Phys.\ Rev.\ D {\bf 15} (1977) 2929
  Erratum: [Phys.\ Rev.\ D {\bf 16} (1977) 1248].
  %doi:10.1103/PhysRevD.15.2929, 10.1103/PhysRevD.16.1248
  %%CITATION = doi:10.1103/PhysRevD.15.2929, 10.1103/PhysRevD.16.1248;%%

%\cite{Callan:1977pt}
\bibitem{Callan:1977pt}
  C.~G.~Callan, Jr. and S.~R.~Coleman,
  %``The Fate of the False Vacuum. 2. First Quantum Corrections,''
  Phys.\ Rev.\ D {\bf 16} (1977) 1762.
  %doi:10.1103/PhysRevD.16.1762
  %%CITATION = doi:10.1103/PhysRevD.16.1762;%%

%\cite{Coleman:1985}
\bibitem{Coleman:1985}
  S.~R.~Coleman,
  %``Aspects of Symmetry: Selected Erice Lectures,''
  Cambridge University Press (1985).
  %doi:10.1017/CBO9780511565045

%\cite{Fubini:1976jm}
\bibitem{Fubini:1976jm}
  S.~Fubini,
  %``A New Approach to Conformal Invariant Field Theories,''
  Nuovo Cim. A \textbf{34} (1976), 521
  %doi:10.1007/BF02785664

%\cite{Lipatov:1976ny}
\bibitem{Lipatov:1976ny}
  L.~Lipatov,
  %``Divergence of the Perturbation Theory Series and the Quasiclassical Theory,''
  Sov. Phys. JETP \textbf{45} (1977), 216-223
  LENINGRAD-76-255.
  
%\cite{Loran:2006sf}
\bibitem{Loran:2006sf}
  F.~Loran,
  %``Fubini vacua as a classical de Sitter vacua,''
  Mod. Phys. Lett. A \textbf{22} (2007), 2217-2235
  %doi:10.1142/S0217732307022426
  \arxiv{hep-th/0612089}[arXiv:hep-th/0612089 [hep-th]].  
  
%\cite{Lee:1985uv}
\bibitem{Lee:1985uv} 
  K.~M.~Lee and E.~J.~Weinberg,
  %``Tunneling Without Barriers,''
  Nucl.\ Phys.\ B {\bf 267}, 181 (1986).
  %doi:10.1016/0550-3213(86)90150-1
  %%CITATION = doi:10.1016/0550-3213(86)90150-1;%%  
  
%\cite{FerrazdeCamargo:1982sk}
\bibitem{FerrazdeCamargo:1982sk} 
  A.~Ferraz de Camargo, R.~C.~Shellard and G.~C.~Marques,
  %``Vacuum Decay in a Soluble Model,''
  Phys.\ Rev.\ D {\bf 29}, 1147 (1984).
  %doi:10.1103/PhysRevD.29.1147
  %%CITATION = doi:10.1103/PhysRevD.29.1147;%%

%\cite{Aravind:2014pva}
\bibitem{Aravind:2014pva}
  A.~Aravind, B.~S.~DiNunno, D.~Lorshbough and S.~Paban,
  %``Analyzing multifield tunneling with exact bounce solutions,''
  Phys.\ Rev.\ D {\bf 91} (2015) no.2,  025026
  %doi:10.1103/PhysRevD.91.025026
  \arxiv{1412.3160}[arXiv:1412.3160 [hep-th]].
  %%CITATION = doi:10.1103/PhysRevD.91.025026;%%  

%\cite{Adams:1993zs}
\bibitem{Adams:1993zs}
  F.~C.~Adams,
  %``General solutions for tunneling of scalar fields with quartic potentials,''
  Phys.\ Rev.\ D {\bf 48} (1993) 2800
  %doi:10.1103/PhysRevD.48.2800
  \arxiv{hep-ph/9302321}[hep-ph/9302321].
  %%CITATION = doi:10.1103/PhysRevD.48.2800;%%

%\cite{Sarid:1998sn}
\bibitem{Sarid:1998sn}
  U.~Sarid,
  %``Tools for tunneling,''
  Phys.\ Rev.\ D {\bf 58} (1998) 085017
  %doi:10.1103/PhysRevD.58.085017
  \arxiv{hep-ph/9804308}[hep-ph/9804308].
  %%CITATION = doi:10.1103/PhysRevD.58.085017;%%

%\cite{Dasgupta:1996qu}
\bibitem{Dasgupta:1996qu}
  I.~Dasgupta,
  %``Estimating vacuum tunneling rates,''
  Phys.\ Lett.\ B {\bf 394} (1997) 116
  %doi:10.1016/S0370-2693(96)01685-1
  \arxiv{hep-ph/9610403}[hep-ph/9610403].
  %%CITATION = doi:10.1016/S0370-2693(96)01685-1;%%

%\cite{Aravind:2014aza}
\bibitem{Aravind:2014aza}
  A.~Aravind, D.~Lorshbough and S.~Paban,
  %``Lower bound for the multifield bounce action,''
  Phys.\ Rev.\ D {\bf 89} (2014) no.10,  103535
  %doi:10.1103/PhysRevD.89.103535
  \arxiv{1401.1230}[arXiv:1401.1230 [hep-th]].
  %%CITATION = doi:10.1103/PhysRevD.89.103535;%%

%\cite{Sato:2017iga}
\bibitem{Sato:2017iga}
  R.~Sato and M.~Takimoto,
  %``Absolute Lower Bound on the Bounce Action,''
  Phys.\ Rev.\ Lett.\  {\bf 120} (2018) no.9,  091802
  %doi:10.1103/PhysRevLett.120.091802
  \arxiv{1707.01099}[arXiv:1707.01099 [hep-ph]].
  %%CITATION = doi:10.1103/PhysRevLett.120.091802;%%

%\cite{Brown:2017cca}
\bibitem{Brown:2017cca}
  A.~R.~Brown,
  %``Thin-wall approximation in vacuum decay: A lemma,''
  Phys.\ Rev.\ D {\bf 97} (2018) no.10,  105002
  %doi:10.1103/PhysRevD.97.105002
  \arxiv{1711.07712}[arXiv:1711.07712 [hep-th]].
  %%CITATION = doi:10.1103/PhysRevD.97.105002;%%

%\cite{Espinosa:2018hue}
\bibitem{Espinosa:2018hue}
  J.~R.~Espinosa,
  %``A Fresh Look at the Calculation of Tunneling Actions,''
  JCAP {\bf 1807} (2018) 036
  %doi:10.1088/1475-7516/2018/07/036
  \arxiv{1805.03680}[arXiv:1805.03680 [hep-th]].
  %%CITATION = doi:10.1088/1475-7516/2018/07/036;%%

%\cite{Jinno:2018dek}
\bibitem{Jinno:2018dek}
  R.~Jinno,
  %``Machine learning for bounce calculation,''
  \Arxiv{1805.12153}[arXiv:1805.12153 [hep-th]].
  %%CITATION = ARXIV:1805.12153;%%

%\cite{Andreassen:2016cff}
\bibitem{Andreassen:2016cff}
  A.~Andreassen, D.~Farhi, W.~Frost and M.~D.~Schwartz,
  %``Direct Approach to Quantum Tunneling,''
  Phys. Rev. Lett. \textbf{117} (2016) no.23, 231601
  %doi:10.1103/PhysRevLett.117.231601
  \arxiv{1602.01102}[arXiv:1602.01102 [hep-th]].  

%\cite{Ai:2019fri}
\bibitem{Ai:2019fri}
  W.~Y.~Ai, B.~Garbrecht and C.~Tamarit,
  %``Functional methods for false vacuum decay in real time,''
  JHEP \textbf{12} (2019), 095
  %doi:10.1007/JHEP12(2019)095
  \arxiv{1905.04236}[arXiv:1905.04236 [hep-th]].
  
%\cite{Hertzberg:2019wgx}
\bibitem{Hertzberg:2019wgx}
  M.~P.~Hertzberg and M.~Yamada,
  %``Vacuum Decay in Real Time and Imaginary Time Formalisms,''
  Phys. Rev. D \textbf{100} (2019) no.1, 016011
  %doi:10.1103/PhysRevD.100.016011
  \arxiv{1904.08565}[arXiv:1904.08565 [hep-th]].
  
%\cite{Braden:2018tky}
\bibitem{Braden:2018tky}
  J.~Braden, M.~C.~Johnson, H.~V.~Peiris, A.~Pontzen and S.~Weinfurtner,
  %``New Semiclassical Picture of Vacuum Decay,''
  Phys. Rev. Lett. \textbf{123} (2019) no.3, 031601
  %doi:10.1103/PhysRevLett.123.031601
  \arxiv{1806.06069}[arXiv:1806.06069 [hep-th]].

%\cite{Kusenko:1995jv}
\bibitem{Kusenko:1995jv}
  A.~Kusenko,
  %``Improved action method for analyzing tunneling in quantum field theory,''
  Phys.\ Lett.\ B {\bf 358} (1995) 51
  %doi:10.1016/0370-2693(95)00994-V
  \arxiv{hep-ph/9504418}[hep-ph/9504418].
  %%CITATION = doi:10.1016/0370-2693(95)00994-V;%%

%\cite{Chigusa:2019wxb}
\bibitem{Chigusa:2019wxb}
  S.~Chigusa, T.~Moroi and Y.~Shoji,
  %``Bounce Configuration from Gradient Flow,''
  Phys.\ Lett.\ B {\bf 800} (2020) 135115
  %doi:10.1016/j.physletb.2019.135115
  \arxiv{1906.10829}[arXiv:1906.10829 [hep-ph]].
  %%CITATION = doi:10.1016/j.physletb.2019.135115;%%
  
%\cite{Sato:2019axv}
\bibitem{Sato:2019axv}
  R.~Sato,
  %``Simple Gradient Flow Equation for the Bounce Solution,''
  Phys.\ Rev.\ D {\bf 101} (2020) no.1,  016012
  %doi:10.1103/PhysRevD.101.016012
  \arxiv{1907.02417}[arXiv:1907.02417 [hep-ph]].
  %%CITATION = doi:10.1103/PhysRevD.101.016012;%%
  
%\cite{John:1998ip}
\bibitem{John:1998ip}
  P.~John,
  %``Bubble wall profiles with more than one scalar field: A Numerical approach,''
  Phys.\ Lett.\ B {\bf 452} (1999) 221
  %doi:10.1016/S0370-2693(99)00272-5
  \arxiv{hep-ph/9810499}[hep-ph/9810499].
  %%CITATION = doi:10.1016/S0370-2693(99)00272-5;%%
  
%\cite{Cline:1999wi}
\bibitem{Cline:1999wi}
  J.~M.~Cline, G.~D.~Moore and G.~Servant,
  %``Was the electroweak phase transition preceded by a color broken phase?,''
  Phys.\ Rev.\ D {\bf 60} (1999) 105035
  %doi:10.1103/PhysRevD.60.105035
  \arxiv{hep-ph/9902220}[hep-ph/9902220].
  %%CITATION = doi:10.1103/PhysRevD.60.105035;%%

%\cite{Wainwright:2011kj}
\bibitem{Wainwright:2011kj}
  C.~L.~Wainwright,
  %``CosmoTransitions: Computing Cosmological Phase Transition Temperatures and Bubble Profiles with Multiple Fields,''
  Comput.\ Phys.\ Commun.\  {\bf 183} (2012) 2006
  %doi:10.1016/j.cpc.2012.04.004
  \arxiv{1109.4189}[arXiv:1109.4189 [hep-ph]].
  %%CITATION = doi:10.1016/j.cpc.2012.04.004;%%
  
%\cite{Konstandin:2006nd}
\bibitem{Konstandin:2006nd}
  T.~Konstandin and S.~J.~Huber,
  %``Numerical approach to multi dimensional phase transitions,''
  JCAP {\bf 0606} (2006) 021
  %doi:10.1088/1475-7516/2006/06/021
  \arxiv{hep-ph/0603081}[hep-ph/0603081].
  %%CITATION = doi:10.1088/1475-7516/2006/06/021;%%

%\cite{Park:2010rh}
\bibitem{Park:2010rh}
  J.~h.~Park,
  %``Constrained potential method for false vacuum decays,''
  JCAP {\bf 1102} (2011) 023
  %doi:10.1088/1475-7516/2011/02/023
  \arxiv{1011.4936}[arXiv:1011.4936 [hep-ph]].
  %%CITATION = doi:10.1088/1475-7516/2011/02/023;%%

%\cite{Akula:2016gpl}
\bibitem{Akula:2016gpl}
  S.~Akula, C.~Bal\'azs and G.~A.~White,
  %``Semi-analytic techniques for calculating bubble wall profiles,''
  Eur.\ Phys.\ J.\ C {\bf 76} (2016) no.12,  681
  %doi:10.1140/epjc/s10052-016-4519-5
  \arxiv{1608.00008}[arXiv:1608.00008 [hep-ph]].
  %%CITATION = doi:10.1140/epjc/s10052-016-4519-5;%%

%\cite{Masoumi:2016wot}
\bibitem{Masoumi:2016wot}
  A.~Masoumi, K.~D.~Olum and B.~Shlaer,
  %``Efficient numerical solution to vacuum decay with many fields,''
  JCAP {\bf 1701} (2017) no.01,  051
  %doi:10.1088/1475-7516/2017/01/051
  \arxiv{1610.06594}[arXiv:1610.06594 [gr-qc]].
  %%CITATION = doi:10.1088/1475-7516/2017/01/051;%%

%\cite{Espinosa:2018szu}
\bibitem{Espinosa:2018szu}
  J.~R.~Espinosa and T.~Konstandin,
  %``A Fresh Look at the Calculation of Tunneling Actions in Multi-Field Potentials,''
  JCAP {\bf 1901} (2019) 051
  %doi:10.1088/1475-7516/2019/01/051
  \arxiv{1811.09185}[arXiv:1811.09185 [hep-th]].
  %%CITATION = doi:10.1088/1475-7516/2019/01/051;%%

%\cite{Athron:2019nbd}
\bibitem{Athron:2019nbd}
  P.~Athron, C.~Balázs, M.~Bardsley, A.~Fowlie, D.~Harries and G.~White,
  %``BubbleProfiler: finding the field profile and action for cosmological phase transitions,''
  Comput.\ Phys.\ Commun.\  {\bf 244} (2019) 448
  %doi:10.1016/j.cpc.2019.05.017
  \arxiv{1901.03714}[arXiv:1901.03714 [hep-ph]].
  %%CITATION = doi:10.1016/j.cpc.2019.05.017;%%
  
%\cite{Sato:2019wpo}
\bibitem{Sato:2019wpo} 
  R.~Sato,
  %``SimpleBounce : a simple package for the false vacuum decay,''
  \Arxiv{1908.10868}[arXiv:1908.10868 [hep-ph]].
  %%CITATION = ARXIV:1908.10868;%%
  %1 citations counted in INSPIRE as of 29 Oct 2019

%\cite{Piscopo:2019txs}
\bibitem{Piscopo:2019txs}
  M.~L.~Piscopo, M.~Spannowsky and P.~Waite,
  %``Solving differential equations with neural networks: Applications to the calculation of cosmological phase transitions,''
  Phys.\ Rev.\ D {\bf 100} (2019) no.1,  016002
  %doi:10.1103/PhysRevD.100.016002
  \arxiv{1902.05563}[arXiv:1902.05563 [hep-ph]].
  %%CITATION = doi:10.1103/PhysRevD.100.016002;%%
    
%\cite{Coleman:1977th}
\bibitem{Coleman:1977th}
  S.~R.~Coleman, V.~Glaser and A.~Martin,
  %``Action Minima Among Solutions to a Class of Euclidean Scalar Field Equations,''
  Commun.\ Math.\ Phys.\  {\bf 58} (1978) 211.
  %doi:10.1007/BF01609421
  %%CITATION = doi:10.1007/BF01609421;%%

%\cite{Blum:2016ipp}
\bibitem{Blum:2016ipp}
  K.~Blum, M.~Honda, R.~Sato, M.~Takimoto and K.~Tobioka,
  %``O(\(N\)) Invariance of the Multi-Field Bounce,''
  JHEP {\bf 1705} (2017) 109
  Erratum: [JHEP {\bf 1706} (2017) 060]
  %doi:10.1007/JHEP05(2017)109, 10.1007/JHEP06(2017)060
  \arxiv{1611.04570}[arXiv:1611.04570 [hep-th]].
  %%CITATION = doi:10.1007/JHEP05(2017)109, 10.1007/JHEP06(2017)060;%%

%\cite{Linde:1980tt}
\bibitem{Linde:1980tt}
  A.~D.~Linde,
  %``Fate of the False Vacuum at Finite Temperature: Theory and Applications,''
  Phys.\ Lett.\  {\bf 100B} (1981) 37.
  %doi:10.1016/0370-2693(81)90281-1
  %%CITATION = doi:10.1016/0370-2693(81)90281-1;%%
  %``Decay of the False Vacuum at Finite Temperature,''
  Nucl.\ Phys.\ B {\bf 216} (1983) 421
  Erratum: [Nucl.\ Phys.\ B {\bf 223} (1983) 544].
  %doi:10.1016/0550-3213(83)90293-6, 10.1016/0550-3213(83)90072-X
  %%CITATION = doi:10.1016/0550-3213(83)90293-6, 10.1016/0550-3213(83)90072-X;%%
  
%\cite{Witten:1984rs}
\bibitem{Witten:1984rs}
  E.~Witten,
  %``Cosmic Separation of Phases,''
  Phys.\ Rev.\ D {\bf 30} (1984) 272.
  %doi:10.1103/PhysRevD.30.272
  %%CITATION = doi:10.1103/PhysRevD.30.272;%%

%\cite{Hogan:1986qda}
\bibitem{Hogan:1986qda}
  C.~J.~Hogan,
  %``Gravitational radiation from cosmological phase transitions,''
  Mon.\ Not.\ Roy.\ Astron.\ Soc.\  {\bf 218} (1986) 629.
  %%CITATION = MNRAA,218,629;%%

%\cite{Duncan:1992ai}
\bibitem{Duncan:1992ai}
  M.~J.~Duncan and L.~G.~Jensen,
  %``Exact tunneling solutions in scalar field theory,''
  Phys.\ Lett.\ B {\bf 291} (1992) 109.
  %doi:10.1016/0370-2693(92)90128-Q
  %%CITATION = doi:10.1016/0370-2693(92)90128-Q;%%

%\cite{Dutta:2012qt}
\bibitem{Dutta:2012qt}
  K.~Dutta, C.~Hector, T.~Konstandin, P.~M.~Vaudrevange and A.~Westphal,
  %``Validity of the kink approximation to the tunneling action,''
  Phys.\ Rev.\ D {\bf 86} (2012) 123517
  %doi:10.1103/PhysRevD.86.123517
  \arxiv{1202.2721}[arXiv:1202.2721 [hep-th]].
  %%CITATION = doi:10.1103/PhysRevD.86.123517;%%

%\cite{Masoumi:2017trx}
\bibitem{Masoumi:2017trx}
  A.~Masoumi, K.~D.~Olum and J.~M.~Wachter,
  %``Approximating tunneling rates in multi-dimensional field spaces,''
  JCAP {\bf 1710} (2017) no.10,  022
  %doi:10.1088/1475-7516/2017/10/022
  \arxiv{1702.00356}[arXiv:1702.00356 [gr-qc]].
  %%CITATION = doi:10.1088/1475-7516/2017/10/022;%%

%\cite{Dutta:2011rc}
\bibitem{Dutta:2011rc}
  K.~Dutta, C.~Hector, P.~M.~Vaudrevange and A.~Westphal,
  %``More Exact Tunneling Solutions in Scalar Field Theory,''
  Phys.\ Lett.\ B {\bf 708} (2012) 309
  %doi:10.1016/j.physletb.2012.01.026
  \arxiv{1110.2380}[arXiv:1110.2380 [hep-th]].
  %%CITATION = doi:10.1016/j.physletb.2012.01.026;%%

%\cite{Pastras:2011zr}
\bibitem{Pastras:2011zr}
  G.~Pastras,
  %``Exact Tunneling Solutions in Minkowski Spacetime and a Candidate for Dark Energy,''
  JHEP {\bf 1308} (2013) 075
  %doi:10.1007/JHEP08(2013)075
  \arxiv{1102.4567}[arXiv:1102.4567 [hep-th]].
  %%CITATION = doi:10.1007/JHEP08(2013)075;%%

%\cite{Guada:2018jek}
\bibitem{Guada:2018jek}
  V.~Guada, A.~Maiezza and M.~Nemev\v{s}ek,
  %``Multifield Polygonal Bounces,''
  Phys.\ Rev.\ D {\bf 99} (2019) no.5,  056020
  %doi:10.1103/PhysRevD.99.056020
  \arxiv{1803.02227}[arXiv:1803.02227 [hep-th]].

%\cite{Kosowsky:1992rz}
\bibitem{Kosowsky:1992rz}
  A.~Kosowsky, M.~S.~Turner and R.~Watkins,
  %``Gravitational waves from first order cosmological phase transitions,''
  Phys.\ Rev.\ Lett.\  {\bf 69} (1992) 2026.
  %doi:10.1103/PhysRevLett.69.2026
  %%CITATION = doi:10.1103/PhysRevLett.69.2026;%%
  Phys.\ Rev.\ D {\bf 45} (1992) 4514.
  %doi:10.1103/PhysRevD.45.4514
  %%CITATION = doi:10.1103/PhysRevD.45.4514;%%
  
%\cite{Grojean:2006bp}
\bibitem{Grojean:2006bp}
  C.~Grojean and G.~Servant,
  %``Gravitational Waves from Phase Transitions at the Electroweak Scale and Beyond,''
  Phys.\ Rev.\ D {\bf 75} (2007) 043507
  %doi:10.1103/PhysRevD.75.043507
  \arxiv{hep-ph/0607107}[hep-ph/0607107].
  %%CITATION = doi:10.1103/PhysRevD.75.043507;%%

%\cite{Hindmarsh:2013xza}
\bibitem{Hindmarsh:2013xza}
  M.~Hindmarsh, S.~J.~Huber, K.~Rummukainen and D.~J.~Weir,
  %``Gravitational waves from the sound of a first order phase transition,''
  Phys.\ Rev.\ Lett.\  {\bf 112} (2014) 041301
  %doi:10.1103/PhysRevLett.112.041301
  \arxiv{1304.2433}[arXiv:1304.2433 [hep-ph]].
  %%CITATION = doi:10.1103/PhysRevLett.112.041301;%%
  %M.~Hindmarsh, S.~J.~Huber, K.~Rummukainen and D.~J.~Weir,
  %``Numerical simulations of acoustically generated gravitational waves at a first order phase transition,''
  Phys.\ Rev.\ D {\bf 92} (2015) no.12,  123009
  %doi:10.1103/PhysRevD.92.123009
  \arxiv{1504.03291}[arXiv:1504.03291 [astro-ph.CO]].
  %%CITATION = doi:10.1103/PhysRevD.92.123009;%%
  %M.~Hindmarsh, S.~J.~Huber, K.~Rummukainen and D.~J.~Weir,
  %``Shape of the acoustic gravitational wave power spectrum from a first order phase transition,''
  Phys.\ Rev.\ D {\bf 96} (2017) no.10,  103520
  %doi:10.1103/PhysRevD.96.103520
  \arxiv{1704.05871}[arXiv:1704.05871 [astro-ph.CO]].
  %%CITATION = doi:10.1103/PhysRevD.96.103520;%%

%\cite{Cutting:2018tjt}
\bibitem{Cutting:2018tjt}
  D.~Cutting, M.~Hindmarsh and D.~J.~Weir,
  %``Gravitational waves from vacuum first-order phase transitions: from the envelope to the lattice,''
  Phys.\ Rev.\ D {\bf 97} (2018) no.12,  123513
  %doi:10.1103/PhysRevD.97.123513
  \arxiv{1802.05712}[arXiv:1802.05712 [astro-ph.CO]] and
  %%CITATION = doi:10.1103/PhysRevD.97.123513;%%
  %``Vorticity, kinetic energy, and suppressed gravitational wave production in strong first order phase transitions,''
  \Arxiv{1906.00480}[arXiv:1906.00480 [hep-ph]].
  %%CITATION = ARXIV:1906.00480;%%
        
%\cite{Caprini:2019egz}
\bibitem{Caprini:2019egz}
  C.~Caprini {\it et al.},
  %``Science with the space-based interferometer eLISA. II: Gravitational waves from cosmological phase transitions,''
  JCAP {\bf 1604} (2016) 001
  %doi:10.1088/1475-7516/2016/04/001
  \arxiv{1512.06239}[arXiv:1512.06239 [astro-ph.CO]],
  JCAP \textbf{03} (2020), 024
  %doi:10.1088/1475-7516/2020/03/024
  \arxiv{1910.13125}[arXiv:1910.13125 [astro-ph.CO]].
  %%CITATION = ARXIV:1910.13125;%%
  
%\cite{Bochkarev:1990fx}
\bibitem{Bochkarev:1990fx}
  A.~I.~Bochkarev, S.~V.~Kuzmin and M.~E.~Shaposhnikov,
  %``Electroweak baryogenesis and the Higgs boson mass problem,''
  Phys.\ Lett.\ B {\bf 244} (1990) 275.
  %doi:10.1016/0370-2693(90)90069-I
  %%CITATION = doi:10.1016/0370-2693(90)90069-I;%%

%\cite{Cohen:1990py}
\bibitem{Cohen:1990py}
  A.~G.~Cohen, D.~B.~Kaplan and A.~E.~Nelson,
  %``Weak Scale Baryogenesis,''
  Phys.\ Lett.\ B {\bf 245} (1990) 561.
  %doi:10.1016/0370-2693(90)90690-8
  %%CITATION = doi:10.1016/0370-2693(90)90690-8;%%
% %\cite{Cohen:1990it}
%\bibitem{Cohen:1990it}
%  A.~G.~Cohen, D.~B.~Kaplan and A.~E.~Nelson,
  %``Baryogenesis at the weak phase transition,''
  Nucl.\ Phys.\ B {\bf 349} (1991) 727.
  %doi:10.1016/0550-3213(91)90395-E
  %%CITATION = doi:10.1016/0550-3213(91)90395-E;%%  
  
%\cite{Turok:1990zg}
\bibitem{Turok:1990zg}
  N.~Turok and J.~Zadrozny,
  %``Electroweak baryogenesis in the two doublet model,''
  Nucl.\ Phys.\ B {\bf 358} (1991) 471.
  %doi:10.1016/0550-3213(91)90356-3
  %%CITATION = doi:10.1016/0550-3213(91)90356-3;%%
  %\cite{Turok:1990in}
%  %``Dynamical generation of baryons at the electroweak transition,''
  Phys.\ Rev.\ Lett.\  {\bf 65} (1990) 2331.
  %doi:10.1103/PhysRevLett.65.2331
  %%CITATION = doi:10.1103/PhysRevLett.65.2331;%%
 
%\cite{Morrissey:2012db}
\bibitem{Morrissey:2012db}
  D.~E.~Morrissey and M.~J.~Ramsey-Musolf,
  %``Electroweak baryogenesis,''
  New J.\ Phys.\  {\bf 14} (2012) 125003
  %doi:10.1088/1367-2630/14/12/125003
  \arxiv{1206.2942}[arXiv:1206.2942 [hep-ph]].
  %%CITATION = doi:10.1088/1367-2630/14/12/125003;%%
   
%\cite{Vachaspati:1991nm}
\bibitem{Vachaspati:1991nm}
  T.~Vachaspati,
  %``Magnetic fields from cosmological phase transitions,''
  Phys.\ Lett.\ B {\bf 265} (1991) 258.
  %doi:10.1016/0370-2693(91)90051-Q
  %%CITATION = doi:10.1016/0370-2693(91)90051-Q;%%
  
%\cite{Sigl:1996dm}
\bibitem{Sigl:1996dm}
  G.~Sigl, A.~V.~Olinto and K.~Jedamzik,
  %``Primordial magnetic fields from cosmological first order phase transitions,''
  Phys.\ Rev.\ D {\bf 55} (1997) 4582
  %doi:10.1103/PhysRevD.55.4582
  \arxiv{astro-ph/9610201}[astro-ph/9610201].
  %%CITATION = doi:10.1103/PhysRevD.55.4582;%%
  
%\cite{DeSimone:2011ek}
\bibitem{DeSimone:2011ek}
  A.~De Simone, G.~Nardini, M.~Quiros and A.~Riotto,
  %``Magnetic Fields at First Order Phase Transition: A Threat to Electroweak Baryogenesis,''
  JCAP {\bf 1110} (2011) 030
  %doi:10.1088/1475-7516/2011/10/030
  \arxiv{1107.4317}[arXiv:1107.4317 [hep-ph]].
  %%CITATION = doi:10.1088/1475-7516/2011/10/030;%%
  
  %\cite{Tevzadze:2012kk}
\bibitem{Tevzadze:2012kk}
  A.~G.~Tevzadze, L.~Kisslinger, A.~Brandenburg and T.~Kahniashvili,
  %``Magnetic Fields from QCD Phase Transitions,''
  Astrophys.\ J.\  {\bf 759} (2012) 54
  %doi:10.1088/0004-637X/759/1/54
  \arxiv{1207.0751}[arXiv:1207.0751 [astro-ph.CO]].
  %%CITATION = doi:10.1088/0004-637X/759/1/54;%%
  
%\cite{Ellis:2019tjf}
\bibitem{Ellis:2019tjf}
  J.~Ellis, M.~Fairbairn, M.~Lewicki, V.~Vaskonen and A.~Wickens,
  %``Intergalactic Magnetic Fields from First-Order Phase Transitions,''
  JCAP {\bf 1909} (2019) no.09,  019
  %doi:10.1088/1475-7516/2019/09/019
  \arxiv{1907.04315}[arXiv:1907.04315 [astro-ph.CO]].
  %%CITATION = doi:10.1088/1475-7516/2019/09/019;%%    

 %\cite{Derrick:1964ww}
\bibitem{Derrick:1964ww}
  G.~H.~Derrick,
  %``Comments on nonlinear wave equations as models for elementary particles,''
  J.\ Math.\ Phys.\  {\bf 5} (1964) 1252.
  %doi:10.1063/1.1704233
  %%CITATION = doi:10.1063/1.1704233;%%

%\cite{Coleman:1980aw}
\bibitem{Coleman:1980aw}
  S.~R.~Coleman and F.~De Luccia,
  %``Gravitational Effects on and of Vacuum Decay,''
  Phys.\ Rev.\ D {\bf 21} (1980) 3305.
  %doi:10.1103/PhysRevD.21.3305
  %%CITATION = doi:10.1103/PhysRevD.21.3305;%%

%\cite{Isidori:2007vm}
\bibitem{Isidori:2007vm}
  G.~Isidori, V.~S.~Rychkov, A.~Strumia and N.~Tetradis,
  %``Gravitational corrections to standard model vacuum decay,''
  Phys.\ Rev.\ D {\bf 77} (2008) 025034
  %doi:10.1103/PhysRevD.77.025034
  \arxiv{0712.0242} [arXiv:0712.0242 [hep-ph]].
  %%CITATION = doi:10.1103/PhysRevD.77.025034;%%
  A.~Salvio, A.~Strumia, N.~Tetradis and A.~Urbano,
  %``On gravitational and thermal corrections to vacuum decay,''
  JHEP {\bf 1609} (2016) 054
  %doi:10.1007/JHEP09(2016)054
  \arxiv{1608.02555} [arXiv:1608.02555 [hep-ph]].
  %%CITATION = doi:10.1007/JHEP09(2016)054;%%

%\cite{Graham:2015cka}
\bibitem{Graham:2015cka}
  P.~W.~Graham, D.~E.~Kaplan and S.~Rajendran,
  %``Cosmological Relaxation of the Electroweak Scale,''
  Phys. Rev. Lett. \textbf{115} (2015) no.22, 221801
  %doi:10.1103/PhysRevLett.115.221801
  \arxiv{504.07551}[arXiv:1504.07551 [hep-ph]].

%\cite{Espinosa:2015eda}
\bibitem{Espinosa:2015eda}
  J.~Espinosa, C.~Grojean, G.~Panico, A.~Pomarol, O.~Pujolàs and G.~Servant,
  %``Cosmological Higgs-Axion Interplay for a Naturally Small Electroweak Scale,''
  Phys. Rev. Lett. \textbf{115} (2015) no.25, 251803
  %doi:10.1103/PhysRevLett.115.251803
  \arxiv{1506.09217}[arXiv:1506.09217 [hep-ph]].

%\cite{Hardy:2015laa}
\bibitem{Hardy:2015laa}
  E.~Hardy,
  %``Electroweak relaxation from finite temperature,''
  JHEP \textbf{11} (2015), 077
  %doi:10.1007/JHEP11(2015)077
  \arxiv{1507.07525}[arXiv:1507.07525 [hep-ph]].

%\cite{Patil:2015oxa}
\bibitem{Patil:2015oxa}
  S.~P.~Patil and P.~Schwaller,
  %``Relaxing the Electroweak Scale: the Role of Broken dS Symmetry,''
  JHEP \textbf{02} (2016), 077
  %doi:10.1007/JHEP02(2016)077
  \arxiv{1507.08649}[arXiv:1507.08649 [hep-ph]].

%\cite{Jaeckel:2015txa}
  \bibitem{Jaeckel:2015txa}
  J.~Jaeckel, V.~M.~Mehta and L.~T.~Witkowski,
  %``Musings on cosmological relaxation and the hierarchy problem,''
  Phys. Rev. D \textbf{93} (2016) no.6, 063522
  %doi:10.1103/PhysRevD.93.063522
  \arxiv{1508.03321}[arXiv:1508.03321 [hep-ph]].

%\cite{Gupta:2015uea}
\bibitem{Gupta:2015uea}
  R.~S.~Gupta, Z.~Komargodski, G.~Perez and L.~Ubaldi,
  %``Is the Relaxion an Axion?,''
  JHEP \textbf{02} (2016), 166
  %doi:10.1007/JHEP02(2016)166
  \arxiv{1509.00047}[arXiv:1509.00047 [hep-ph]].

%\cite{Batell:2015fma}
\bibitem{Batell:2015fma}
  B.~Batell, G.~F.~Giudice and M.~McCullough,
  %``Natural Heavy Supersymmetry,''
  JHEP \textbf{12} (2015), 162
  %doi:10.1007/JHEP12(2015)162
  \arxiv{1509.00834}[arXiv:1509.00834 [hep-ph]].

%\cite{Jaeckel:2016qjp}
\bibitem{Jaeckel:2016qjp}
  J.~Jaeckel, V.~M.~Mehta and L.~T.~Witkowski,
  %``Monodromy Dark Matter,''
  JCAP \textbf{01} (2017), 036
  %doi:10.1088/1475-7516/2017/01/036
  \arxiv{1605.01367}[arXiv:1605.01367 [hep-ph]].

%\cite{Fonseca:2019lmc}
\bibitem{Fonseca:2019lmc}
  N.~Fonseca, E.~Morgante, R.~Sato and G.~Servant,
  %``Relaxion Fluctuations (Self-stopping Relaxion) and Overview of Relaxion Stopping Mechanisms,''
  \Arxiv{1911.08473}[arXiv:1911.08473 [hep-ph]].
  %%CITATION = ARXIV:1911.08473;%%
  
%\cite{Brown:2011um}
\bibitem{Brown:2011um}
  A.~R.~Brown and A.~Dahlen,
  %``The Case of the Disappearing Instanton,''
  Phys.\ Rev.\ D {\bf 84} (2011) 105004
  %doi:10.1103/PhysRevD.84.105004
   \arxiv{1106.0527}[arXiv:1106.0527 [hep-th]].
  %%CITATION = doi:10.1103/PhysRevD.84.105004;%%

%\cite{Darme:2017wvu}
\bibitem{Darme:2017wvu}
  L.~Darm\'e, J.~Jaeckel and M.~Lewicki,
  %``Towards the fate of the oscillating false vacuum,''
  Phys.\ Rev.\ D {\bf 96} (2017) no.5,  056001
  %doi:10.1103/PhysRevD.96.056001
  \arxiv{1704.06445}[arXiv:1704.06445 [hep-ph]].
  %%CITATION = doi:10.1103/PhysRevD.96.056001;%%
  
  %\cite{Sarangi:2007jb}
\bibitem{Sarangi:2007jb} 
  S.~Sarangi, G.~Shiu and B.~Shlaer,
  %``Rapid Tunneling and Percolation in the Landscape,''
  Int.\ J.\ Mod.\ Phys.\ A {\bf 24}, 741 (2009)
  %doi:10.1142/S0217751X09042529
  \arxiv{0708.4375}[arXiv:0708.4375 [hep-th]].
  %%CITATION = doi:10.1142/S0217751X09042529;%%
  
  %\cite{Saffin:2008vi}
\bibitem{Saffin:2008vi} 
  P.~M.~Saffin, A.~Padilla and E.~J.~Copeland,
  %``Decay of an inhomogeneous state via resonant tunnelling,''
  JHEP {\bf 0809}, 055 (2008)
  %doi:10.1088/1126-6708/2008/09/055
  \arxiv{0804.3801}[arXiv:0804.3801 [hep-th]].
  %%CITATION = doi:10.1088/1126-6708/2008/09/055;%%
  
  %\cite{Tye:2009rb}
\bibitem{Tye:2009rb} 
  S.-H.~H.~Tye and D.~Wohns,
  %``Resonant Tunneling in Scalar Quantum Field Theory,''
  \Arxiv{0910.1088}[arXiv:0910.1088 [hep-th]].
  %%CITATION = ARXIV:0910.1088;%%
  
%\cite{Darme:2019ubo}
\bibitem{Darme:2019ubo}
  L.~Darmé, J.~Jaeckel and M.~Lewicki,
  %``Generalized escape paths for dynamical tunneling in QFT,''
  Phys.\ Rev.\ D {\bf 100} (2019) no.9,  096012
  %doi:10.1103/PhysRevD.100.096012
  \arxiv{1907.04865}[arXiv:1907.04865 [hep-th]].
  %%CITATION = doi:10.1103/PhysRevD.100.096012;%%
  
%\cite{Guada:2020}
\bibitem{Guada:2020}
  V.~Guada, M.~Nemev\v{s}ek and L.~Ubaldi,
  work in progress.

%\cite{Strumia:1998nf}
\bibitem{Strumia:1998nf}
  A.~Strumia and N.~Tetradis,
  %``A Consistent calculation of bubble nucleation rates,''
  Nucl.\ Phys.\ B {\bf 542} (1999) 719
  %doi:10.1016/S0550-3213(98)00804-9
  \arxiv{hep-ph/9806453}[hep-ph/9806453].
  %%CITATION = doi:10.1016/S0550-3213(98)00804-9;%%

%\cite{Munster:1999hr}
\bibitem{Munster:1999hr}
  G.~Munster and S.~Rotsch,
  %``Analytical calculation of the nucleation rate for first order phase transitions beyond the thin wall approximation,''
  Eur.\ Phys.\ J.\ C {\bf 12} (2000) 161
  %doi:10.1007/s100529900242
  \arxiv{cond-mat/9908246}[cond-mat/9908246].
  %%CITATION = doi:10.1007/s100529900242;%%

%\cite{Baacke:2003uw}
\bibitem{Baacke:2003uw}
  J.~Baacke and G.~Lavrelashvili,
  %``One loop corrections to the metastable vacuum decay,''
  Phys.\ Rev.\ D {\bf 69} (2004) 025009
  %doi:10.1103/PhysRevD.69.025009
  \arxiv{hep-th/0307202}[hep-th/0307202].
  %%CITATION = doi:10.1103/PhysRevD.69.025009;%%

%\cite{Dunne:2005rt}
\bibitem{Dunne:2005rt}
  G.~V.~Dunne and H.~Min,
  %``Beyond the thin-wall approximation: Precise numerical computation of prefactors in false vacuum decay,''
  Phys.\ Rev.\ D {\bf 72} (2005) 125004
  %doi:10.1103/PhysRevD.72.125004
  \arxiv{hep-th/0511156}[hep-th/0511156].
  %%CITATION = doi:10.1103/PhysRevD.72.125004;%%

%\cite{Andreassen:2016cvx}
\bibitem{Andreassen:2016cvx}
  A.~Andreassen, D.~Farhi, W.~Frost and M.~D.~Schwartz,
  %``Precision decay rate calculations in quantum field theory,''
  Phys.\ Rev.\ D {\bf 95} (2017) no.8,  085011
  %doi:10.1103/PhysRevD.95.085011
  \arxiv{1604.06090}[arXiv:1604.06090 [hep-th]].
  %%CITATION = doi:10.1103/PhysRevD.95.085011;%%

  %\cite{Degrassi:2012ry}
\bibitem{Degrassi:2012ry} 
  G.~Degrassi, S.~Di Vita, J.~Elias-Miro, J.~R.~Espinosa, G.~F.~Giudice, G.~Isidori and A.~Strumia,
  %``Higgs mass and vacuum stability in the Standard Model at NNLO,''
  JHEP {\bf 1208}, 098 (2012)
  %doi:10.1007/JHEP08(2012)098
  \arxiv{1205.6497}[arXiv:1205.6497 [hep-ph]].
  %%CITATION = doi:10.1007/JHEP08(2012)098;%%
  
  %\cite{Buttazzo:2013uya}
\bibitem{Buttazzo:2013uya} 
  D.~Buttazzo, G.~Degrassi, P.~P.~Giardino, G.~F.~Giudice, F.~Sala, A.~Salvio and A.~Strumia,
  %``Investigating the near-criticality of the Higgs boson,''
  JHEP {\bf 1312}, 089 (2013)
  %doi:10.1007/JHEP12(2013)089
  \arxiv{1307.3536}[arXiv:1307.3536 [hep-ph]].
  %%CITATION = doi:10.1007/JHEP12(2013)089;%%
  
  %\cite{Chigusa:2017dux}
\bibitem{Chigusa:2017dux} 
  S.~Chigusa, T.~Moroi and Y.~Shoji,
  %``State-of-the-Art Calculation of the Decay Rate of Electroweak Vacuum in the Standard Model,''
  Phys.\ Rev.\ Lett.\  {\bf 119}, no. 21, 211801 (2017)
  %doi:10.1103/PhysRevLett.119.211801
  \arxiv{1707.09301}[arXiv:1707.09301 [hep-ph]].
  %%CITATION = doi:10.1103/PhysRevLett.119.211801;%%

%\cite{Andreassen:2017rzq}
\bibitem{Andreassen:2017rzq} 
  A.~Andreassen, W.~Frost and M.~D.~Schwartz,
  %``Scale Invariant Instantons and the Complete Lifetime of the Standard Model,''
  Phys.\ Rev.\ D {\bf 97}, no. 5, 056006 (2018)
  %doi:10.1103/PhysRevD.97.056006
  \arxiv{1707.08124}[arXiv:1707.08124 [hep-ph]].
  %%CITATION = doi:10.1103/PhysRevD.97.056006;%%

%\cite{Chigusa:2018uuj}
\bibitem{Chigusa:2018uuj}
  S.~Chigusa, T.~Moroi and Y.~Shoji,
  %``Decay Rate of Electroweak Vacuum in the Standard Model and Beyond,''
  Phys.\ Rev.\ D {\bf 97} (2018) no.11,  116012
  %doi:10.1103/PhysRevD.97.116012
  \arxiv{1803.03902}[arXiv:1803.03902 [hep-ph]].
  %%CITATION = doi:10.1103/PhysRevD.97.116012;%%

%\cite{Espinosa:2019hbm}
\bibitem{Espinosa:2019hbm} 
  J.~R.~Espinosa,
  %``Tunneling without Bounce,''
  Phys.\ Rev.\ D {\bf 100}, no. 10, 105002 (2019)
  %doi:10.1103/PhysRevD.100.105002
  \arxiv{1908.01730}[arXiv:1908.01730 [hep-th]].
  %%CITATION = doi:10.1103/PhysRevD.100.105002;%%

%%%Two Higgs Double Model========
%\cite{Branco:2011iw}
\bibitem{Branco:2011iw}
  G.~Branco, P.~Ferreira, L.~Lavoura, M.~Rebelo, M.~Sher and J.~P.~Silva,
  %``Theory and phenomenology of two-Higgs-doublet models,''
  Phys. Rept. \textbf{516} (2012), 1-102
  %doi:10.1016/j.physrep.2012.02.002
  \arxiv{1106.0034}[arXiv:1106.0034 [hep-ph]].

%\cite{Huber:2000mg}
\bibitem{Huber:2000mg} 
  S.~J.~Huber and M.~G.~Schmidt,
  %``Electroweak baryogenesis: Concrete in a SUSY model with a gauge singlet,''
  Nucl.\ Phys.\ B {\bf 606}, 183 (2001)
  %doi:10.1016/S0550-3213(01)00250-4
  \arxiv{hep-ph/0003122}[hep-ph/0003122].
  %%CITATION = doi:10.1016/S0550-3213(01)00250-4;%%

%\cite{Huber:2006wf}
\bibitem{Huber:2006wf} 
  S.~J.~Huber, T.~Konstandin, T.~Prokopec and M.~G.~Schmidt,
  %``Electroweak Phase Transition and Baryogenesis in the nMSSM,''
  Nucl.\ Phys.\ B {\bf 757}, 172 (2006)
  %doi:10.1016/j.nuclphysb.2006.09.003
  \arxiv{hep-ph/0606298}[hep-ph/0606298].
  %%CITATION = doi:10.1016/j.nuclphysb.2006.09.003;%%
  
%\cite{Patel:2011th}
\bibitem{Patel:2011th} 
  H.~H.~Patel and M.~J.~Ramsey-Musolf,
  %``Baryon Washout, Electroweak Phase Transition, and Perturbation Theory,''
  JHEP {\bf 1107}, 029 (2011)
  %doi:10.1007/JHEP07(2011)029
  \arxiv{1101.4665}[arXiv:1101.4665 [hep-ph]].
  %%CITATION = doi:10.1007/JHEP07(2011)029;%%
    
%\cite{Cline:2012hg}
\bibitem{Cline:2012hg} 
  J.~M.~Cline and K.~Kainulainen,
  %``Electroweak baryogenesis and dark matter from a singlet Higgs,''
  JCAP {\bf 1301}, 012 (2013)
  %doi:10.1088/1475-7516/2013/01/012
  \arxiv{1210.4196}[arXiv:1210.4196 [hep-ph]].
  %%CITATION = doi:10.1088/1475-7516/2013/01/012;%%

%\cite{Chala:2016ykx}
\bibitem{Chala:2016ykx} 
  M.~Chala, G.~Nardini and I.~Sobolev,
  %``Unified explanation for dark matter and electroweak baryogenesis with direct detection and gravitational wave signatures,''
  Phys.\ Rev.\ D {\bf 94}, no. 5, 055006 (2016)
  %doi:10.1103/PhysRevD.94.055006
  \arxiv{1605.08663}[arXiv:1605.08663 [hep-ph]].
  %%CITATION = doi:10.1103/PhysRevD.94.055006;%%

%%%Dark matter====================  

%\cite{Burgess:2000yq}
\bibitem{Burgess:2000yq} 
  C.~P.~Burgess, M.~Pospelov and T.~ter Veldhuis,
  %``The Minimal model of nonbaryonic dark matter: A Singlet scalar,''
  Nucl.\ Phys.\ B {\bf 619}, 709 (2001)
  %doi:10.1016/S0550-3213(01)00513-2
  \arxiv{hep-ph/0011335}[hep-ph/0011335].
  %%CITATION = doi:10.1016/S0550-3213(01)00513-2;%%
 
%\cite{McDonald:2001vt}
\bibitem{McDonald:2001vt} 
  J.~McDonald,
  %``Thermally generated gauge singlet scalars as selfinteracting dark matter,''
  Phys.\ Rev.\ Lett.\  {\bf 88}, 091304 (2002)
  %doi:10.1103/PhysRevLett.88.091304
  \arxiv{hep-ph/0106249}[hep-ph/0106249].
  %%CITATION = doi:10.1103/PhysRevLett.88.091304;%%

%\cite{Gonderinger:2009jp}
\bibitem{Gonderinger:2009jp} 
  M.~Gonderinger, Y.~Li, H.~Patel and M.~J.~Ramsey-Musolf,
  %``Vacuum Stability, Perturbativity, and Scalar Singlet Dark Matter,''
  JHEP {\bf 1001}, 053 (2010)
  %doi:10.1007/JHEP01(2010)053
  \arxiv{0910.3167}[arXiv:0910.3167 [hep-ph]].
  %%CITATION = doi:10.1007/JHEP01(2010)053;%%
       
%\cite{Cline:2013gha}
\bibitem{Cline:2013gha} 
  J.~M.~Cline, K.~Kainulainen, P.~Scott and C.~Weniger,
  %``Update on scalar singlet dark matter,''
  Phys.\ Rev.\ D {\bf 88}, 055025 (2013)
  Erratum: [Phys.\ Rev.\ D {\bf 92}, no. 3, 039906 (2015)]
  %doi:10.1103/PhysRevD.92.039906, 10.1103/PhysRevD.88.055025
  \arxiv{1306.4710}[arXiv:1306.4710 [hep-ph]].
  %%CITATION = doi:10.1103/PhysRevD.92.039906, 10.1103/PhysRevD.88.055025;%%

%%%Gravitational waves=================

%\cite{Caprini:2009yp}
\bibitem{Caprini:2009yp} 
  C.~Caprini, R.~Durrer and G.~Servant,
  %``The stochastic gravitational wave background from turbulence and magnetic fields generated by a first-order phase transition,''
  JCAP {\bf 0912}, 024 (2009)
  %doi:10.1088/1475-7516/2009/12/024
  \arxiv{0909.0622}[arXiv:0909.0622 [astro-ph.CO]].
  %%CITATION = doi:10.1088/1475-7516/2009/12/024;%%

%\cite{Espinosa:2010hh}
\bibitem{Espinosa:2010hh} 
  J.~R.~Espinosa, T.~Konstandin, J.~M.~No and G.~Servant,
  %``Energy Budget of Cosmological First-order Phase Transitions,''
  JCAP {\bf 1006}, 028 (2010)
  %doi:10.1088/1475-7516/2010/06/028
  \arxiv{1004.4187}[arXiv:1004.4187 [hep-ph]].
  %%CITATION = doi:10.1088/1475-7516/2010/06/028;%%
    
%\cite{Binetruy:2012ze}
\bibitem{Binetruy:2012ze} 
  P.~Binetruy, A.~Bohe, C.~Caprini and J.~F.~Dufaux,
  %``Cosmological Backgrounds of Gravitational Waves and eLISA/NGO: Phase Transitions, Cosmic Strings and Other Sources,''
  JCAP {\bf 1206}, 027 (2012)
  %doi:10.1088/1475-7516/2012/06/027
  \arxiv{1201.0983}[arXiv:1201.0983 [gr-qc]].
  %%CITATION = doi:10.1088/1475-7516/2012/06/027;%%

%\cite{Thrane:2013oya}
\bibitem{Thrane:2013oya} 
  E.~Thrane and J.~D.~Romano,
  %``Sensitivity curves for searches for gravitational-wave backgrounds,''
  Phys.\ Rev.\ D {\bf 88}, no. 12, 124032 (2013)
  %doi:10.1103/PhysRevD.88.124032
 \arxiv{1310.5300}[arXiv:1310.5300 [astro-ph.IM]].
  %%CITATION = doi:10.1103/PhysRevD.88.124032;%%
      
%\cite{Ellis:2019oqb}
\bibitem{Ellis:2019oqb} 
  J.~Ellis, M.~Lewicki, J.~M.~No and V.~Vaskonen,
  %``Gravitational wave energy budget in strongly supercooled phase transitions,''
  JCAP {\bf 1906}, 024 (2019)
  %doi:10.1088/1475-7516/2019/06/024
  \arxiv{1903.09642}[arXiv:1903.09642 [hep-ph]].
  %%CITATION = doi:10.1088/1475-7516/2019/06/024;%%
  
%%%Finite Temperature===================

  %\cite{Dolan:1973qd}
\bibitem{Dolan:1973qd} 
  L.~Dolan and R.~Jackiw,
  %``Symmetry Behavior at Finite Temperature,''
  Phys.\ Rev.\ D {\bf 9}, 3320 (1974).
  %doi:10.1103/PhysRevD.9.3320
  %%CITATION = doi:10.1103/PhysRevD.9.3320;%%
  
  %\cite{Weinberg:1974hy}
\bibitem{Weinberg:1974hy} 
  S.~Weinberg,
  %``Gauge and Global Symmetries at High Temperature,''
  Phys.\ Rev.\ D {\bf 9}, 3357 (1974).
  %doi:10.1103/PhysRevD.9.3357
  %%CITATION = doi:10.1103/PhysRevD.9.3357;%%
  
  %\cite{Quiros:1999jp}
\bibitem{Quiros:1999jp} 
  M.~Quiros,
  %``Finite temperature field theory and phase transitions,''
  \arxiv{hep-ph/9901312}[hep-ph/9901312].
  %%CITATION = HEP-PH/9901312;%%

  %\cite{Kapusta:2006pm}
\bibitem{Kapusta:2006pm} 
  J.~I.~Kapusta and C.~Gale,
  %``Finite-temperature field theory: Principles and applications,''
  doi:10.1017/CBO9780511535130.
  %%CITATION = doi:10.1017/CBO9780511535130;%%
  
%%==========================================  
  
%\cite{Fowlie:2018eiu}
\bibitem{Fowlie:2018eiu} 
  A.~Fowlie,
  %``A fast C++ implementation of thermal functions,''
  Comput.\ Phys.\ Commun.\  {\bf 228}, 264 (2018)
  %doi:10.1016/j.cpc.2018.02.015
  \arxiv{1802.02720}[arXiv:1802.02720 [hep-ph]].
  %%CITATION = doi:10.1016/j.cpc.2018.02.015;%%
  
  %\cite{Espinosa:2011ax}
\bibitem{Espinosa:2011ax} 
  J.~R.~Espinosa, T.~Konstandin and F.~Riva,
  %``Strong Electroweak Phase Transitions in the Standard Model with a Singlet,''
  Nucl.\ Phys.\ B {\bf 854}, 592 (2012)
  %doi:10.1016/j.nuclphysb.2011.09.010
  \arxiv{1107.5441}[arXiv:1107.5441 [hep-ph]].
  %%CITATION = doi:10.1016/j.nuclphysb.2011.09.010;%%
  
  %\cite{Anderson:1991zb}
\bibitem{Anderson:1991zb} 
  G.~W.~Anderson and L.~J.~Hall,
  %``The Electroweak phase transition and baryogenesis,''
  Phys.\ Rev.\ D {\bf 45}, 2685 (1992).
  %doi:10.1103/PhysRevD.45.2685
  %%CITATION = doi:10.1103/PhysRevD.45.2685;%%
    
  %\cite{Camargo-Molina:2013qva}
\bibitem{Camargo-Molina:2013qva}
  J.~E.~Camargo-Molina, B.~O'Leary, W.~Porod and F.~Staub,
  %``$\mathbf{Vevacious}$: A Tool For Finding The Global Minima Of One-Loop Effective Potentials With Many Scalars,''
  Eur.\ Phys.\ J.\ C {\bf 73} (2013) no.10,  2588
  %doi:10.1140/epjc/s10052-013-2588-2
  \arxiv{1307.1477}[arXiv:1307.1477 [hep-ph]].
  %%CITATION = doi:10.1140/epjc/s10052-013-2588-2;%%    
    
  %\cite{Konoplich:1986zp}
\bibitem{Konoplich:1986zp}
  R.~V.~Konoplich and S.~G.~Rubin,
  %``Decay Probability For Metastable Vacuum In Scalar Theory. (in Russian),''
  Yad.\ Fiz.\  {\bf 42} (1985) 1282.
  %%CITATION = YAFIA,42,1282;%%
  R.~V.~Konoplich,
  %``Calculation of Quantum Corrections to Nontrivial Classical Solutions by Means of the Zeta Function,''
  Theor.\ Math.\ Phys.\  {\bf 73} (1987) 1286
   [Teor.\ Mat.\ Fiz.\  {\bf 73} (1987) 379].
  %doi:10.1007/BF01041911
  %%CITATION = doi:10.1007/BF01041911;%%
    
  %\cite{Coleman:1973jx}
\bibitem{Coleman:1973jx} 
  S.~R.~Coleman and E.~J.~Weinberg,
  %``Radiative Corrections as the Origin of Spontaneous Symmetry Breaking,''
  Phys.\ Rev.\ D {\bf 7}, 1888 (1973).
  %doi:10.1103/PhysRevD.7.1888
  %%CITATION = doi:10.1103/PhysRevD.7.1888;%%
  
  %\cite{Jackiw:1974cv}
\bibitem{Jackiw:1974cv} 
  R.~Jackiw,
  %``Functional evaluation of the effective potential,''
  Phys.\ Rev.\ D {\bf 9}, 1686 (1974).
  %doi:10.1103/PhysRevD.9.1686
  %%CITATION = doi:10.1103/PhysRevD.9.1686;%% 
  
  %\cite{Iliopoulos:1974ur}
\bibitem{Iliopoulos:1974ur} 
  J.~Iliopoulos, C.~Itzykson and A.~Martin,
  %``Functional Methods and Perturbation Theory,''
  Rev.\ Mod.\ Phys.\  {\bf 47}, 165 (1975).
  %doi:10.1103/RevModPhys.47.165
  %%CITATION = doi:10.1103/RevModPhys.47.165;%%
  
  %\cite{Andreassen:2014eha}
\bibitem{Andreassen:2014eha} 
  A.~Andreassen, W.~Frost and M.~D.~Schwartz,
  %``Consistent Use of Effective Potentials,''
  Phys.\ Rev.\ D {\bf 91}, no. 1, 016009 (2015)
  %doi:10.1103/PhysRevD.91.016009
  \arxiv{1408.0287}[arXiv:1408.0287 [hep-ph]],
  %%CITATION = doi:10.1103/PhysRevD.91.016009;%%
  %``Consistent Use of the Standard Model Effective Potential,''
  Phys.\ Rev.\ Lett.\  {\bf 113}, no. 24, 241801 (2014)
  %doi:10.1103/PhysRevLett.113.241801
  \arxiv{1408.0292}[arXiv:1408.0292 [hep-ph]].
  %%CITATION = doi:10.1103/PhysRevLett.113.241801;%%
    
\end{thebibliography}
\def\arxiv#1[#2]{\href{http://arxiv.org/abs/#1}{[#2]}}
\def\Arxiv#1[#2]{\href{http://arxiv.org/abs/#1}{#2}}

\end{document}